\pdfoutput=1
\PassOptionsToPackage{dvipsnames}{xcolor}
\documentclass[11pt,a4paper]{article}
\usepackage[utf8]{inputenc}  
\usepackage{url}
\usepackage{jcappub}
\usepackage{aas_macros}

\usepackage{amsfonts}
\usepackage{subfigure}
\usepackage{appendix}
\usepackage{underscore}
\usepackage{breakcites}

\usepackage{epsfig}  
\usepackage{graphicx}   
\usepackage{slashed}       
\usepackage{tikz}
\usepackage{multirow}
\usepackage{comment}
\usepackage{amssymb}
\usepackage[capitalize]{cleveref}
\usepackage{bm}
\usepackage{soul}
\usepackage{orcidlink}
\usepackage[normalem]{ulem}
\usepackage{relsize}

\DeclareMathOperator{\cm}{cm}

\DeclareMathOperator{\MeV}{MeV}

\DeclareMathOperator{\erg}{erg}

\newcommand{\Mo}{^{92,94}{\rm Mo}}
\newcommand{\Ru}{^{96,98}{\rm Ru}}
\newcommand{\Nb}{^{92}{\rm Nb}}
\newcommand{\Tc}{^{98}{\rm Tc}}

\newcommand{\Otsuki}{\emph{Otsuki+}}
\newcommand{\Thompson}{\emph{Thompson+}}
\newcommand{\CF}{\emph{Cardall}\&\emph{Fuller}}

\allowdisplaybreaks

\bibpunct{[}{]}{,}{n}{}{,}
\definecolor{ForestGreen}{RGB}{34,139,34}

\begin{document}

\begin{flushright} 
\noindent SLAC-PUB-250711
\end{flushright}

\title{$\nu p$-process in Core-Collapse Supernovae: Imprints of General Relativistic Effects}

\author[a]{Alexander Friedland~\orcidlink{0000-0002-5047-4680},\,}
\emailAdd{alexfr@slac.stanford.edu}

\author[a,b]{Derek J. Li~\orcidlink{0000-0001-6040-251X},\,}
\emailAdd{djpli@stanford.edu}

\author[a]{Giuseppe Lucente~\orcidlink{0000-0003-1530-4851},\,}
\emailAdd{lucenteg@slac.stanford.edu}

\author[a,c,d]{Ian Padilla-Gay~\orcidlink{0000-0003-2472-3863},\,}
\emailAdd{ianpaga@berkeley.edu}

\author[a,e,f,g]{and Amol V.\ Patwardhan~\orcidlink{0000-0002-2281-799X}\,}
\emailAdd{apatwardhan@reed.edu}

\affiliation[a]{SLAC National Accelerator Laboratory, Stanford University, Menlo Park, CA 94025}

\affiliation[b]{Leinweber Institute for Theoretical Physics, Stanford University, Stanford, CA 94305}

\affiliation[c]{Department of Physics, University of California Berkeley, Berkeley, CA 94720}

\affiliation[d]{Department of Physics, University of California San Diego, 
La Jolla, CA 92093}

\affiliation[e]{%
School of Physics and Astronomy, University of Minnesota, Minneapolis, MN 55455
}
\affiliation[f]{%
Department of Physics, New York Institute of Technology, New York, NY 10023
}
\affiliation[g]{%
Department of Physics, Reed College, Portland, OR 97202
}

\abstract{\textcolor{black}{The origin of a number of proton-rich isotopes in the solar system has been a long-standing puzzle. A promising explanation is the $\nu p$-process, which is posited to operate in the neutrino-driven outflows that form inside core-collapse supernovae after shock revival. While recent studies have analyzed several relevant physical effects that influence the efficiency of this process, the impact of General Relativity (GR) on it remains unexplored. We perform a comparative analysis of the time-integrated $\nu p$-process yields in Newtonian and fully GR calculations, using detailed models of time-evolving outflow profiles. The GR effects are seen to suppress the production of seed nuclei, significantly boosting the resulting $p$-nuclide abundances. Our reference GR model, with an 18~$M_\odot$ progenitor, reproduces both the relative and absolute solar system abundances of the entire set of the $p$ nuclides in the mass range $74\leq A\leq102$. The yields are suboptimal in our 12.75~$M_\odot$ GR model, where the outflow transitions to the supersonic regime several seconds into the explosion, suppressing further $p$-nuclide production. In both models, most of the production of the crucial $^{92,94}{\rm Mo}$ and $^{96,98}{\rm Ru}$ $p$ isotopes occurs relatively early, 1--3 seconds after shock revival. In contrast, a large fraction of the shielded isotope $^{92}{\rm Nb}$ is produced in the subsequent ejecta. The impact of GR on this isotope is especially large, with its final abundance boosted by a factor of 25 compared to a Newtonian calculation.  In summary, with the GR effects taken into account, the $\nu p$-process in a sufficiently massive progenitor can provide a unifying explanation for the origin of all $p$ nuclei in the solar system up to $^{102}$Pd. }}

\maketitle
\date{\today}

\section{\textcolor{black}{Introduction}}

A vast majority of the stable nuclides in the Universe are produced via \lq\lq slow\rq\rq\ neutron captures ($s$-process) or \lq\lq rapid\rq\rq\ neutron captures ($r$-process) in astrophysical sources~\cite{Burbidge:1957vc, Cameron:1958vx, Arnould:2007gh, Cowan:2019pkx, Kajino:2019abv, Kaeppeler:2010kk, Lugaro:2023qli, Thielemann:2007, Sneden:2008, Nomoto:2013, Arcones:2023, Fischer:2024}. Most of these nuclides lie on the neutron-rich side of the valley of stability in the nuclide chart. Proton-rich nuclides ($p$ nuclides) on the other hand are more challenging to produce because they cannot be synthesized directly through a sequence of neutron captures interspersed with beta decays: they are shielded from beta decays by stable isotopes and are thus inaccessible via the nucleosynthesis paths of the $s$- or $r$-processes~\cite{ArnouldGoriely2003}. 

Such proton-rich isotopes are observed in nature, for example, in meteoritic samples from our solar system. Roughly 35 $p$ nuclides have been identified~\cite{Anders:1989, Lodders:2003}, with $^{74}{\rm Se}$ being the lightest and $^{196}{\rm Hg}$ the heaviest. 
While most $p$ nuclides are 1--2 orders of magnitude less abundant than nearby $s$- and $r$-process nuclides~\cite{Meyer:1994}, some---most famously $\Mo$ and $\Ru$---contribute significant fractions to their overall elemental abundances~\cite{ArnouldGoriely2003, Rauscher:2013}. Historically, a majority of the attempts to reconcile observed $p$-nuclide abundances have considered a secondary process---the $\gamma$ process---involving photodisintegration of $s$-process seeds into $p$-rich nuclides via neutron removal~\cite{Woosley:1978, Howard:1991, Travaglio:2011, Travaglio:2014, Rauscher:2014fea}. However, these proposals generally fall short of being able to account for the entirety of the observed Mo and Ru $p$-rich isotopic abundances in the solar system~\cite{Rauscher:2013, Rauscher:2014fea}. 

These difficulties have motivated investigations of an alternative framework, involving a primary process that functions like the isospin mirror of the $r$-process, namely, a sequence of rapid proton captures and inverse beta decays. 
The most straightforward realization of this framework, known as the $rp$-process~\cite{Schatz:1998zz}, brings up the following considerations: (i) proton captures need to overcome Coulomb repulsion which halts their absorption below a certain temperature; (ii) on the other hand, the temperature needs to be below $\lesssim 3$ GK, to avoid the conditions of quasi-statistical equilibrium, at which the seeds heavier than the iron group are photodisintegrated; and (iii) certain nuclides along the $rp$-process reaction pathway have long beta decay lifetimes (e.g., $^{64}$Ge takes approximately 63 seconds to decay) and hence act as \lq\lq waiting points\rq\rq. The difficulty arises because the time-scales available in astrophysical systems that transition through the desired temperature window are too short to accommodate the waiting points.

Here, we will focus on another mechanism, which we regard to be the most promising solution to this long-standing problem: the $\nu p$-process~\cite{Frohlich:2005ys,Wanajo:2006rp,Pruet:2005qd}. This process occurs in the presence of a strong flux of neutrinos and antineutrinos. The antineutrinos play a crucial role: their capture on free protons furnishes a neutron supply in the otherwise proton-rich medium. The resulting neutrons immediately enter the neutron-proton $(n,p)$ exchange reactions with seed nuclei, circumventing the need to wait for the beta decays~\cite{Frohlich:2005ys,Wanajo:2006rp,Pruet:2005qd,Wanajo:2010mc}. 

The $\nu p$-process has been proposed to occur in a core-collapse supernova (CCSN) explosion, in neutrino-driven outflows (NDOs) from the surface of the proto-neutron star (PNS). Such outflows form in the hot bubble region surrounding the PNS after the shock revival~\cite{Duncan1986,Qian:1996xt,Otsuki:1999kb,Thompson:2001} and their nature depends on the interplay between the PNS gravity, the neutrino heating near the PNS surface and the confining pressure of the matter surrounding the hot bubble~\cite{Friedland:2020ecy}. The outflow lasts for the duration of the neutrino burst, which is determined by the Helmholtz cooling time of the PNS and is typically 10--15 seconds~\cite{Mirizzi:2015eza}. To compute the nucleosynthetic yields, one has to model the time evolution of this system, run the reaction network calculation on the time-dependent particle trajectories and integrate the results over time. This can be done either by taking a full simulation of the explosion and postprocessing it, or by creating a realistic, semi-analytic time-evolving model of the outflows. The latter approach has the advantage that it allows one to explore the sensitivity of the yields on various physics ingredients.

Over time, several such physical ingredients have been identified. The first and most obvious is the electron fraction in the NDO, $Y_e$. One must have at least $Y_e > 0.5$ to assure proton-rich conditions, with values $Y_e \gtrsim 0.55$ practically required to have a sufficient number fraction of free protons to drive proton-rich nucleosynthesis. Fortunately, such numbers appear in many recent simulations, with details controlled by the luminosities and mean energies of $\nu_e$ and $\bar{\nu}_e$. Other relevant quantities are the lengths of time spent in the seed nuclei formation window (3--6 GK) and in the window following freeze-out from nuclear quasi-statistical equilibrium (QSE) between the iron group seeds and heavier nuclei ($1.5\mbox{--}3\ $GK)~\cite{Wanajo:2010mc}. Generally, a shorter duration for the former helps the yields, as does a longer duration for the latter. Still  another important quantity is the value of entropy per baryon in the outflow, which strongly depends on the properties of the PNS~\cite{Qian:1996xt}. This quantity controls the efficiency of the triple-$\alpha$ reaction, which is the bottleneck of the seed formation. Entropy values above $\gtrsim 65\mbox{--}70$ for the radiation component in the outflow are generally favored.

In recent years, a more detailed understanding of the $\nu p$-process has continued to emerge~\cite{Wanajo:2010mc,Arcones:2011zj,Eichler:2017kvd,Nishimura:2019jlh,Rauscher:2019mcn,Fujibayashi:2015rma,Sasaki:2017jry,Xiong:2020ntn,Sasaki:2021ffa,Sasaki:2023ysp,Fisker:2009,Bliss:2014qiz,Bliss:2018djg,Jin:2020,Friedland:2023kqp,Nevins:2024dkr}. For example, it was shown that in-matter collisions~\cite{Beard:2017jpg} can have a large effect on the rate of the triple-$\alpha$ reaction and hence on the $\nu p$-process yields~\cite{Jin:2020}. A very important role was found to be played by the hydrodynamics of the NDO, with subsonic profiles giving much larger $\nu p$ yields than the supersonic one~\cite{Friedland:2023kqp}. While the in-matter effects tend to suppress the yields~\cite{Jin:2020}, with self-consistently computed subsonic profiles the yields were still found acceptable, especially with the PNS mass in the range $\gtrsim 1.7 \,M_\odot$~\cite{Friedland:2023kqp}. Such large values of the PNS mass are naturally found in modern simulations of the explosions of massive progenitor stars, which are the same stars that are expected to have subsonic NDOs. Thus, the present situation for the $\nu p$-process appears very promising, at least until other relevant physical effects are identified that could upset the present understanding.

What other relevant physical phenomena should one consider? The next natural target for investigation are the effects of general relativity (GR), which are the focus of the present paper. While the GR effects are expected to be small at the radii of $\sim 200\mbox{--}2000$ km, where the proton and neutron captures of the $\nu p$-process take place, this is not necessarily so at the radii of 20--50 km, where the engine driving the outflow is located. In fact, with the gravitational potential values of order $10^{-1}$ there and the dominant heating rates dependent on the sixth power of the neutrino energy, one may expect at least the neutrino redshift/blueshift GR corrections to be substantial. 

Several classical papers~\cite{Qian:1996xt,Cardall:1997bi, Otsuki:1999kb, Thompson:2001} investigated the effects of GR on the hydrodynamics of NDOs in a core-collapse supernova. Their results cannot be immediately applied to our problem, for two immediate reasons: (i) they do not deal with the $\nu p$-process at all (they predate its discovery), and (ii) they assume either a subsonic or a transonic outflow, rather than solving for the nature of the outflow self-consistently. On the other hand, a useful connection to the present work can still be made: Refs.~\cite{Cardall:1997bi, Otsuki:1999kb, Thompson:2001} do provide relativistic steady-state equations and, using them, find rather significant effects---notably, increases in the entropy per baryon by as much as 30--40 units, compared to their Newtonian calculations. Interestingly,  Ref.~\cite{Friedland:2023kqp}, which implemented the GR equations from~\cite{Cardall:1997bi}, did not report a comparably large change in entropy. This apparent contradiction needs to be resolved. Furthermore, the equations in~\cite{Cardall:1997bi, Otsuki:1999kb, Thompson:2001} are stated in rather different forms, and the later works have not examined their equivalence. For all these reasons, a systematic analysis of both the formalism and results of GR hydrodynamics in connection to their Newtonian counterparts appears motivated.

Our goal in the present paper is to undertake such a systematic study of the GR effects on the outflows themselves, and on the ensuing $\nu p$-process. We intend to not only carefully incorporate the fully relativistic physics, as previous works have done, but also analyze the relative import of individual GR corrections, e.g., strengthening of gravity, enthalpic mass, neutrino blueshift, etc.---an analysis hitherto absent from the literature. For this, we present a derivation of the NDO equations in a form where the identification and physical content of the relativistic hydrodynamic corrections are transparent. We then take the relativistic outflows as input to our nucleosynthesis calculations and quantify the impacts of the relativistic corrections on the production of different $p$-isotopes. We also present results of GR outflows with termination shocks and assess the effect of the supersonic transition on nucleosynthesis. Following this, we explore the dependence of our conclusions on the relevant physics inputs, particularly the progenitor mass. 

To model the yields, we compute the trajectories of tracer particles for a number of time snapshots and then integrate the yields over time. We use our own model of the NDO, which incorporates all GR effects, but can also perform calculations in the fully Newtonian limit or with some of the GR effects turned on and off at will. In each snapshot, the calculation gives the thermodynamic conditions as a function of time and distance to the PNS. Using these results, we post-process the trajectories with the nuclear reaction network {\tt SkyNet}~\cite{Lippuner:2017tyn}.

The structure of this paper is as follows. Section~\ref{sec:statement} provides an overview of the physical problem under consideration. In Sec.~\ref{sec:outflow}, we present the steady-state NDO equations in a consistent general relativistic treatment. In Sec.~\ref{sec:model}, we describe the steady-state outflow solutions and the associated tracer trajectories used to compute the nucleosynthetic yields, while in Sec.~\ref{sec:nuflux} we discuss the neutrino physics inputs required to model the outflows. In Sec.~\ref{sec:outflow:results}, we assess how different relativistic corrections and the choice of the equation of state modify the resulting outflow properties. From that point onward, the focus shifts to nucleosynthesis: Sec.~\ref{sec:nup-stages} summarizes the stages of the $\nu p$-process, and Sec.~\ref{sec:nup-method} details our setup for the nucleosynthesis calculations. Section~\ref{sec:8cases} shows the implications of the various relativistic corrections on the $\nu p$-process yields, and Sec.~\ref{sec:glob-comp} compares time-integrated nucleosynthesis yields between GR and Newtonian calculations, for our benchmark 18\,$M_\odot$ progenitor model. Then, we present in Sec.~\ref{sec:prog-mass} results for lighter progenitors (12.75\,$M_\odot$ and 9\,$M_\odot$). Final remarks and conclusions are given in Sec.~\ref{sec:conclusion}. For interested readers, an extensive set of Appendices provides the detailed calculations needed to reproduce the neutrino outflows and isotopic abundances presented in this work, including comparisons with the existing literature and possibilities for additional refinements, with the aim of making this paper self-contained.

\section{\textcolor{black}{Statement of the Problem}}
\label{sec:statement}

We will be interested in the physical conditions and processes that take place around the PNS several seconds after the core bounce. By this time, the explosion shockwave is revived and is expanding outward through the stellar envelope. At the same time, in the region immediately surrounding the PNS, a low-density bubble is formed. The bubble contains high-entropy matter that is heated by streaming neutrinos near the PNS surface and is launched radially outward. The outflow slows down as it runs into the surrounding low-entropy material, which is itself expanding, albeit more slowly. As the material cools from temperatures of several MeV to hundreds of keV and then even further, various stages of proton-rich nucleosynthesis take place. 

Since this nucleosynthesis process involves essential out-of-equilibrium dynamics, various timescales play crucial roles, and it is important to faithfully model the hydrodynamics of the outflow. This includes both the acceleration and deceleration stages, merger into the surrounding medium, and the subsequent evolution. Here, an important consideration is that it is not known a priori whether the outflow will reach supersonic speeds and form a termination shock upon interaction with the surrounding material, or if it will remain subsonic throughout. The system is peculiarly near-critical~\cite{Friedland:2020ecy}, with the physical character of the outflow dictated by the interplay of neutrino energy deposition, PNS properties, and the details of the density profile in the progenitor envelope. Hence, rather than starting with a shocked or unshocked ansatz, as done in some of the earlier literature on the subject, we self-consistently solve the outflow equations according to the boundary conditions determined by the progenitor profile. 

As the front shock moves outward, the confining pressure around the hot bubble drops. The neutrino fluxes also change with time, as does the PNS radius. As a result, the hot bubble region slowly evolves, on a $\mathcal{O}(1\,{\rm s})$ timescale.\footnote{The edge of the hot bubble is determined by the boundary between the high-entropy, neutrino-heated material, and the surrounding low-entropy material in the stellar envelope.} To compute the integrated nucleosynthetic yields, we create a time series of snapshots. We then construct trajectories of matter particles in these snapshots to be used for the nucleosynthesis calculations.

The basic steps of this approach have been discussed in \cite{Friedland:2023kqp}. Here, we implement a number of substantial enhancements to this earlier treatment. In addition to the accurate GR effects, which are the principle focus of this paper, we include careful matching of the outflow onto the slowly expanding material, a more complete set of neutrino-matter reactions, evolving protoneutron star radius and numerous other enhancements, as described in detail below. While not invalidating the earlier results, the present treatment is a considerable improvement, both conceptually and quantitatively.

\section{\textcolor{black}{General-Relativistic Steady-State Hydrodynamic Equations}} \label{sec:outflow}

We begin by presenting the general-relativistic, spherically symmetric, steady-state hydrodynamic equations, used to model the first segment of the NDO in the hot bubble; see App.~\ref{app:outflow:equations} for their derivation. Next, their reduction to Newtonian hydrodynamics is discussed.
Finally, we compare our equations with those in the existing literature. Throughout, we adopt natural units with $\hbar=c=k_B=1$.

\subsection{\textcolor{black}{Equations of evolution in spherical symmetry}}
\label{sec:outflow:equations}
We model the NDO, from just above the PNS to its merger into the homologously expanding surrounding material, under the simplifying assumptions of spherical symmetry, vacuum Schwarzschild spacetime, and steady flow. Under these common assumptions, the system resembles other relativistic flows studied in the literature.
Here, we adopt as our starting point the approach, including the choice of using covariant variables, of Ref.~\cite{Shapiro:1983du}, which, as we shall later see, renders the content of the relativistic corrections transparent. Our physical system is, nevertheless, different in significant respects from spherical accretion (which Ref.~\cite{Shapiro:1983du} treats): 
the neutrino-heated outflow is not adiabatic; the equations of state (EoS) appropriate for characterizing the NDO are different; and the boundary conditions differ essentially between outflow and accretion. Thus, although proceeding from the same fundamental principles, the differential equations of hydrodynamics are, as expected, different from those in~\cite{Shapiro:1983du}.

The three governing principles are  baryon number conservation, momentum conservation, and the first law of thermodynamics, as familiar from fluid mechanics. Baryon number conservation is expressed by the covariant continuity equation  
\begin{equation} \label{eq:fluid:continuity}
        \nabla_\mu (nu^\mu)=0\,,
\end{equation} 
where $n$ denotes the baryon number density in the fluid’s rest frame, and $u^\mu$ is the bulk four-velocity of the baryon–radiation plasma. Momentum conservation of the plasma, or the Euler equation, is given as
\begin{equation} \label{eq:fluid:energy-momentum-conservation}
\left(\delta_\nu^j+u^ju_\nu\right)\nabla_\mu T^{\mu\nu}=0\,,
\end{equation}
where the energy–momentum tensor of a perfect fluid is given by $T^{\mu\nu}\equiv (\rho+P)u^\mu u^\nu + P g^{\mu\nu}$, with $\rho$ and $P$ representing the total energy density and pressure in the rest frame, respectively. Finally, the first law of thermodynamics reads
\begin{equation} \label{eq:thermo:id}
        d\left(\frac{\rho}{n}\right)=dq-Pd\left(\frac{1}{n}\right)\,,
\end{equation}
where $dq$ denotes specific heat deposition per baryon. 

As mentioned, $dq$ is nonzero for the NDO, prominently in the region close to the PNS, unlike in adiabatic accretion. The first law can be equivalently written in terms of the entropy \textit{per baryon} $S$, which in steady state is
\begin{equation} \label{eq:thermo:dS}
u\frac{dS}{dr} = \frac{\dot{q}}{T}.
\end{equation}
Here $u \equiv dr/d\tau$ is the radial component of the coordinate four-velocity, and $\dot{q} \equiv dq/d\tau$ denotes the net rate of heating and cooling per baryon per proper time (see Sec.~\ref{sec:heatcool} for the collection of the contributing processes we implement). Neutrino heating introduces an entropy gradient in the outflow, which plays a dominant role in driving the expansion. 

Having defined the fluid dynamic variables, we specify our thermodynamic ensemble. The baryon-radiation plasma is an ensemble of non-relativistic baryon gas and radiation consisting of photons, electrons, and positrons in thermal equilibrium. Traditionally, NDOs in a SN are often modeled under the approximation that all mass resides in baryons, and all pressure and entropy are due to radiation (e.g., \cite{Qian:1996xt}). Also, a fixed number of relativistic degrees of freedom (RDF) in the radiation is {sometimes} assumed. Here, we present hydrodynamic equations which go beyond these simplifications: (1) the baryon gas component is included in the thermal ensemble, characterized by its entropy $S_b$, pressure $P_b$, and internal energy $\bar{\rho}_b$; (2) variable number of effective RDF is implemented, since the temperature in the outflow varies from well above the electron mass close to the PNS to significantly below it near the merger. This is characterized by three dimensionless functions, $g_*^\rho$, $g_*^P$, and $g_*^S$, in terms of which the state functions of total radiation, ($\gamma, e^+,e^-$), are \cite{Kolb:1990vq, Husdal:2016haj}:
\begin{equation} \label{eq:radeos}
    \rho_{\rm r} = \frac{\pi^2}{30}g_*^\rho(T) T^4, \quad P_{\rm r} = \frac{\pi^2}{90}g_*^P(T) T^4, \quad S_{\rm r}=\frac{2\pi^2}{45} g_*^S(T) \frac{T^3}{n}\,.
\end{equation}
The corresponding expressions for the Fermi-Dirac integrals, i.e., $g_*$'s, are given in App.~\ref{app:gstar}.  With these, all thermal quantities in the three fluid equations are expressed as functions of $n$ and $T$. Note that the radiation's enthalpy density $\rho_{\rm r}+P_{\rm r}$ also contributes to the inertia of the fluid, as necessary for relativistic hydrodynamics.

Under these conditions, Eqs.~(\ref{eq:fluid:continuity})--(\ref{eq:thermo:id}) can be reformulated as three coupled differential equations for the evolution of the baryon number density $n$, the temperature $T$,  and the velocity $u$ as functions of radius.
A detailed derivation with commentary is given in App.~\ref{app:outflow:equations}.

\begin{enumerate}
\item Density evolution from the continuity equation:
\begin{equation} \label{eq:std:dn}
    \boxed{\frac{1}{n}\frac{dn}{dr} = -\frac{1}{u}\frac{du}{dr} - \frac{2}{r}\,.}
\end{equation}
\item Temperature evolution from the thermodynamic identity:
\begin{equation} \label{eq:std:dT}
    \boxed{\left(4\beta_*P_{\rm r}+\frac{3}{2}P_b\right)\frac{dT}{dr} = \left(\frac{4g_*^S}{g_*^P}P_{\rm r}+P_b\right)\frac{T}{n}\frac{dn}{dr} + nT\frac{\dot{q}}{u}\,,}
\end{equation}
where $\beta_*\equiv g_*^S(3+\tilde{g}_*^S)/g_*^P$ is a combination of the number of effective RDF, and $\tilde{g}_* \equiv d\ln g_*/d\ln T$'s for the state functions respectively; $P_b = nT$ is the baryonic gas pressure. 
\item Velocity evolution from the Euler equation
\begin{equation} \label{eq:std:du}
    \boxed{\left(\frac{\mu v_s^2}{u}-u\right)\frac{du}{dr} = \frac{GM}{r^2} - \frac{2\mu v_s^2}{r} + \frac{\mu n}{\rho+P}\Pi_1^{{\rm r} b}\frac{\dot{q}}{u}\, .}
\end{equation}
This form bears much resemblance to the Newtonian velocity equation yet, formally, contains \textit{all} relativistic hydrodynamic corrections. Here, 
\begin{equation} \label{eq:mu}
    \mu\equiv 1 + u^2 -\frac{2GM}{r}
\end{equation}
is a recurring factor of GR hydrodynamics in spherical symmetry, with $M$ the PNS mass. The factor $\mu$ connects the coordinate velocity $u$ to the physical velocity $v$ measured by a stationary observer in Schwarzschild spacetime:
\begin{equation}
    v=\frac{u}{\sqrt{\mu}}\,,
\end{equation} 
and therefore encapsulates relativistic velocity and gravity corrections.\footnote{When the choice of velocity variable is $v$, the symbol $y(r,v)\equiv\sqrt{(1-2GM/r)/(1-v^2)} =\sqrt{\mu(r,v)}$ has been used previously in the literature, so that $u=y(r,v)\,v$~\cite{Cardall:1997bi, Thompson:2001}.} 

Next, $v_s$ is the adiabatic sound speed given as
\begin{equation}
    \label{eq:state:vs2:form}
    v_s^2 = v_{\rm r}^2 + v_b^2 = \frac{1}{\rho+P}\left[(4+\tilde{g}_*^P)\Pi_2^{{\rm r} b}P_{\rm r} + (1+\Pi_2^{{\rm r} b})P_b\right]\,,
\end{equation} 
where $P = P_{\rm r} + P_b$ is the total pressure, and 
$\rho = \rho_b + \rho_{\rm r} + \bar{\rho}_b$ is the total energy density of the fluid, with $\rho_b = m_N n$ and $\bar{\rho}_b = 3 n T/2$ the energy densities of the baryon mass and thermal non-relativistic baryon gas.
Also, there are two coefficients involving components of pressure:
\begin{equation}\label{eq:pi2gammab}
    \Pi_1^{{\rm r} b}\equiv \frac{(4+\tilde{g}_*^P)P_{\rm r}+P_b}{4\beta_*P_{\rm r}+\frac{3}{2}P_b}, \quad\quad \Pi_2^{{\rm r} b}\equiv \frac{4\frac{g_*^S}{g_*^P}P_{\rm r} + P_b}{4\beta_* P_{\rm r}+ \frac{3}{2}P_b}\,.
\end{equation}
The former is a coefficient of the heating term in Eq.~\eqref{eq:std:du}, while the latter comes from the isentropic evolution in Eq.~\eqref{eq:std:dT}. 

\end{enumerate}

The differential equations \eqref{eq:std:dT}--\eqref{eq:std:du} determine the evolution of the steady-state NDO over radius, and they are in the \textit{standard form} of ${\frac{d}{dr}(u,T,n) = f(u,T,n)}$.
This form of equations is being presented here for the first time, containing explicit expressions related to the equations of state. We compare them with the hydrodynamic NDO equations in the existing literature in Sec.~\ref{sec:comparison-equations}.

For a radiation-dominated atmosphere, a special case often assumed in existing studies, the outflow equations simplify. In this approximation, we neglect the contribution of the non-relativistic baryon gas to the thermal ensemble, i.e. pressure, internal energy, and entropy: $P_b=\frac{2}{3}\bar{\rho}_b\to 0$, $S_b\to 0$. For instance, to obtain the sound speed under this approximation, we simply drop all $P_b$ terms in Eqs.~\eqref{eq:state:vs2:form} and \eqref{eq:pi2gammab}, while also taking $\rho+P\to \rho_b + \rho_{\rm r} + P_{\rm r}$.

In the NDO, the outflow rate of mass $\dot{M}$ is, by the assumption of steady flow, an integral of motion:
\begin{equation}
\dot{M} = 4\pi r^2 \rho_b u\,.
\end{equation}
Additionally, in the absence of heating (or when it becomes negligible), a second conserved quantity \cite{Shapiro:1983du} is
\begin{equation} \label{eq:fluid:conserved-I}
    I \equiv \left(\frac{\rho+P}{n}\right)^2 \left(1+u^2-\frac{2GM}{r}\right)\,,
\end{equation}
a relativistic version of the Bernoulli function. The incorporation of neutrino heating alters $I$ as
\begin{equation} \label{eq:fluid:conserved-I:q}
    \frac{dI}{dr} = \frac{\rho+P}{n}\frac{2\mu}{u} \dot{q}\,.
\end{equation}
While $\dot{M}$ is always conserved,  Eq.~\eqref{eq:fluid:conserved-I} is effectively constant only once the NDO escapes the neutrino heating region.

\subsection{\textcolor{black}{Reduction to Newtonian hydrodynamics}}
\label{sec:outflow:Newtonian}

Our relativistic steady-state equations \eqref{eq:std:dT}--\eqref{eq:std:du} reduce to the familiar Newtonian ones under four simple prescriptions, classified as either pertaining to the form of the hydrodynamic equations or to the heating terms. For hydrodynamics, the two reductions are
\begin{enumerate}
    \item $\mu\rightarrow 1$, which identifies the coordinate and physical velocities, i.e. $u\to v$, and ignores corrections for gravitation and relativistic velocities. 
    \item
    $\rho+P\rightarrow \rho_b$, i.e. neglecting thermal energy (and pressure) in relation to mass energy, which, for instance, 
    modifies the form of adiabatic sound speed and overestimates it.
\end{enumerate}
For convenience, we refer to the second Newtonian approximation as the omission of `radiation mass' ($m_\gamma\to 0$), since $\rho+P$ appears in place of $\rho_b$ in Eq.~\eqref{eq:std:du}. The corresponding sound speed under this approximation is readily acquired by taking $\rho+P\to \rho_b$ only in the denominator of Eq.~\eqref{eq:state:vs2:form}.\footnote{This follows from the definition of the sound speed (see Eq.~\eqref{eq:state:vs2:form} and Eq.~\eqref{eq:app:state:vs2}), and the form of Eq.~\eqref{eq:thermo:dS} for entropy change, which is independent of whether baryon mass dominates over internal energies.
}

An exception to these simple prescriptions is for the conserved quantity $I$ in Eq.~\eqref{eq:fluid:conserved-I}, which does depend on the non-trivial form of $\mu$ in Eq.~\eqref{eq:mu}. For outflows with partial GR corrections, i.e. either $\mu\rightarrow 1$ or ignoring radiation mass, it is not immediately clear what the forms of the conserved quantity, if any, are. Another exception is for the thermodynamic identity \eqref{eq:thermo:id}, which involves $\rho+P$, but there $\rho+P$ cannot be simply replaced. 

If we adopt both these reductions, which encompass (i) weak gravity, (ii) non-relativistic bulk flow, and (iii) domination of baryon mass over thermal energy $\bar{\rho}=\rho_{\rm r}+\bar{\rho}_b$ and pressure, the conserved quantity in Eq.~(\ref{eq:fluid:conserved-I}) becomes
\begin{equation}
\begin{aligned}
    I &= m_N^2\left(1 + \frac{\bar{\rho}+P}{\rho_b}\right)^2\left(1+u^2-\frac{2GM}{r}\right) \\
    &\approx 2 m_N^2\left(\frac{1}{2} + \frac{v^2}{2} - \frac{GM}{r} + \frac{h}{m_N} \right)\,,
\end{aligned}
\end{equation}
in terms of the specific enthalpy $h\equiv (\bar{\rho}+P)/n$. Dropping the constant mass term, we recover the non-relativistic (NR) form of the invariant, i.e. the Bernoulli quantity: 
\begin{equation}\label{eq:fluid:conserved-I:NR}
\begin{aligned}
    I_{\mathrm{NR}} &=  \frac{v^2}{2} - \frac{GM}{r} + \frac{h}{m_N} \\
    &=  \frac{v^2}{2} - \frac{GM}{r} + \frac{TS+\mu_b}{m_N}\,,
\end{aligned}
\end{equation}
where $\mu_b$ is the chemical potential of the (non-relativistic) monatomic baryon gas and $\mu_{e^\pm}=0$ is assumed. This is a specific (per baryon) energy, and it coincides with the expression in Eq.~(25) of Ref.~\cite{Qian:1996xt} when assuming radiation domination.
As the radius tends to infinity, the non-relativistic conditions (i)--(iii) become sufficiently satisfied. Then, for subsonic outflows, the enthalpy term alone remains and, for a given temperature $T_\infty$, the quantity $m_NI_{\mathrm{NR}}-\mu_b$ is a direct measure of $S_\infty$. This energy $I_{\rm NR}$ is indeed conserved by the Newtonian hydrodynamic equations, i.e. in the Newtonian limit $\mu\to 1$ and $\rho+P\to\rho_b$.

Relativistic corrections can also be included with an effective in-matter gravitational `potential' that can be derived from the Tolman-Oppenheimer-Volkoff (TOV) equation of hydrostatic equilibrium \cite{Shapiro:1983du, Qian:1996xt}:
\begin{equation} \label{eq:fluid:Newton:Veff}
    \frac{dV_{\rm eff}}{dr} \approx \frac{G m(r)}{r^2} \left(1-\frac{2Gm(r)}{r}\right)^{-1}\left(1+\frac{P}{\rho_b}+\frac{4\pi r^3 P}{m(r)}\right)\,,
\end{equation}
where $m(r)$ is the total mass enclosed by radius $r$. This expression encompasses a relativistic deepening of the gravitational well, thermal contributions to mass, as well as gravitation from the enclosed mass $m(r)>M$ (which our foregoing GR treatment does not incorporate). Unlike the Newtonian potential, $V_{\rm eff}$ is a `potential' only in a loose sense since in spherical symmetry all dynamic quantities, including $P$, are functions of radius, and Eq.~\eqref{eq:fluid:Newton:Veff} can be formally integrated.

In this context, we follow Ref.~\cite{Qian:1996xt} and adopt the state variables of $v$, $T$, and $S$ here to write the differential equations describing the outflow evolution. Then, entropy deposition is described by Eq.~\eqref{eq:thermo:dS}, which is unaffected (up to $u\to v$), while the other two differential equations are
\begin{align} \label{eq:fluid:Newton:Veff:dvdq}
    \left(\frac{v_s^2}{v}-v\right)\frac{dv}{dr} &= \frac{dV_{\rm eff}}{dr}  -\frac{2 v_s^2}{r} + \Pi_1^{{\rm r} b}\frac{\dot{q}}{m_N v} \,,\\
    \frac{\dot{q}}{m_N} &= v\frac{d}{dr}\left(\frac{v^2}{2} +V_{\rm eff} + \frac{h}{m_N}\right).
\end{align}
The second equation identifies the modified conserved quantity $I_{\rm eff}$. Evaluating the radial derivative and employing Eq.~(\ref{eq:fluid:Newton:Veff}), one can obtain a differential equation of $dT/dr$ to evolve temperature. We note that Ref.~\cite{Qian:1996xt} only modifies the velocity equation and retains the heating equation with $I_{\rm NR}$ unchanged. Instead, we include $V_{\rm eff}$ in both of the above equations to ensure consistency with the basic fluid equations. We observe that Eq.~\eqref{eq:fluid:Newton:Veff} is mainly a gravitational correction, which includes enthalpic (radiation) mass only in enhancing gravity. Crucially, it does not revise the form of sound speed, as a consistent incorporation of radiation mass would.

\subsection{\textcolor{black}{Comparison with previous relativistic hydrodynamic equations}}
\label{sec:comparison-equations}

Previously, \CF~\cite{Cardall:1997bi}, \Otsuki~\cite{Otsuki:1999kb} and \Thompson~\cite{Thompson:2001} have employed steady-state GR equations to model neutrino-driven outflows. The differential equations for the relativistic outflow in these references assume rather different forms. Our approach aligns with that of \Otsuki~
in adopting the same fundamental equations [their Eqs.~(1--3)] from Ref.~\cite{Shapiro:1983du}, in which form the physical principles are transparent. However,
\Otsuki~ do not present their differential equations in the standard form of ${du}/{dr}$ and ${dT}/{dr}$, which would, for instance, show their form of sound speed as the singularity in the velocity equation. This prevents us from performing an explicit comparison with our hydrodynamic equations.
\Thompson~state equivalent fundamental equations of an alternate form and subsequently also state equations in standard form. We checked that their equations agree with those presented here for the EoS we adopt.

\CF~directly state their outflow equations in standard form, using the state variables $v$,
$T$, $S$ in continuity with the corresponding Newtonian equations in Ref.~\cite{Qian:1996xt}.  As further discussed in App.~\ref{app:outflowDES}, there are some missing terms in the differential equation of velocity in \CF~[Eq.~(1) therein], leading to a violation of the outflow mass-rate conservation. In App.~\ref{app:outflowDES}, we derive the corrected forms of the equations in \CF~ and show that these modified equations conserve $\dot{M}$ numerically and can boost the outflow's peak subsonic velocity by up to $\sim 50\%$ compared to using Eq.~(1) of \CF. On the other hand, the corresponding entropy change using our corrected equation is modest---only about $\sim 5$ per baryon, relative to using theirs.

At this point, we \textit{emphasize} that these hydrodynamic equations themselves are \textit{not} the cause of the difference in entropy enhancement (relative to Newtonian outflows) found between \CF,~\Otsuki,~\Thompson~on one hand and Ref.~\cite{Friedland:2023kqp} on the other.

\section{\textcolor{black}{Modeling the Outflow and Particle Trajectories}}
\label{sec:model}

In this Section we construct the steady-state outflow solutions and the associated tracer trajectories that will later be used to compute the nucleosynthetic yields, with the steady-state NDO providing the first segment of each trajectory. Section~\ref{sec:outflow:bvp} presents the prescription used to solve the outflow equations introduced above, formulated as a boundary value problem requiring the boundary conditions specified in Sec.~\ref{sec:model:BC}. As emphasized earlier, CCSN outflows operate in a near-critical regime; consequently, we solve the hydrodynamic equations self-consistently within our setup, determining for each case whether the flow remains subsonic or undergoes a transonic transition. In the latter case, the fully relativistic Rankine–Hugoniot conditions derived in Sec.~\ref{sec:outflow:RH} are applied, improving upon the Newtonian treatments adopted in earlier work. Section~\ref{sec:traj} outlines how these outflow solutions are used to build the tracer trajectories, consisting of (i) an outward acceleration from the PNS surface, followed by deceleration through interaction with the surrounding slowly expanding medium (modeled through the steady-state NDO) and (ii) a merger into the homologous expansion behind the front shock (FS). Finally, Sec.~\ref{sec:model:comparison} compares our setup with those adopted in earlier Newtonian and GR wind studies. In contrast to prior work, which imposed identical neutrino heating parameters at the neutrinosphere across the Newtonian and GR cases, our analysis makes use of luminosities and spectra provided at large radii by SN simulations, leading to qualitatively different behavior near the PNS and substantially different outflow solutions.

Readers whose primary interest lies in the resulting outflows and nucleosynthesis rather than the construction of the outflow solutions may proceed directly to Sec.~\ref{sec:outflow:results}.

\subsection{\textcolor{black}{Boundary value problem}}
\label{sec:outflow:bvp}

In order to model the NDOs and assess the impact of GR effects, one must first specify the conditions under which the outflow equations are solved and compared. The most straightforward method of comparison is as an ``initial" value problem, where the state is specified at some radius (e.g., at the PNS surface).  Such an approach often leads to qualitatively different GR and Newtonian outflows, typically with a subsonic Newtonian solution and a supersonic GR one, making the comparison physically misleading. A more robust characterization of the outflow is provided instead by the thermal conditions (temperature and entropy) at the neutrinosphere and by the far pressure $P_f$ of the cold material expanding behind the front shock \cite{Duncan1986}. For this reason, we choose to compare outflows using a boundary-value problem (BVP), for which the effect of relativistic corrections on the outflow is less obvious a priori.

Importantly, relativistic hydrodynamic corrections preserve the qualitative structure of the Newtonian outflow equations, which contain a vacuum solution and two regimes of subsonic and transonic outflows demarcated by a critical (subsonic) solution. Consequently, we can apply the unified method of solving the BVP presented in Ref.~\cite{Friedland:2020ecy}, using as boundary conditions the temperature and entropy at the neutrinosphere, and the far pressure $P_f$.

We impose the far pressure at the radius $r_g$ where the NDO merges with the expanding material behind the FS. This material, far from the PNS, undergoes an approximately homologous, i.e., Hubble-like, expansion~\cite{Friedland:2023kqp}, whose velocity profile can be described by
\begin{equation} \label{eq:FS:homologous}
    v_h(r) = \frac{v_{\rm FS}}{R_{\rm FS}} \, r\,,
\end{equation}
once the FS radius $R_{\rm FS}$ and velocity $v_{\rm FS}$ are specified.\footnote{With the motion of the front shock, the physical system is truly time-dependent, and the problem arises of how to model time dependence with a steady-state description. We address this in Sec.~\ref{sec:traj}.} To identify the radius where the steady-state NDO runs into this homologous material, we adopt a heuristic ``gluing'' prescription, and we define the merging radius $r_g$ as the point where
\begin{equation}
v(r_g) = v_h(r_g)\,.
\label{eq:FS:rg}
\end{equation}
We therefore define a BVP by imposing the far pressure $P_f$, which is continuous across the merging point, at the radius $r_g$. The determination of $r_g$ and the matching of $P_f$ are implemented self-consistently within the same iterative routine. This represents an improvement over the simplified treatments in Refs.~\cite{Friedland:2020ecy,Mukhopadhyay:2022yrd,Friedland:2023kqp} (see App.~\ref{app:outflow:bvp} for more details), and ensures a more accurate connection between the steady-state and homologous regimes.

Once the boundary conditions are specified, we solve the BVP following the shooting method of Ref.~\cite{Friedland:2020ecy}. If $P_f > P_f^c$ (the gluing pressure of the critical subsonic solution), the outflow is \emph{subsonic}, and we determine it by root-finding on the initial velocity $v_0$ at the gain radius: larger $v_0$ yields a larger $r_g$ and a smaller resulting $P_f$. For $P_f<P_f^c$, the outflow becomes \emph{supersonic} and we adjust the termination-shock radius $R$. Greater $R$ (more violent shocks) corresponds monotonically to a larger $r_g$ and smaller $P_f$. The critical solution itself, as a limiting subsonic solution, can be found by applying a bisection-type search on $v_0$ to determine the critical velocity.

Having outlined the procedure to solve the BVP and where the boundary conditions are imposed, we now specify their explicit form in Sec.~\ref{sec:model:BC}, and the additional conditions required in the presence of a termination shock in Sec.~\ref{sec:outflow:RH}.

\subsection{\textcolor{black}{Boundary conditions}}
\label{sec:model:BC}

\begin{figure} 
\includegraphics[width=0.6\textwidth]{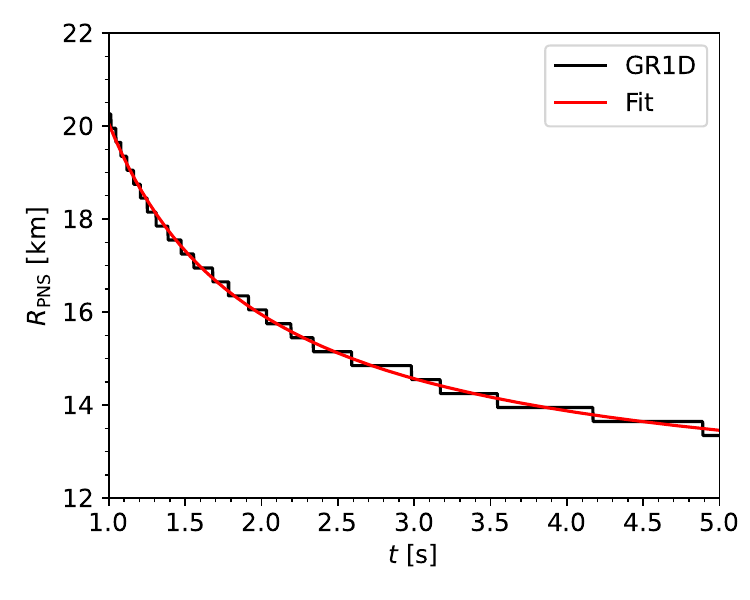}
\centering
\caption{Radius of the PNS from the 1D CCSN code {\tt GR1D}~\cite{OConnor:2014sgn} (black) and our fit defined in Eq.~\eqref{eq:pns-fit} (red) as functions of the time after bounce. The simulation is for an artificially exploding CCSN of a $15~M_{\odot}$ progenitor leaving a $1.8~M_\odot$ PNS. The crucial input to determining the PNS mass is the EoS which in this case is the modern SFHo~\cite{Steiner:2012rk,Hempel:2009mc}.}
\label{fig:pns-gr1d}
\end{figure}

The inner and outer boundary conditions of the outflows are determined by the properties of the PNS and the outgoing FS correspondingly.
To build our models, we employ a subset of the progenitor profiles of Ref.~\cite{Sukhbold:2015}. Our benchmark model is an $18~M_\odot$ star, but we also consider lighter progenitors with $12.75~M_\odot$ and $9~M_\odot$ masses. For the two more massive progenitors, we assume that the SN explosion leads to the formation of a PNS with gravitational mass $M=1.8~M_\odot$, consistent with modern SN simulations, which report PNS masses in the range $1.6$--$2.0$ $M_\odot$ for progenitors of 13--20 $M_\odot$~\cite{Vartanyan:2018iah,Burrows:2020qrp, Burrows:2019zce}. For the PNS radius, we take into account its time dependence using our parametrization:
\begin{equation}\label{eq:pns-fit}
    R_{\rm PNS}(t) = (11.77 + 8.28~t^{-0.987})~{\rm km}\,,
\end{equation}
where $t$ is the post-bounce time in seconds. This time evolution, shown in Fig.~\ref{fig:pns-gr1d}, is obtained using our CCSN simulation performed with {\tt GR1D}~\cite{OConnor:2014sgn, OConnor:2018sti}. It captures the evolution of the PNS radius in recent CCSNe hydrodynamics simulations in spherical symmetry~\cite{OConnor:2018sti} and represents an improvement over the constant PNS radius assumed in Ref.~\cite{Friedland:2023kqp}.

We assume that this PNS radius coincides with the gain radius, where heating balances cooling, and we accordingly set the outflow temperature there to balance nucleon heating and cooling, allowing a small ($1\%$) heating excess. We also assume for all models a radiation entropy $S_{\rm r}=6$ at the gain radius, from which the baryon density can be inferred [using $S_{\rm r}$ from Eq.~\eqref{eq:radeos}] once the temperature has been determined.

The outer boundary conditions can be inferred from the FS position. We assume the FS moves at constant velocity, with its position given by 
\begin{equation}
    R_{\rm FS}(t)= v_{\rm FS}\,t\,,
\label{eq:RFS}
\end{equation}
where we take $v_{\rm FS}=6000$~km/s as a benchmark value for the 18 $M_\odot$ and larger values for lighter progenitors, as shall be reported. Typical shock velocities seen in SN simulations of progenitor masses greater than $10\,M_\odot$ are in the range of $6000\text{--}8000$ km/s (e.g., \cite{Burrows:2019rtd, Burrows:2019zce, Bollig:2020phc}). From the FS position we can approximate the density of the homologously expanding material behind it as
\begin{equation}
    \rho_s(t)  = \frac{M_{\rm plow}(t)}{4\pi/3\,R_{\rm FS}^3(t)}\,,
\label{eq:rhos}
\end{equation}
where $M_{\rm plow}(t)$ is the total mass swept up by the FS: ${M_{\rm plow}=M_{\rm prog}(R_{\rm FS})-M_{\rm PNS}}$, with $M_{\rm prog}(R_{\rm FS})$ the enclosed progenitor mass within the radius $R_{\rm FS}$~\cite{Sukhbold:2015}. 
The homologously expanding material is characterized by approximately uniform density, temperature and entropy~\cite{Friedland:2023kqp}. In our analysis we again assume its radiation entropy to be $S_{\rm r}=6$. From the entropy and the baryon density, we can infer the temperature and then the pressure using the EoS [$S_{\rm r}$ and $P_{\rm r}$ from Eq.~\eqref{eq:radeos}]. Between the NDO and this expanding material, there is a discontinuity in temperature and density. Yet, we assume no pressure gradient exists across this boundary. Thus, the pressure of the homologous material, computed as described here, is the far pressure $P_f$ which we impose as our outer boundary condition for our steady-state NDO described in the previous section.

\subsection{\textcolor{black}{Discontinuity across a termination shock}}
\label{sec:outflow:RH}

When the far pressure satisfies $P_f<P_f^c$, the steady-state BVP discussed in the previous section admits no purely subsonic solution. In this case, the outflow must undergo a transition to a supersonic branch, and the steady-state solution is completed by introducing a termination shock located at some radius $R$ beyond the sonic point. The physical state upstream and downstream of this shock must satisfy a generalized version of the Rankine–Hugoniot (RH) relations, accounting for relativistic corrections and variable RDF. As shown in Ref.~\cite{Friedland:2020ecy} for Newtonian solutions, steady-state solutions with termination shocks can be obtained by placing the shock after the sonic point and enforcing the RH conditions at its location.

For relativistic outflows, the generalized RH relations still follow from the conservation of baryon number, momentum, and energy fluxes across the discontinuity. Baryon mass continuity (c.f. Eq.~(\ref{eq:fluid:continuity})) requires 
\begin{equation}
    \label{eq:RH:continuity}
    n_1 u_1 = n_2 u_2 \,,
\end{equation}
using subscripts 1, 2 to denote quantities before and after the shock respectively. The conservation of the flow of momentum across the shock yields 
\begin{equation}
\label{eq:RH:momentum}
    (\rho_1+P_1)u_1^2 + \left(1-\frac{2GM}{r}\right)P_1 = (\rho_2+P_2)u_2^2 + \left(1-\frac{2GM}{r}\right)P_2\,,
\end{equation}
where $r$ is the position of the shock. Third, Eq.~(\ref{eq:fluid:conserved-I:q}) indicates that $I$ should be conserved since any particle moves across the shock radius in infinitesimal time:
\begin{equation}
\label{eq:RH:energy}
    \left(\frac{\rho_1+P_1}{n_1}\right)^2\left(1+u_1^2-\frac{2GM}{r}\right) = \left(\frac{\rho_2+P_2}{n_2}\right)^2\left(1+u_2^2-\frac{2GM}{r}\right).
\end{equation}

As further discussed in App.~\ref{app:outflow:RH}, these three equations determine the downstream state $(u_2,n_2,T_2)$ from the upstream quantities $(u_1,n_1,T_1)$. It is important to emphasize that including relativistic corrections in the RH relations, together with a consistent treatment of three species of effective RDF  (see App.~\ref{app:gstar} for more details), is essential for obtaining physically meaningful shocks. This approach extends beyond previous treatments~\cite{Friedland:2020ecy,Friedland:2023kqp} that relied on Newtonian RH conditions. Neglecting these relativistic and thermodynamic corrections can lead to unphysical results, such as positive velocity jumps or negative baryon-density jumps, especially when the termination shock forms close to the sonic point in the near-critical regime. Since CCSN outflows typically operate in precisely such near-critical conditions \cite{Friedland:2020ecy}, an accurate and fully relativistic implementation of the RH conditions is indispensable.

\subsection{\textcolor{black}{Synthesis of tracer trajectories}}
\label{sec:traj}

The outflow solution obtained from the BVP defines the first segment of the tracer trajectories used to compute the nucleosynthetic yields. To obtain these trajectories, we follow the general strategy of Ref.~\cite{Friedland:2023kqp}, but with several important improvements detailed below. Each tracer trajectory naturally divides into two phases. In the first, the tracer accelerates outward from the PNS surface and decelerates due to interactions with the surrounding slowly expanding material; in the second, it joins the homologous expansion of the material behind the FS, with which it then co-moves. As discussed in Sec.~\ref{sec:nup-stages}, the early stages of the $\nu p$-process occur during the first phase, whereas the late-time evolution takes place during the second phase, which extends over several seconds~\cite{Wanajo:2010mc,Fischer:2009}.

Because the duration of the first phase is $\Delta t \sim \mathcal{O}(1)$~s, we construct this portion of the trajectory directly from the steady-state NDO solution of the BVP. 
For each tracer, we use the outflow corresponding to the boundary conditions at the merging time, i.e., the moment when the tracer joins the homologous expansion. This choice ensures a smooth and physically consistent transition to the second segment of the trajectory.\,\footnote{Operationally, this corresponds to assuming that the FS position does not change appreciably between launch and merger, such that the outer boundary condition $P_f$ remains constant. Using launch-time conditions, which better represent the earliest part of the evolution, while allowing the FS to move during $\Delta t$ would produce a discontinuity at the merging, because $P_f$ decreases as the FS expands.}

This approach highlights the limitation of the steady-state approximation: the outflow conditions evolve over $\Delta t$ but the model cannot capture this time dependence, so one must select the effective time at which the steady-state profile is evaluated; any time within $\Delta t$ is formally allowable. Choosing the merger time yields the most consistent overall trajectory, at the cost of neglecting short-term variations. A fully time-dependent treatment of the NDO, requiring the solution of the relevant partial differential equations, is left to future work.

The second phase begins at the gluing radius $r_g$, determined as described in Sec.~\ref{sec:outflow:bvp}, where the outflow velocity matches the homologous velocity profile. Beyond this point, the tracer co-moves with the expanding material swept up by the FS. The velocity is held fixed at its value at $r_g$, and the entropy $S$ is fixed to the value inherited from the steady-state outflow immediately prior to gluing. From the tracer entropy $S(n,T)$ and the far pressure $P_f(n,T)$, which evolves according to the progenitor structure (see Sec.~\ref{sec:model:BC}), we determine the temperature $T(t)$ and baryon density $n(t)$ using the EoS. This procedure yields a consistent time evolution of $T(t)$ and $n(t)$ in the second segment. The continuity of $T(t)$ and $n(t)$ across the two segments is ensured by the continuity of $P_f(n,T)$ at the merging radius $r_g$, together with our choice of imposing boundary conditions at the merging time. This procedure improves upon the treatment of Ref.~\cite{Friedland:2023kqp}, where continuity at $r_g$ was imposed by construction using simple scaling relations $n\propto \rho_s$ and $T\propto \rho_s^{1/3}$, where $\rho_s$ is the density of the surrounding material defined in Eq.~\eqref{eq:rhos}.

\subsection{\textcolor{black}{Comparison with outflow modeling in previous studies}}
\label{sec:model:comparison}
The BVP framework developed above allows us to model both subsonic and supersonic outflow solutions from first principles, following a procedure analogous to Ref.~\cite{Friedland:2023kqp}, but incorporating the refinements discussed in the previous sections. This approach goes beyond earlier studies that considered only one of the two regimes: \CF~and \Otsuki~analyzed exclusively subsonic NDOs, while \Thompson~focused solely on the steady-state transonic wind problem. These papers predate the $\nu p$-process and were concerned with the $r$-process. In our case, such a priori assumption should not be made, since the $\nu p$-process yields crucially depend on the outflow regime. Since the system is near-critical~\cite{Friedland:2020ecy}, results depend on the details of the physical system and it is essential to solve the outflow equations self-consistently according to the boundary conditions set by the progenitor profile.

Another notable difference from previous literature concerns the careful modeling of the transition between the steady-state outflows and the surrounding homologously expanding material. This treatment is essential for our purposes, since along each tracer trajectory the $\nu p$-process continues for more than 1~s after the tracer is launched. This aspect was not addressed by previous works on NDOs, which focused primarily on the early expansion phase.

Despite these differences, which make our setup substantially distinct from previous literature, it is possible to compare the relativistic effects obtained with our framework to those reported by \CF, \Otsuki, and \Thompson. Our approach, including the BVP solution and the construction of tracer trajectories, enables a direct and consistent comparison between GR and Newtonian outflows.
Neither of the two hydrodynamic corrections from relativity discussed in Sec.~\ref{sec:outflow:Newtonian} requires modification of the (thermal) boundary conditions of $T$ and $S_{\rm r}$ at the gain radius and $P_f$ at $r_g$. So we may impose the same boundary values across outflows including these corrections. What remains to be specified are the parameters of neutrino heating, in particular the luminosities and energy moments. For relativistic outflows, which experience gravitational redshift, neutrino energies must be specified together with the radius. In the studies of \CF,~\Otsuki,~\Thompson~
(as well as in the post-Newtonian example discussed in Ref.~\cite{Qian:1996xt}), Newtonian and GR outflows are prescribed the same neutrino luminosities and energy moments at the neutrinosphere radius. As a result, large enhancements of entropy ($\Delta S\sim 30$) were obtained. In contrast, our analysis uses neutrino luminosities and spectra predicted by SN simulations---reported far from the PNS, e.g. at 500~km---as external input \cite{Huedepohl:2009wh}, 
following Ref.~\cite{Friedland:2023kqp}. Therefore, we assume the same neutrino luminosity for both GR and Newtonian outflows at this large radius and \textit{not} at the PNS. This means that, for a Newtonian outflow, the luminosity at the neutrinosphere is the same as at large radii; while for a relativistic outflow, neutrino energies must be \emph{blueshifted} from the far radius to the neutrinosphere. As a result, while heating rates are similar at large distances, they differ significantly near the PNS, where the outflows originate. This difference in setup from the previous GR studies of \CF,~\Otsuki,~\Thompson~leads to significantly different results (see Sec.~\ref{sec:outflow:results:subsonic}), revealing the importance of an accurate characterization of boundary conditions and neutrino-heating parameters.

\section{\textcolor{black}{Neutrino Physics Inputs}}
\label{sec:nuflux}

In this Section we specify the neutrino luminosities and spectra required to solve the outflow equations and to quantify the impact of relativistic corrections.

\subsection{\textcolor{black}{Processes of neutrino heating and cooling}}\label{sec:heatcool}

Processes of neutrino heating and cooling have a formative role in determining the outflows, and they enter the equations of Sec.~\ref{sec:outflow:equations} via the heating rate $\dot{q}$. Here, we follow Refs.~\cite{Qian:1996xt, Otsuki:1999kb} and implement the dominant processes of free nucleon heating and cooling $\nu_e+n\leftrightarrow p + e^-$ and $\bar{\nu}_e+p\leftrightarrow n + e^+$, heating from elastic neutrino-electron scattering $e+\nu\rightarrow e+\nu$, and cooling through $e^+e^-$ annihilation $e^-+e^+\rightarrow \nu+\bar{\nu}$, using the rates reported in the literature~\cite{Qian:1996xt, Otsuki:1999kb}. The largest contribution to heating is given by the (anti)neutrino absorption by free nucleons $\nu_e+n\to p + e^-$ and $\bar{\nu}_e+p\to n + e^+$, which was the only heating process considered in Ref.~\cite{Friedland:2023kqp}. Here, we consider the additional contribution from elastic neutrino-electron scattering $e+\nu\rightarrow e+\nu$, which increases heating by tens of percent and the maximum velocity of the outflow by a factor of 2. For the conditions relevant to our analysis, neutrino-antineutrino pair annihilation into electron positron pairs $\nu+\bar{\nu}\rightarrow e^-+e^+ $ can be neglected, since its contribution accounts for only a few percent of the total heating rate at the PNS surface and declines very rapidly with radius. We verified that the inclusion of this last process would leave the outflow nearly unchanged.

As shown by their full expressions reported in App.~\ref{app:heatcool}, we compute heating rates using neutrino luminosities and spectra predicted by SN simulations, reported at a far radius $R_{\rm far}=500$~km from the PNS and we include two relativistic corrections, namely gravitational blueshift/redshift and bending of neutrino geodesic. The latter yields a geometrical factor $1-g_{1}(r)$, with $g_{1}(r)$ given by
\begin{equation}
    g_1(r) = \left[1-\left(\frac{R_\nu}{r}\right)^2\frac{1-2GM/r}{1-2GM/R_\nu}\right]^{1/2}\,,
\end{equation}
where the ratio $(1-2GM/r)/(1-2GM/R_\nu)$ quantifies the effect of bending and would be 1 in Newtonian geometry. 
Additionally, the blueshift factor in the Schwarzschild geometry is given by
\begin{equation}\label{eq:Phi}
    \Phi(r) = \sqrt{\frac{1-2GM/R_{\rm ref}}{1-2GM/r}}\,,
\end{equation}
which becomes unity in the Newtonian geometry.

Close to the PNS, the total heating rate is dominated by the cooling processes because of the
strong dependence on temperature, i.e. $\sim T^6$ for electron capture on nucleons and $\sim T^9$ for electron-positron pair annihilation. Outside the gain radius, temperature becomes lower and neutrino heating processes become dominant. Therefore, matter has to flow to remove the deposited heat, resulting in the formation of the NDO.

\subsection{Properties of neutrino emission}\label{sec:nuprop}

Heating and cooling rates can be computed once neutrino luminosities and spectra are specified. \textcolor{black}{Here we use the same model and parameter values as in Ref.~\cite{Friedland:2023kqp}. However, we explicitly discuss our choices to underline their impact on the electron fraction within the NDO. In particular,} we assume in our model a ``pinched Fermi-Dirac'' parametrization for the neutrino energy distributions~\cite{Keil:2002in}
\begin{equation}
    f_{\nu_i}\propto\frac{1}{e^{E_{\nu_i}/T_{\nu_i}-\eta_{\nu_i}}}\,,
\end{equation}
where $T_{\nu_i}$ is the effective neutrino temperature and $\eta_{\nu_i}$ is the degeneracy parameter, related to the first and second moment of the energy distribution,
\begin{equation}
    \langle E_{\nu_i} \rangle = T_{\nu_i} \, \frac{F_3(\eta_{\nu_i})}{F_2(\eta_{\nu_i})}, \hspace{5mm}
    \epsilon_{\nu_i} = T_{\nu_i} \, \frac{F_4(\eta_{\nu_i})}{F_3(\eta_{\nu_i})} \ ,
\end{equation}
respectively, where $F_n(\eta_{\nu_i})$ are the Fermi integrals $F_n(\eta_{\nu_i}) \equiv \int_0^\infty dx~x^n / (e^{x-\eta_{\nu_i}}+1)$ \textcolor{black}{and $\epsilon_{\nu_i} = \langle E_{\nu_i}^2 \rangle / \langle E_{\nu_i} \rangle$.} We assume both $T_{\nu_i}$ and $\eta_{\nu_i}$ to be time independent for all (anti)neutrino species, with values given in Tab.~\ref{tab:nupar}. 

\begin{table}[t!]
\centering
\begin{tabular}{|c c|c|c|c|}
\hline\hline
Properties & & $\nu_e$ & $\bar{\nu}_e$ &  $\nu_x$ \\
\hline
$L_{\nu_i}(1~{\rm s})$& [erg/s] & $7\times 10^{51}$ & $5.74 \times 10^{51}$ & $7\times 10^{51}$ \\
$T_{\nu_i}$ & [MeV] & $2.67$ & $3.39$ & $3.42$ \\
$\eta_{\nu_i}$&  & $2.1$ & $1.5$ & $0.4$ \\
$\epsilon_{\nu_i}$ & [MeV] & $11.98$ & $14.72$ & $14.22$ \\
$\varepsilon_{\nu_i}$& [MeV] & $13.08$ & $16.16$ & $15.77$ \\
\hline\hline
\end{tabular}
\caption{Benchmark values for the neutrino properties used throughout this work. We consider three species of neutrinos, where $\nu_x$ represents the admixture of heavy-lepton (anti)neutrinos. These values are consistent with the values reported in Ref.~\cite{Huedepohl:2009wh}.}
\label{tab:nupar}
\end{table}
On the other hand, we model the time evolution of (anti)neutrino luminosities in the cooling phase as 
\begin{equation}
    L_{\nu_i}(t)=L_{\nu_i}(t_0)\,e^{-(t-t_0)/t_d}\,,
\label{eq:lnu}
\end{equation}
\textcolor{black}{where $t_0 = 1\,$s is a fixed reference time, $t_d = 3$~s is the exponential decay rate of the luminosities in the cooling phase, and $L_{\nu_i}(t_0)$ a normalization constant, which would be the same for all the flavors if one assumes equipartition. Neutrino luminosities and spectra determine the equilibrium value of the electron fraction in the NDO $Y_{e0}$. This has to be computed including weak magnetism (WM) and recoil corrections to the charged-current opacities for $\nu_e$ and $\bar{\nu}_e$, which reduce the cross-section for $\bar\nu_e$ on $p$ and slightly increase that for $\nu_e$ on $n$, leading to~\cite{Horowitz:1999fe, Burrows:2002jv} }
\begin{equation}\label{eq:ye-wm}
  Y_{e}  \approx \left(1 + \frac{L_{\bar{\nu}_e}}{L_{\nu_e}} \frac{\epsilon_{\bar{\nu}_e}-2\Delta+1.2\Delta^2/\epsilon_{\bar{\nu}_e} - 7.1 {\varepsilon_{\bar\nu_e}^2}/{m_N} }{\epsilon_{\nu_e}+2\Delta+1.2\Delta^2/\epsilon_{\nu_e} + 1.1 {\varepsilon_{\nu_e}^2}/{m_N}} \right)^{-1}\,,
\end{equation}
\textcolor{black}{in weak equilibrium, with $\Delta = 1.293$~MeV the neutron-proton mass difference and $\varepsilon_{\nu_i} = (\langle E_{\nu_i}^3\rangle/\langle E_{\nu_i}\rangle)^{1/2}$. This gives ${Y_e}_0\approx0.6$, when $L_{\nu_e}=L_{\bar{\nu}_e}$ and neutrino energy moments in Tab.~\ref{tab:nupar} are assumed. Such a large value of $Y_e$ is favorable for the synthesis of heavy $p$ nuclides via the $\nu p$-process. If we neglect WM/recoil, the electron fraction is given by ~\cite{Qian:1996xt}}
\begin{equation}\label{eq:ye-nowm}
  Y_e  \approx \left(1 + \frac{L_{\bar{\nu}_e}}{L_{\nu_e}} \frac{\epsilon_{\bar{\nu}_e}-2\Delta+1.2\Delta^2/\epsilon_{\bar{\nu}_e}}{\epsilon_{\nu_e}+2\Delta+1.2\Delta^2/\epsilon_{\nu_e}} \right)^{-1}\,,
\end{equation}
\textcolor{black}{yielding ${Y_e}_0=0.55$ if we assume $L_{\nu_e}=L_{\bar{\nu}_e}$ and $\epsilon_{\nu_i}$ in Tab.~\ref{tab:nupar}. For the prospects of the $\nu p$-process, having $Y_{e0} \approx 0.60$ instead of $0.55$ makes a significant difference; therefore, neglecting WM/recoil corrections is not advisable. 
However, since the open-source reaction network {\tt SkyNet}~\cite{Lippuner:2017tyn,skynet}, used in our nucleosynthesis calculations, does not include these corrections (see Sec.~\ref{sec:nup-method} and App.~\ref{app:setup-skynet}), we approximately mimic them by using a scaled-down value of $L_{\bar\nu_e}$ relative to $L_{\nu_e}$. Specifically, we use $L_{\bar{\nu}_e}=L_{\nu_e}/1.22$ in our case. This choice reproduces an equilibrium electron fraction of $Y_{e0}=0.6$ when Eq.~\eqref{eq:ye-nowm} is evaluated with the neutrino parameters listed in Tab.~\ref{tab:nupar}.
}

\section{Outflow Results: Effects of GR \textcolor{black}{and Equations of State}}
\label{sec:outflow:results}

\textcolor{black}{Here we present numerical results of relativistic outflows, modeling stellar and neutrino properties as specified above.} We consider two progenitor models---18 $M_\odot$ and 12.75 $M_\odot$---both with an assumed PNS mass of 1.8 $M_\odot$. More massive progenitors tend to produce more subsonic outflows with a higher far pressure, while a lighter progenitor, coupled with a relatively fast out-moving front shock, allows the appearance of transonic outflows with termination shocks. \textcolor{black}{We examine the $18\,M_\odot$ model in Sec.~\ref{sec:outflow:results:subsonic} and the $12.75~M_\odot$ case in Sec.~\ref{sec:outflow:results:supersonic}.}
For continuity with earlier works~\cite{Qian:1996xt, Cardall:1997bi, Friedland:2020ecy, Friedland:2023kqp}, we \textit{assume radiation domination} in our thermal ensemble, \textcolor{black}{modeled by incorporating  a variable number of effective RDF. The impact of the assumed EoS is assessed in Sec.~\ref{sec:outflow:gstar}, with a detailed discussion on the role of temperature-dependent RDF in App.~\ref{sec:corr:gstar} and the analysis of the non-relativistic baryon-gas component in App.~\ref{sec:baryonic-gas}.}

\textcolor{black}{As we will show}, the considerable enhancement by GR in entropy \textcolor{black}{per baryon} of $\sim 30$ units reported in the literature \cite{Qian:1996xt, Cardall:1997bi, Otsuki:1999kb, Thompson:2001} is due to the convention of prescribing the same neutrino spectra to relativistic and Newtonian outflows at the neutrinosphere rather than at some far radius, where simulations and observations report luminosities. Relying on such inputs from simulations prompts us to adopt the latter convention where luminosities are prescribed the same values at a far radius for both relativistic and Newtonian outflows. In this case, Newtonian and GR outflows have a much more subdued entropy difference, such as reported in \cite{Friedland:2023kqp}. A fully consistent approach would involve modeling the PNS dynamics together with the outflow in both GR and Newtonian frameworks, which is beyond the scope of the present paper. 

\subsection{GR effects on subsonic outflows} \label{sec:outflow:results:subsonic}

First, we study outflows that remain subsonic, which are expected to be conducive to the $\nu p$-process~\cite{Friedland:2023kqp}. For this purpose, we employ our benchmark  $18~M_\odot$ progenitor model. We use parameters of the PNS and the progenitor at $t=2.5$~s as a representative example.\footnote{As shown in Sec.~\ref{sec:time-ev}, this is within the optimal production time window for $\nu p$ yields.}

\begin{figure} 
\includegraphics[width=\textwidth]{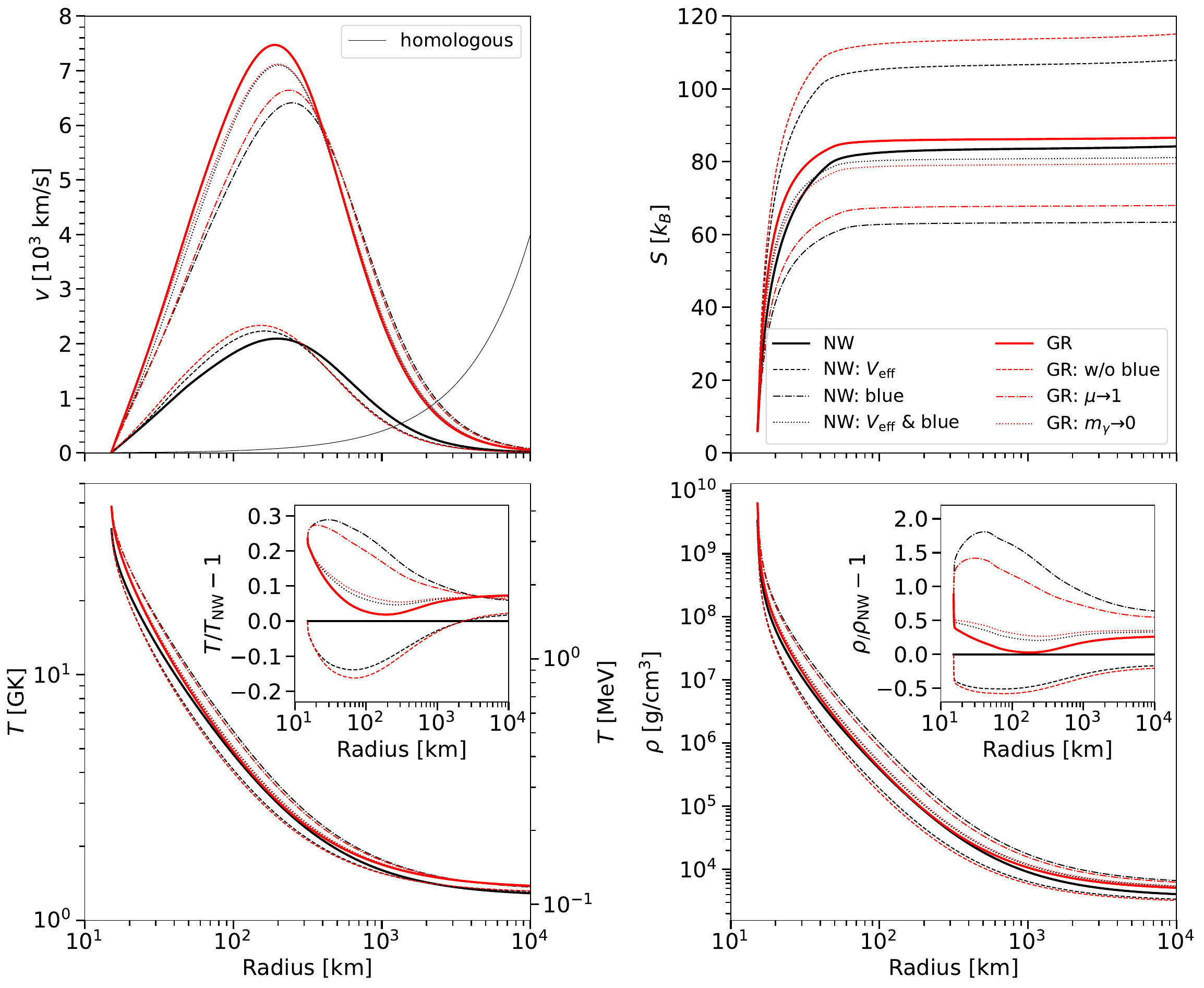}
\centering
\caption{Steady-state outflows with various degrees of relativistic corrections, plotted over radius at $t=2.5$~s for our benchmark $18~M_\odot$ progenitor model. For the temperature and density profiles we also show the relative difference (with the fully Newtonian as reference) in the inset. Newtonian profiles are in black and GR in red. Two classes of GR corrections are shown: those for the hydrodynamic equations and those for neutrino heating (redshift and geodesic bending together). All GR outflows, if not labeled otherwise, use the correct heating terms (with an initial blueshift from 500 km to the neutrinosphere radius), while Newtonian outflows using fully GR heating terms are labeled with `blue'. Also shown is the homologous expansion behind the front shock, $v_h(r)=v_{\mathrm{FS}}~r/R_{\mathrm{FS}}$ (thin black line), used to determine the location where $P_f$ is imposed. 
} 
\label{fig:outflows:gr}
\end{figure}

The steady-state equations of Sec.~\ref{sec:outflow:equations} for general-relativistic outflows are solved, including with the non-relativistic reductions identified in Sec.~\ref{sec:outflow:Newtonian}. We also compare with Newtonian outflows, with and without the effective potential $V_{\mathrm{eff}}$ and GR corrections to heating terms. For each type of outflow, the temperature at the PNS radius is set so that the rates of nucleon heating and cooling are balanced, allowing a sliver of excess in heating. The results are shown in Fig.~\ref{fig:outflows:gr}, where the outflows are classified as either general relativistic (labeled GR---in red) or Newtonian (NW---in black), based on the form of the hydrodynamic equations. Newtonian outflows assume both $\mu=1$ and $m_\gamma = 0$, and the effective potential $V_{\mathrm{eff}}$ of Sec.~\ref{sec:outflow:Newtonian} may be added separate from these reductions. We do not disentangle redshift from bending and have implemented $\dot{q}$ either with fully GR corrections (blueshift from 500 km to the neutrinosphere and subsequent redshift together with geodesic-bending) or none. 

The effect of employing the relativistic heating terms, in particular the blueshift to the neutrinosphere radius, as we have separately verified, is apparent from the plot of velocities (upper left panel), accounting for the discrepancy between the two groups of curves. The peak subsonic velocities can differ by as large as a factor of 3. This drastic acceleration due to blueshift propels the outflow rapidly through the region of neutrino heating, and it prevails over the increase of heating itself (which also increases temperature) to effectively curtail the development of entropy (c.f. Eq.~\eqref{eq:thermo:dS}). Without it, we observe a high entropy from GR hydrodynamics (dashed red), with an enhancement greater than 30 compared to the Newtonian (solid black), as similarly reported by Refs.~\cite{Cardall:1997bi, Otsuki:1999kb, Thompson:2001}. Albeit, in our framework utilizing simulation outputs, we regard as the fully relativistic outflow the GR solution with blueshift (solid red) and the fully Newtonian outflow the NW one without blueshift (solid black). Comparing these two cases, a considerable acceleration and a modicum of entropy gain (here $\sim$ 2) are observed from GR corrections. For the BVP, blueshift also raises both temperature and density, with the factor of increase of $g_*^ST^3$ smaller than that of $\rho_b$ to yield a lower entropy when combined. As shown in the lower panels of Fig.~\ref{fig:outflows:gr}, including blueshift under Newtonian hydrodynamics (dot-dashed black) increases temperature and density by almost $30\%$ and a factor 3, respectively, close to the PNS, with smaller differences at larger radii (compared to solid black). Similarly, neglecting blueshift under GR hydrodynamics (dashed red) reduces temperature by around $10\%$ and density by about $50\%$ almost globally (compared to solid red). 

Having commented on the effect of blueshift, we now turn to examine the relativistic hydrodynamic corrections, when the heating terms are identical. Here we may take the Newtonian outflow with relativistic heating terms as the baseline (dot-dashed black). Of the two GR corrections of $\mu$ and radiation mass, the former's contribution is greater: the outflow assuming $m_\gamma\to 0$ (dotted red) is much more proximate to the fully GR outflow (solid red) than the $\mu\to 1$ outflow (dot-dashed red) is. Indeed, in all four panels, it is clear that $\mu$ accounts for most of the relativistic hydrodynamic correction, specifically in providing the bulk of the GR enhancement in entropy. For reference, using the fully GR outflow as the standard, the component of radiation energy density (and pressure) is always less than $\sim 10\%$ of $\rho_b$. The reduction of adiabatic sound speed reaches $9\%$ at 20 km and then subsides. On the other hand, the relativistic factor $|\mu-1|$ ranges from 0.35 to 0.05, with a steady diminution, in the first 100 km. It is notable that the relativistic correction by $\mu$ (dotted red) on the outflow is well-modeled by the TOV potential (dotted black) for all the state variables. Although $V_{\mathrm{eff}}$ incorporates the radiation component of energy, it does so only to correct the gravitational term (i.e. sound speed is not corrected), rendering it essentially a gravitational correction, as is $\mu$.

Finally, we point out in passing that the decrease in both temperature and density observed by \Otsuki~is also present in these results. Both the incorporation of radiation mass and the relativistic velocity correction $\mu$ contribute to these trends, but again, the lower panels of Fig.~\ref{fig:outflows:gr} show that $\mu$ has the greater impact. Coincidentally, at $r\sim 100$~km, the effect of GR hydrodynamic corrections (mainly $\mu$) and that of blueshift largely cancel, producing similar $T$ and $\rho_b$. One may also notice that at the tail end of the temperature plot, there are intersections of different curves. For the current example, this is a consequence of our definition of the BVP where the radius at which we impose $P_f$ is not fixed. Had we imposed $P_f$ at a fixed, distant radius, there would be no intersection; and by construction under radiation domination, the temperatures would match at that radius. Nonetheless, as $P_f$ drops and the outflows approach criticality, crossings may indeed occur, and we have no reason a priori against them.

In summary, comparing GR outflows including the full panoply of hydrodynamic and heating corrections with Newtonian outflows with none,
\begin{itemize}
    \item The peak subsonic velocity can be increased by a multiple due to blueshift of neutrino energies.
    \item A modest gain in entropy is observed, typically with $\Delta S \lesssim 5$. The enhancement from the GR gravitational correction is counterbalanced by the reduction from blueshift.
    \item $T$ and $\rho_b$ are significantly modified closer to the PNS, and these corrections persist at large radii.
\end{itemize}

We have checked for different far pressures, progenitor models, and front shock velocities that the qualitative observations of this section are not peculiar to the particular example here but hold rather generally for subsonic outflows.

\subsection{GR effects on transonic outflows} \label{sec:outflow:results:supersonic}
\begin{figure} 
\includegraphics[width=\textwidth]{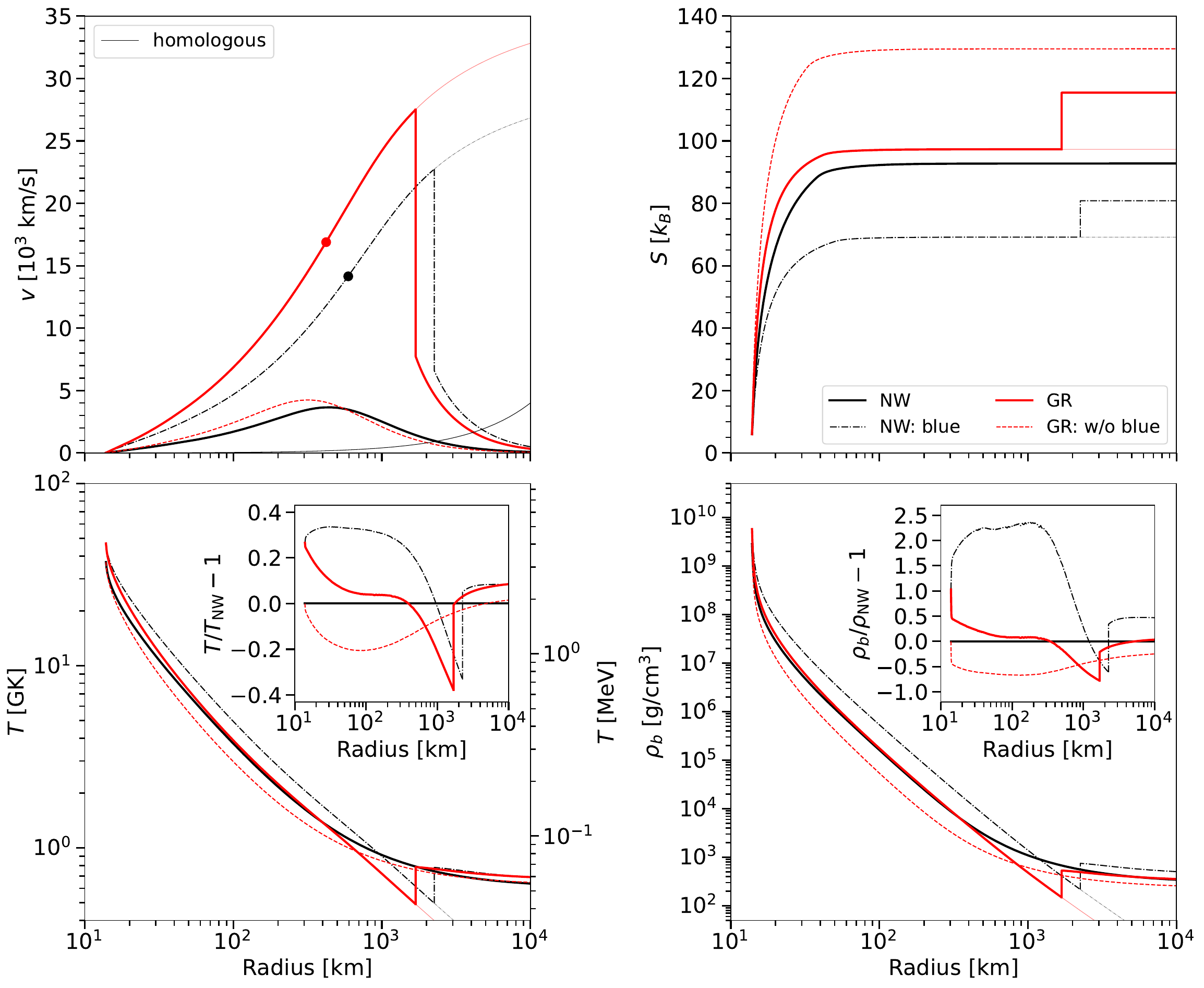}
\centering
\caption{Transonic GR outflow and other selected Newtonian and relativistic outflows for comparison at $t=4$~s for the 12.75 $M_\odot$ progenitor. Outflows with GR hydrodynamic equations are shown in red, NW outflows are in black and the modifier `blue' denotes whether GR heating terms are used. For outflows with termination shocks, the corresponding vacuum solutions are also shown in faded tone and the dots represent the sonic points at 422 km for GR (red) and 594 km for NW with blueshift (black). The thin black line in the upper left panel represents the homologous expansion velocity $v_h$ defined in Eq.~\eqref{eq:FS:homologous}. As shown in the figure, the supersonic solutions arise by the boosted heating from neutrino blueshift to the PNS radius.} 
\label{fig:outflows:supersonic}
\end{figure}

Now we turn to transonic outflows. For a salient example, we consider outflows at $t=4$~s for the 12.75 $M_\odot$ progenitor with FS velocity $v_{\rm FS}=8000$~km/s. As shown in Fig.~\ref{fig:outflows:supersonic}, a sizable termination shock appears in the fully GR profile (solid red). For comparison, a GR (in hydrodynamics) outflow using Newtonian heating terms (dotted red) and a Newtonian (in hydrodynamics) outflow using relativistic heating terms (dot-dashed black) are also displayed, together with a purely Newtonian outflow (solid black). To avoid clutter, we do not show outflows with the intermediate corrections $\mu\to 1$ or $\rho+P\to \rho_b$, for which the Rankine-Hugoniot conditions of Sec.~\ref{sec:outflow:RH} also do not hold exactly.

From the velocity profiles (upper left panel), we clearly see that it is mainly the blueshift of luminosities that causes the supersonic transition. Indeed, transonic outflows are obtained in the fully GR case and in the Newtonian case with blueshift, while the fully Newtonian outflow and the GR one without blueshift remain subsonic.

The qualitative observations made for subsonic outflows also apply to the vacuum solutions, which coincide with the respective transonic outflows up to the shock positions. In particular, before the termination shock, relativistic hydrodynamics considerably increases entropy ($\Delta S\lesssim 30$), and reduces both temperature and baryon density by consistent fractions ($\sim 20\%$ for $T$ and up to $\sim 200\%$ for $\rho_b$) away from the PNS. It is also notable that the hydrodynamic correction to the wind velocity is more persistent than for subsonic outflows, when neutrino blueshift is not considered.

The character of the termination shocks is also affected by relativistic hydrodynamic corrections in an interesting way. The radius of the GR sonic point is smaller than that of the Newtonian case with blueshift, and the relativistic shock appears more pronounced: the entropy gain of the GR outflow (in red) is almost 20, while the Newtonian one (in dot-dashed) is closer to 10. This is due to the difference in the vacuum solutions rather than the modification of the Rankine-Hugoniot conditions. 

Between the fully GR (solid red) and fully Newtonian (solid black) outflows, the supersonic velocity of the former is many times the subsonic velocity of the latter. The modest difference in entropy  ($\Delta S \lesssim 5$) is increased, as expected, after the termination shock. In the temperature and density profiles, there is for each a crossing between Newtonian and GR solutions, and at large radii (post-shock of the GR outflow) their differences are not great.

We did not intend to suggest here that neutrino blueshift from a far radius is the sole cause of supersonic transitions. When the confining pressure $P_f$ becomes sufficiently low, the outflow becomes transonic and forms a termination shock. For example, employing an even lighter ($9\,M_\odot$) progenitor, a lower PNS mass of 1.3~$M_\odot$, and a greater front-shock velocity of $v_\text{FS} = 10^4$ km/s, even the Newtonian outflows (without blueshift) are transonic for $t \gtrsim 1.5$~s. Blueshift for GR outflows induces the transition earlier, at $t \simeq 1.0$~s.

It is worth noting that the outflow solutions obtained with the $12.75\,M_\odot$ progenitor boundary conditions were found to be safely subsonic in Ref.~\cite{Friedland:2023kqp}. With the additional physics refinements introduced in this work, including GR outflow equations with all the appropriate terms, additional heating terms, and the more careful treatment of relativistic degrees of freedom and the far boundary condition, the outflows get faster in general, resulting in the $12.75\,M_\odot$ progenitor conditions being near-critical rather than subsonic.

\subsection{\textcolor{black}{Impact of equation of state}}
\label{sec:outflow:gstar}

The dependence of Eqs.~\eqref{eq:std:dT}–\eqref{eq:std:du} on the variable RDF, $g_{*}(T)$, and on the individual pressure components implies that the evolution of the outflow is sensitive to the assumed EoS. In this context, the literature spans a wide range of approaches: Refs.~\cite{Qian:1996xt,Cardall:1997bi} adopt fixed RDF; Refs.~\cite{Friedland:2020ecy,Friedland:2023kqp} allow for a variable $g_*(T)$ but with an approximate implementation; and Refs.~\cite{Otsuki:1999kb,Thompson:2001} derive equations that, in principle, support a variable $g_*$, although the EoS employed is not readily specified.

\begin{figure}
\includegraphics[width=0.95\textwidth]{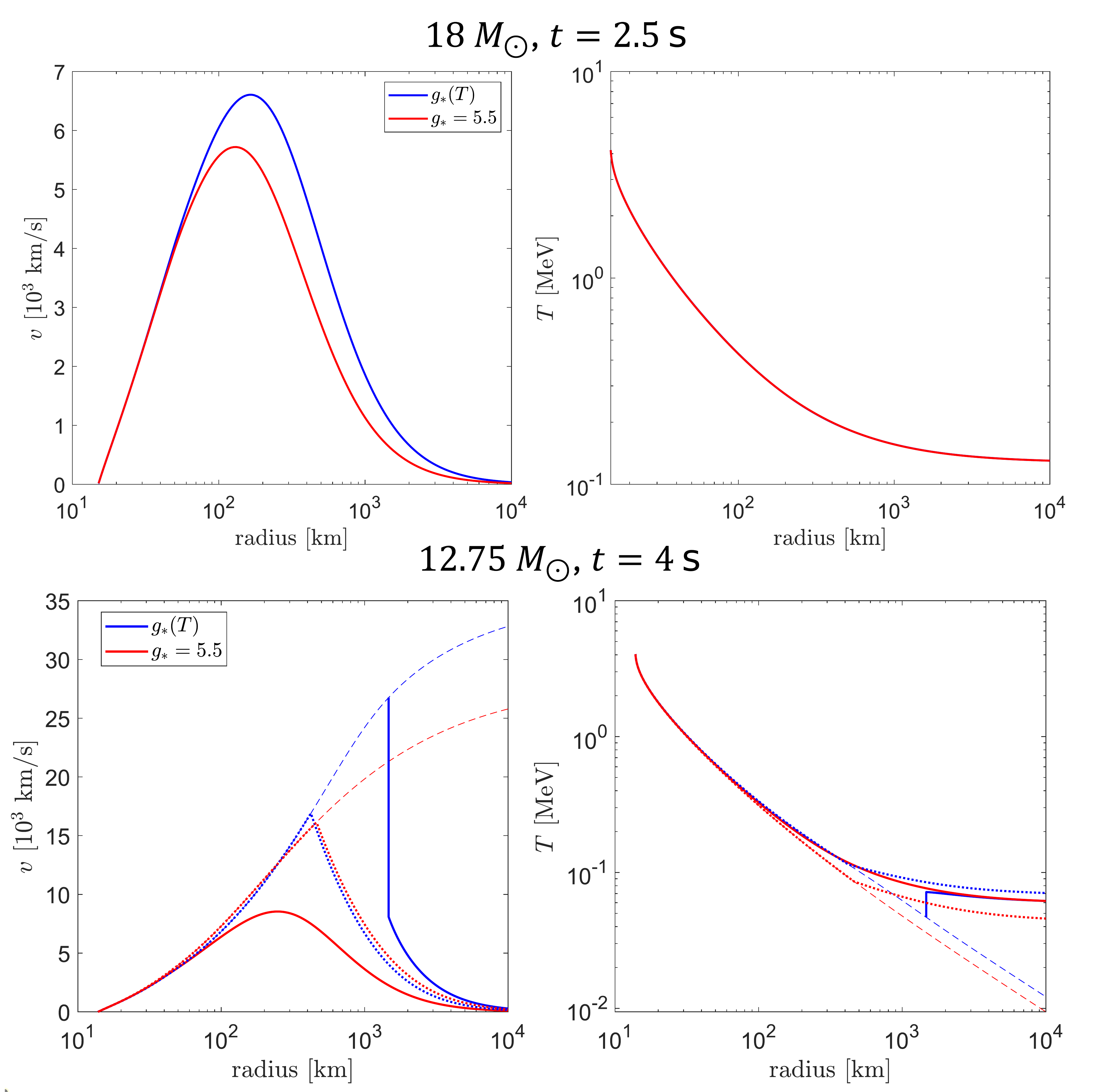}
\centering
\caption{Velocity (left panels) and temperature (right panels) profiles of outflows with constant $g_*=5.5$ (red) versus variable RDF $g_*(T)$ (blue), assuming radiation domination, for the $18$ $M_\odot$ model at $t=2.5$~s (upper panels) and the 12.75 $M_\odot$ model at $t=4$~s (lower panels). In the upper right panel, the temperature profiles overlap. In the lower panels, the outflows with far temperature $T_f$ imposed at $r=10^4$~km are shown in solid lines. For reference, the vacuum solutions (dashed lines) and the critical subsonic solutions (dotted lines) are also shown.} 
\label{fig:gstarbvp-both}
\end{figure}

The assumption of a constant $g_*=5.5$ is not adequate once the temperature drops below the electron mass as the outflow expands. To assess the impact of this assumption, we compare outflows computed with a self-consistent variable $g_*$ to those with a constant one. We adopt the convention that the far temperature $T_f$ under radiation domination is the same in both cases (see App.~\ref{app:gstar} for more details).

For outflows deeply in the subsonic regime (e.g., in the $18\,M_\odot$ model), including $g_*(T)$ leads to a modest increase in the velocity (upper left panel of Fig.~\ref{fig:gstarbvp-both}) and has minimal impact on the temperature (upper right). Entropy profiles remain essentially unchanged, while baryon density is decreased by tens of percent at $r\gtrsim 120$~km. The impact is more pronounced for transonic outflows. As shown in the lower panels of Fig.~\ref{fig:gstarbvp-both}, significant differences occur near the critical point, where the temperature is close to the electron mass and $g_*(T)$ deviates from 5.5. Although the entropy changes only slightly, both velocity (lower left) and temperature (lower right) increase, and the baryon density decreases. For outflows compared assuming the same $T_f$, fixing $g_*=5.5$ overestimates the far pressure by a factor $\sim 2.75$, making NDO solutions more subsonic. Therefore, when outflows are near-critical, replacing variable $g_*(T)$ (blue lines) with $g_*=5.5$ (red lines) can miss the transonic character of the outflow altogether. In fact, for our 12.75 $M_\odot$ model, outflow with constant $g_*=5.5$ remain subsonic at late times. An extended discussion and analysis of the results described here can be found in App.~\ref{sec:corr:gstar}.

Another refinement concerns the baryonic gas contribution to the hydrodynamic equations. While such contribution can be subdominant within the bulk of the NDO region, it is important within the first few kilometers above the PNS and also affects the far boundary conditions---where entropy is low. We found that including the baryonic gas in the EoS effectively preserves the total asymptotic entropy, lowers the radiation entropy, and generally makes the flow more subsonic. While we neglect baryonic gas contributions to the EoS in our main results, its influence is analyzed in App.~\ref{sec:baryonic-gas}.

\section{$\nu p$-process: Stages and Conditions}\label{sec:nup-stages}

In the previous sections we have analyzed how general relativistic corrections affect the NDO. From now on, we will focus on their effects on the production of $p$ nuclides via the $\nu p$-process. This nucleosynthesis process proceeds in a sequence of thermodynamic stages as the outflow from the PNS expands and cools. Each stage is characterized by different nuclear reactions and equilibrium conditions that determine the evolution of the composition. Below, we summarize the key stages of the $\nu p$-process \textcolor{black}{(see also Refs.~\cite{Frohlich:2005ys, Wanajo:2010mc, Friedland:2023kqp}).}

\smallskip
\textbf{\textit{\textcolor{black}{Starting conditions: Weak equilibrium and NSE}}}.--
The nucleosynthesis path starts in the proximity of the PNS, where the temperature is a few MeV ($T\gtrsim 10$~GK) and the density is above $10^9 \ {\rm g/cm^3}$. In these conditions, the plasma is in nuclear statistical equilibrium (NSE), which means that $n$ and $p$ form nuclei which are immediately broken apart, so that the abundances do not change. Additionally, the charged current weak processes $\nu_e \, n \rightleftharpoons e^- \, p$ and $\bar\nu_e \, p \rightleftharpoons e^+ \, n$ are also in equilibrium, so that, locally, the electron fraction $Y_e$ depends on the relative fluxes and energy spectra of $\nu_e$ and $\bar\nu_e$. As the temperature drops to around $T \sim 9\,\mathrm{GK}$, $\alpha$-particles ($^4$He) begin to form in significant amounts. As long as $T > 6\,\mathrm{GK}$, the triple-$\alpha$ reaction ($3\alpha \rightleftharpoons {^{12}\mathrm{C}}$) remains in equilibrium, but little net production of heavier nuclei occurs. 

\smallskip
\textbf{\textit{Stage I: Seed-formation and iron-group QSE}}.-- 
As the temperature decreases below $T < 6\,\mathrm{GK}$, the triple-$\alpha$ reaction falls out of equilibrium, and seed nuclei with mass number $A \gtrsim 12$ begin to form in quasi-statistical equilibrium (QSE). This phase (for ${T>4~{\rm GK}}$) leads to the production of the seed nucleus $^{56}{\rm Ni}$, representing the starting point of the $\nu p$-process. Seeds remain in QSE with heavier nuclei until the temperature drops further (to 3 GK). It is therefore instructive to define the timescale $\tau_1$, i.e. the time spent between $T=6$~GK and $T=3$~GK, representing the duration of the seed production window~\cite{Wanajo:2010mc}. {Another useful indicator is the entropy per baryon of the radiation component during this stage---a higher entropy implies a lower baryon density for a given temperature, and consequently, a slower triple-$\alpha$ reaction, stymieing seed nuclei formation.}

\smallskip
\textbf{\textit{Stage II: Iron group QSE freeze-out and proton + neutron capture}}.-- 
Outflow continues to cool down. When $T < 3\,\mathrm{GK}$, the system freezes out of QSE, and the $\nu p$-process starts, with the nucleosynthesis proceeding beyond $^{56}{\rm Ni}$. In the temperature window $3\,\mathrm{GK} > T > 1.5\,\mathrm{GK}$, a subdominant population of free neutrons is produced via antineutrino captures on free protons, $p(\bar{\nu}_e,e^+)n$. 
This process is facilitated if the outflow has a large number of free protons per $^{56}{\rm Ni}$ seed nucleus at $T=3$~GK. The subdominant population of neutrons that results from neutrino capture triggers $(n,p)$ and $(n,\gamma)$ exchange reactions that allow the nucleosynthetic flow to bypass slow $\beta^+$-decay waiting points. Combined with proton captures $(p,\gamma)$, this enables the synthesis of nuclei along the $rp$-process chain. The longer its duration, the higher the efficacy of this process. Therefore, it is useful to define the timescale $\tau_2$, i.e. the time spent between $T=3$~GK and $T=1.5$~GK, representing the duration of the $\nu p$-process. The ratio $\tau_2/\tau_1$ is a good indicator of the efficacy of the $\nu p$-process. In this context, another useful parameter is the ratio of neutrons produced (by antineutrino capture on protons) to seed nuclei, given by 
\begin{equation}\label{eq:deltan}
    \Delta_n = \frac{Y_p\,n_{\bar{\nu}_e}}{Y_{A\geq12}}\,,
\end{equation}
where $Y_p$ and $Y_{A\geq 12}$ are the proton and the seed fractions at $T=3$~GK, while $n_{\bar{\nu}_e}$ represents the number of $\bar{\nu}_e$ captured per free proton \textcolor{black}{in the temperature range $3\,\mathrm{GK} > T > 1.5\,\mathrm{GK}$}, leading to neutron formation. More explicitly~\cite{Wanajo:2010mc}
\begin{equation}
    n_{\bar{\nu}_e}=\int_{T=3~{\rm GK}}^{T=1.5~{\rm GK}} \lambda_{\bar{\nu}_e} dt \,,
\end{equation}
where $\lambda_{\bar{\nu}_e}$ is the rate for electron antineutrino capture on free protons $p(\bar{\nu}_e,\,e^+)n$. This can be approximately written as~\cite{Pruet:2005qd}
\begin{equation}
    \lambda_{\bar{\nu}_e}\approx 0.06\,\frac{L_{\bar{\nu}_e}}{10^{52}~\erg/\sec}\,\frac{T_{\bar{\nu}_e}}{4~\MeV}\left(\frac{10^8~\cm}{r}\right)^2\,,
\label{eq:lambdabarnu}
\end{equation}
which is a function of the neutrino emission properties and the distance to the neutrinosphere. The dependence of the rate on $r$ in Eq.~\eqref{eq:lambdabarnu} encodes the scaling of the antineutrino flux $\mathcal{F}_{\bar{\nu}_e}\propto r^{-2}$. Therefore, $\lambda_{\bar{\nu}_e}$ will decrease with distance, while  $\Delta_n$ decreases as well making the $\nu p$-process less efficient if the \textcolor{black}{the $3\,\mathrm{GK} > T > 1.5\,\mathrm{GK}$ region} is located farther from the PNS. \textcolor{black}{It is useful to define two characteristic radii, $R_3$ and $R_{1.5}$---corresponding to temperatures of $T = 3$~GK and $T = 1.5$~GK, respectively---which mark the spatial boundaries of the region  where $(p,\gamma)$, $(n,p)$ and $(n,\gamma)$ processes can operate in conjunction to synthesize nuclei beyond the iron group following QSE freeze-out.} 

\smallskip
\textbf{\textit{Stage III: Proton capture freeze-out and late-time neutron captures}}.-- As the temperature falls below $T < 1.5\,\mathrm{GK}$, Coulomb barriers prevent further proton captures via $(p,\gamma)$ reactions. However, free neutron production via neutrino interactions on free protons $p(\bar{\nu}_e,e^+)n$ continues at these late times. The number of neutrons produced from neutrino captures on protons is given by~\cite{Wanajo:2010mc}
\begin{equation}
    \Delta_n^\prime = \frac{Y_p\,n_{\bar{\nu}_e}^\prime}{Y_{A\geq12}}\,,
\label{eq:deltanprime}
\end{equation}
where $n_{\bar{\nu}_e}^{\prime}$ is the number of $\bar{\nu}_e$ captured per free proton in the temperature range $T<1.5$~GK
\begin{equation}
    n_{\bar{\nu}_e}^\prime=\int_{T\leq1.5~{\rm GK}} \lambda_{\bar{\nu}_e} dt \ .
\end{equation}
These late-time neutrons enable additional $(n,\gamma)$ and $(n,p)$ reactions that help push the composition toward the valley of stability. Notably, the metastable isotope $^{92}\mathrm{Nb}$, which is shielded from $\beta^+$ decays of $p$-rich nuclei by the stable $^{92}$Mo, 
can be synthesized in the $\nu p$-process in this way. The late-time availability of a few neutrons per seed nucleus ($\sim$ 3--6) proves to be optimal for the efficient production of these isotopes~\cite{Friedland:2023kqp}.

\smallskip
\textbf{\textit{Stage IV: Late-time decays to stability}}.-- At late times, by $t \sim 10^9\,\mathrm{s}$, the $\beta^+$ decays of unstable nuclei complete the formation of stable isotopes.

\section{\textcolor{black}{Setup of Nucleosynthesis Calculation and Time Evolution}}
\label{sec:nup-method}

To compute the abundances of nuclides synthesized in the $\nu p$-process, we post-process the tracer trajectories described in Sec.~\ref{sec:traj} with the open-source reaction network {\tt SkyNet}~\cite{Lippuner:2017tyn,skynet}. We assume that a tracer trajectory built from an outflow at time $t$ corresponds to a mass element launched at time $t_{\rm launch}\equiv t$---in essence, this means evolving the luminosity for each tracer in accordance with Eq.~\eqref{eq:lnu} starting from $t_\text{launch}$. For each such tracer, we use as {\tt SkyNet} inputs the time evolution of its temperature $T$ and density $\rho$, its location $r$, as well as the time-evolving PNS radius, $R_{\rm PNS}$, and the constant PNS mass, $M_{\rm PNS}$. The inclusion of a realistic, time-varying $R_{\rm PNS}(t)$ represents an improvement over the constant-radius approximation adopted in Ref.~\cite{Friedland:2023kqp}. Moreover, we provide {\tt SkyNet} with the time-dependent electron (anti)neutrino luminosities $L_{\nu_i}$ ($\nu_i = \nu_e,\,\bar{\nu}_e$) and spectral parameters $T_{\nu_i}$ and $\eta_{\nu_i}$, defined in Sec.~\ref{sec:heatcool}. To be consistent with the outflow solutions, we also take into account GR effects on the neutrino temperatures and luminosities when appropriate.\footnote{In {\tt SkyNet}, for each of the eight cases encompassing the different relativistic corrections outlined in Sec.~\ref{sec:outflow:results}, an appropriate binary choice of blueshift vs no-blueshift (consistent with the respective outflow solutions) is made for the neutrino luminosities and temperatures.} Additional details of the {\tt SkyNet} setup are provided in App.~\ref{app:setup-skynet}.

Each {\tt SkyNet} run allows us to compute instantaneous yields produced in a single mass element launched at $t_{\rm launch}$. However, to compare our results with the solar system abundances, the computation of the time-integrated yields is required. Thus, we sequentially solve outflow equations starting at $t_{\rm launch}= 1$~s, and advancing in steps of $\Delta t_{\rm launch} = 0.1$~s. This represents a much finer-grained sequence of trajectories across $t_\text{launch}$, compared to Ref.~\cite{Friedland:2023kqp}, allowing us to investigate the detailed time-dependence of the various isotopic yields of interest, namely, those of $\Mo$, $\Ru$, and $\Nb$---an analysis that was beyond the scope of Ref.~\cite{Friedland:2023kqp}. At each launch time we obtain a distinct outflow, reflecting the evolving boundary conditions and neutrino heating rates governed by the time dependence of $R_{\rm PNS}(t)$ [Eq.~\eqref{eq:pns-fit}], $R_{\rm FS}(t)$ [Eq.~\eqref{eq:RFS}], and $L_{\nu_i}(t)$ [Eq.~\eqref{eq:lnu}]. For each such outflow, we compute the tracer trajectories and provide {\tt SkyNet} with the quantities previously described. In this way, we simulate the SN evolution and the resulting yields during all the cooling phase.

\section{Results: Analysis of GR Effects on Instantaneous Yields}
\label{sec:8cases}

The results in Sec.~\ref{sec:outflow:results:subsonic} show that relativistic corrections to neutrino heating and to the hydrodynamic equations in spherical symmetry affect the evolution of the NDO. Here, we assess the impact of these corrections on the nucleosynthetic yields, by focusing on instantaneous yields at $t_{\rm launch} = 2.5$~s\,\footnote{The same value of $t$ is considered in Sec.~\ref{sec:outflow:results:subsonic} to assess GR effects on the NDO, since it is within the optimal production time window for $\nu p$ yields.}  for the $18~M_\odot$ progenitor model.  We systematically incorporate relativistic corrections one at a time, allowing us to disentangle their effect on the nucleosynthesis yields, which is unknown a priori.

\begin{figure}[t!]
\centering
\includegraphics[width=0.85\columnwidth]{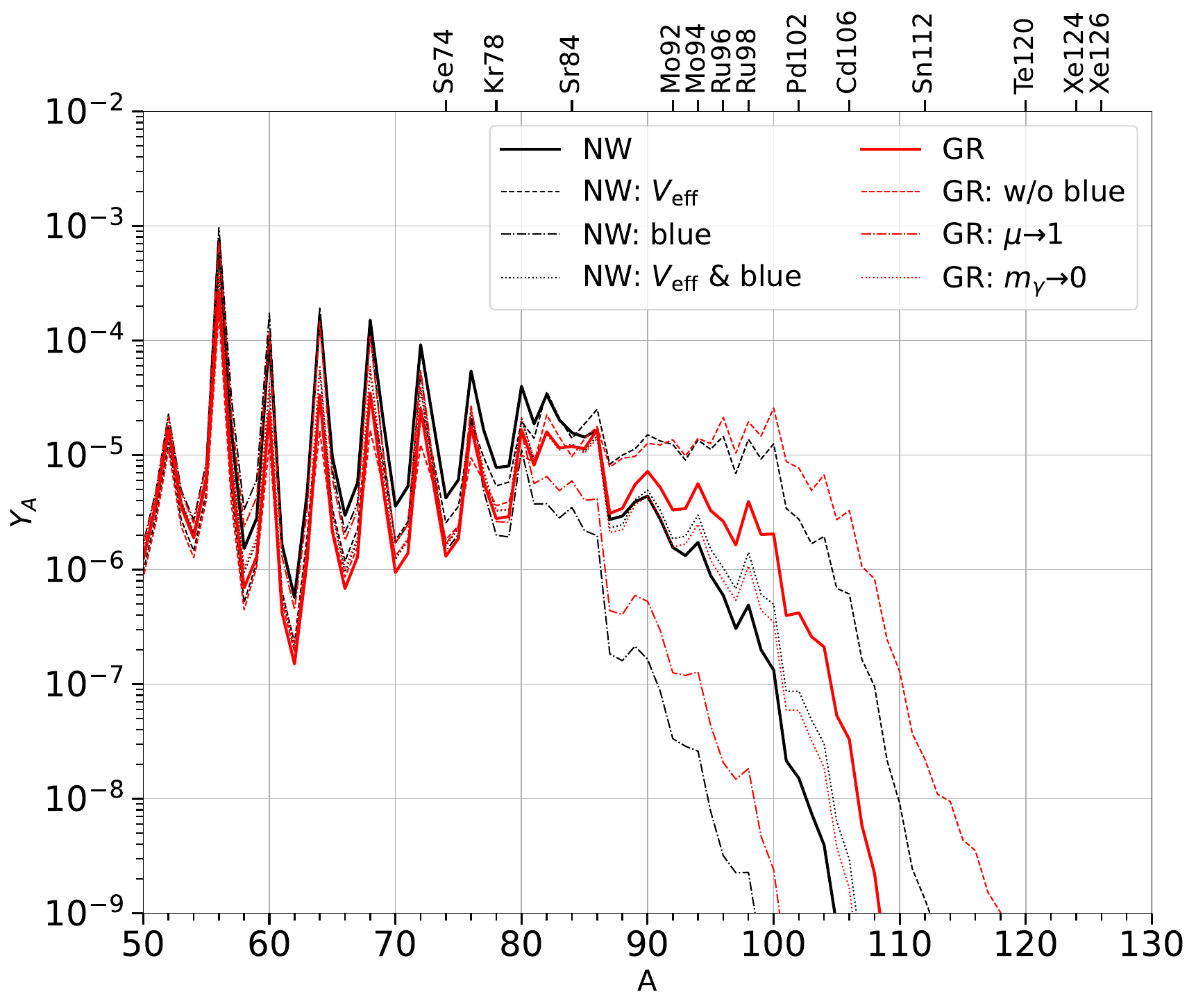}
  \caption{Instantaneous yields at $t_{\rm launch}=2.5$~s for different test cases based on Newtonian (black) and GR (red) equations and different corrections, in our $18~M_{\odot}$ progenitor model. See the caption of Fig.~\ref{fig:outflows:gr} and the main text for more details.}
  \label{fig:cases}
\end{figure}

In Fig.~\ref{fig:cases}, we show the instantaneous abundances $Y_A$ of the nuclides as a function of the mass number $A$ for the different test cases of interest, previously introduced in Sec.~\ref{sec:outflow:results:subsonic}.
We see that the yields obtained in the fully GR case  (solid red line), when considering all the relativistic corrections, are larger than the ones in the fully Newtonian case  (solid black line) for $A\gtrsim 90$. In particular, the production of the $\Mo$ and $\Ru$ isotopes is increased by a factor of few, while for heavier elements the increase is larger than one order of magnitude. Such an increase can be explained by observing the values of the parameters characterizing the efficiency of the $\nu p$-process, shown in Tab.~\ref{tab:cases}. In particular, although $S$ in both cases are comparable, with the entropy being slightly higher in the fully GR case (see Sec.~\ref{sec:outflow:bvp}), the ratio $\tau_2/\tau_1$ is three times larger in GR than in NW. Such a ratio results in $\Delta_n\approx8$ for GR, larger than $\Delta_n \approx 5$ in NW, revealing that the $\nu p$-process is more efficient in the former case. Remarkably, also $\Delta_n^\prime$ is larger in GR, leading to an increase in the $\Nb$ production by a factor of $\sim$5. 

The factor of a few difference in the abundance of elements with $A\approx 90\text{--}100$ comes from the combination of the different corrections to the heating rates and to the hydrodynamic equations, affecting the yields in different ways. Indeed, as shown in Fig.~\ref{fig:cases}, excluding one (or several) of such corrections would lead to completely different yields, with a discrepancy up to 4 orders of magnitude in the  $90 \lesssim A \lesssim 105$ atomic mass window. Interestingly, the case with \textcolor{black}{the largest yields for $A\gtrsim90$ is GR without blueshift} (dashed red), with $Y_A \approx 10^{-5}$ for $A\lesssim 100$, while the \textcolor{black}{smallest ones are obtained when considering NW with blueshift, where $Y_A$ decreases from $\sim 10^{-7}$ at $A=90$ to below $10^{-9}$ at $A=100$.} This reveals the substantial impact of the blueshift-corrected neutrino heating terms on the yields. 

This impact can also be understood from the quantities shown in Tab.~\ref{tab:cases}. 
\begin{table}[t!]
\centering
\begin{tabular}{|c|c|c|c|c|c|c|c|c|c|}
\hline\hline
Case & $S$ & $\tau_1$ & $\tau_2$ & $\tau_2/\tau_1$ & $R_{3}$ & $R_{1.5}$ &$\Delta_n$ & $\Delta_n^\prime$ & $\dot{M}$ \\
 & [$k_B$] & [s] & [s] &  &  [km] &  [km] & &  & [$10^{-5}~M_\odot/{\rm s}$] \\
\hline
NW &  83.6 & 0.064 & 1.208 & 18.92 & 199.6 & 1366 & 5.00 & 0.85 & 4.48  \\
NW: blue &  63.4 & 0.028 & 1.113 & 40.11 & 267.6 & 2413 & 2.00  & 0.35 & 32.6 \\
NW: $V_{\rm eff}$ &  108 & 0.049 & 1.155 & 23.56 & 162.6 & 1212 & 13.9 & 2.43 & 2.49 \\
NW: $V_{\rm eff}$ \& blue &  81.1 & 0.020 & 1.043 &  51.36 & 215.4 & 2130 & 6.00 & 1.09 & 18.4 \\
GR &  86.6 & 0.018 & 1.043 & 56.48 & 206.6 & 2097 & 8.11 &  1.48 & 16.2 \\
GR: w/o blue & 115 & 0.045 & 1.150 & 25.72 & 156.7 & 1195 & 19.1 & 3.29 & 2.22 \\
GR: $\mu\to 1$ & 68.0 & 0.026 & 1.083 & 42.10 & 255.7 & 2338 & 2.67 & 0.48 & 28.3 \\
GR: $m_\gamma \to 0$ & 79.5 & 0.020 & 1.062 & 52.06 & 218.8 & 2161 & 5.62 & 1.01 & 19.1 \\
\hline\hline
\end{tabular}
\caption{Main features of the different cases we consider, involving relativistic corrections to the hydrodynamic equations and neutrino heating, {for a trajectory launched at $t_{\rm launch} = 2.5$~s in our $18\,M_\odot$ progenitor model}. See also Fig.~\ref{fig:outflows:gr} for the NDO solutions and Fig.~\ref{fig:cases} for the associated instantaneous yields.}
\label{tab:cases}
\end{table}
On one hand, models with blueshift have outflows that expand faster ($v\approx6000~{\rm km/s}$) than the ones without it (almost a factor of three lower). Thus, naively one would expect the yields to be \textcolor{black}{greater} in the faster case due to a larger $\tau_2/\tau_1\approx50$. However, blueshift can reduce the entropy by $\Delta S\approx20$ in the NW cases \textcolor{black}{(see ``{NW}'' vs ``{NW: blue}'' and ``{NW: $V_{\rm eff}$}'' vs ``{NW: $V_{\rm eff}$} \& {blue}''),\footnote{\textcolor{black}{Throughout the text, we use the expression ``{A}'' vs ``{B}'' to refer to a comparison between the case labeled as A and the one labeled as B.}}} and $\Delta S\approx30$ in the GR scenario \textcolor{black}{(i.e., ``{GR}'' vs ``{GR:  w/o blue}'')}. Importantly, although $dS/dr \propto \dot{q}$ (see Eq.~\eqref{eq:thermo:dS}) and a blueshift-enhanced $\dot{q}$  would seek to increase $S$, the tracer quickly escapes the region of large heating. As a result, there is less total heating overall leading to a smaller increase in $S$ compared to the runs without blueshift.
This extra push to larger radii results also in smaller antineutrino fluxes during \textcolor{black}{stage II}, since they scale as $r^{-2}$, reducing $\Delta_n$, as discussed in Sec.~\ref{sec:nup-stages}.  Indeed, in all the cases including blueshift, the \textcolor{black}{$3\,\mathrm{GK} > T > 1.5\,\mathrm{GK}$ region} is located farther from the PNS, with $R_{3}\gtrsim 200$~km and $R_{1.5}\gtrsim 2000$~km, as shown in Tab.~\ref{tab:cases} (in contrast with the cases without blueshift, which all have $R_{3} < 200$\,km and $R_{1.5} < 1400$\,km). The combination of all these aspects leads to a value of $\Delta_n \approx 19$ for the GR scenario without blueshift, which is more than a factor 2 larger than $\Delta_n = 8$ for the fully GR case, and almost an order of magnitude larger than $\Delta_n = 2$ for NW with blueshift, resulting in the significant difference in the yields observed in Fig.~\ref{fig:cases}.

Besides blueshift, the relativistic hydrodynamic corrections discussed in Sec.~\ref{sec:outflow:equations} also affect the $\nu p$ yields, due to the changes in the outflows shown in Fig.~\ref{fig:outflows:gr}. As stressed in Sec.~\ref{sec:outflow:results:subsonic}, the largest GR correction stems from setting $\mu \to 1$ (the dot-dashed line in Fig.~\ref{fig:cases}), which reduces the yields by more than one order of magnitude at $A\gtrsim 90$ compared to the fully GR case (solid red). As shown in Tab.~\ref{tab:cases}, taking the limit $\mu \to 1$ decreases the entropy by approximately $\Delta S \approx 20$, reduces the ratio $\tau_2/\tau_1$ by 25\% due to a slightly slower outflow (see the upper-left panel of Fig.~\ref{fig:outflows:gr}), and shifts the \textcolor{black}{$3\,\mathrm{GK} > T > 1.5\,\mathrm{GK}$ region} more than $10\%$ farther out as a result of the increased temperature (lower-left panel of Fig.~\ref{fig:outflows:gr}). The combination of these aspects amounts to decreasing $\Delta_n$ down to $\sim 3$, lowering the yields. 

On the other hand, neglecting the radiation mass (GR: $m_\gamma \rightarrow0$, see the dotted red line in Fig.~\ref{fig:cases}) has a smaller impact on the yields. As shown in Tab.~\ref{tab:cases}, this correction lowers the entropy by $\Delta S \approx 7$ and leads to a negligible change in both $\tau_2/\tau_1$ and the location of the \textcolor{black}{$3\,\mathrm{GK} > T > 1.5\,\mathrm{GK}$ region} compared to the fully GR case. As a result, the yields are reduced by no more than a factor of a few in the atomic mass window $90<A<100$.

As expected, we obtain yields similar to the GR case with $m_\gamma \rightarrow0$ (dotted red line) when we consider NW with the TOV potential $V_{\rm eff}$ and blueshift (dotted black line). 
As discussed in Sec.~\ref{sec:outflow:bvp}, $V_{\rm eff}$ makes the gravitational potential deeper, leading to a larger entropy than the NW one. If we incorporate $V_{\rm eff}$ while neglecting blueshift, we observe an even larger increase in the entropy and the yields are significantly enhanced (see the dashed black line), becoming comparable to the yields in the GR case without blueshift\,\footnote{Even better agreement would be observed between the NW case with $V_{\rm eff}$ and no blueshift and the GR case with $m_\gamma\to0$ and no blueshift; however, the latter is not shown in Fig.~\ref{fig:cases}.} (dashed red line).

As we will further discuss  in Sec.~\ref{sec:time-ev}, the absolute abundance of elements produced in the $\nu p$-process depends not only on the instantaneous yields $Y_A$ but also on the mass outflow rate $\dot{M}$, which determines how much material is ejected by the NDO (as fixed by the outer boundary condition). Therefore, we show in the last column of Tab.~\ref{tab:cases} the different values of $\dot{M}$ in the considered cases, {computed at sufficiently far radius (where $u\approx v$)}. As expected, blueshifted luminosities make the mass outflow rate higher due to more efficient neutrino heating. 
In particular, for each pair of cases where we can compare the result with vs without blueshift (i.e., \textcolor{black}{``{NW}'' vs ``{NW: blue}'', ``{NW: $V_{\rm eff}$}'' vs ``{NW: $V_{\rm eff}$} \& {blue}'', and ``{GR}'' vs ``{GR: blue}''}), we observe an increase in $\dot{M}$ by a factor slightly larger than $\sim 7$. \textcolor{black}{This implies} that the corrections to neutrino heating could also have a positive feedback on the production of the $\nu p$ nuclides (positive feedback on $\dot M$ but negative feedback on instantaneous yields---so the net effect could go either way).

The discussion in this section highlights the importance of self-consistently including all relativistic corrections in order to accurately assess the nucleosynthetic yields. Neglecting one or more of these corrections would result in a significant error in the estimation of the abundances.\,\footnote{Even the smallest GR correction, i.e., $m_\gamma \to 0$, would lead to an underestimation of the yields compared to the fully GR case by a factor of a few in the window $90\lesssim A \lesssim 100$ and by about one order of magnitude for heavier elements.} 

Even though the discussion above was based on a representative snapshot of the outflow, we have verified that different snapshots (i.e. different launch times) qualitatively agree with the results reported in Fig.~\ref{fig:cases}. In the following, we will focus only on the two physically relevant outflow solutions, namely full NW and full GR, while incorporating the notions of time evolution and time-integrated yields.

\section{Results: GR Effects on Integrated Yields in the Cooling Phase}

\label{sec:glob-comp}
In this section, we analyze the time evolution of the abundances produced in the NW and GR cases and identify the key differences that impact the efficacy of the $\nu p$-process. We then examine how these differences shape the yields integrated over time and assess how well they can explain the solar abundances found in meteorites.

\subsection{Time evolution}
\label{sec:time-ev}
 
\begin{figure}[t!]
\centering
  \includegraphics[width=0.8\columnwidth]{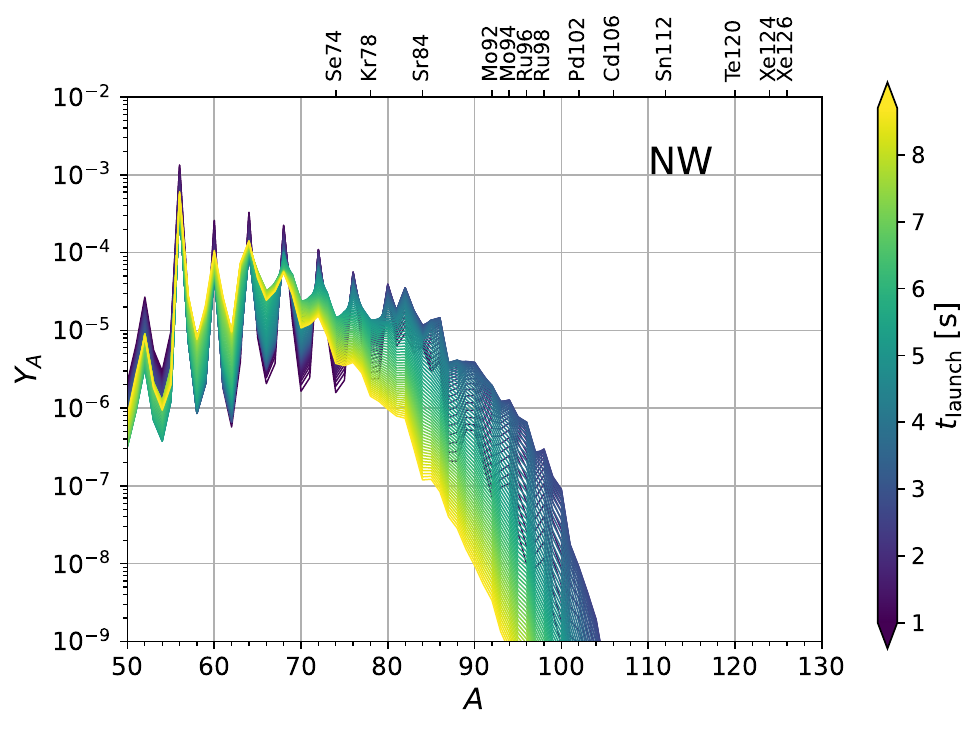}
    \includegraphics[width=0.8\columnwidth]{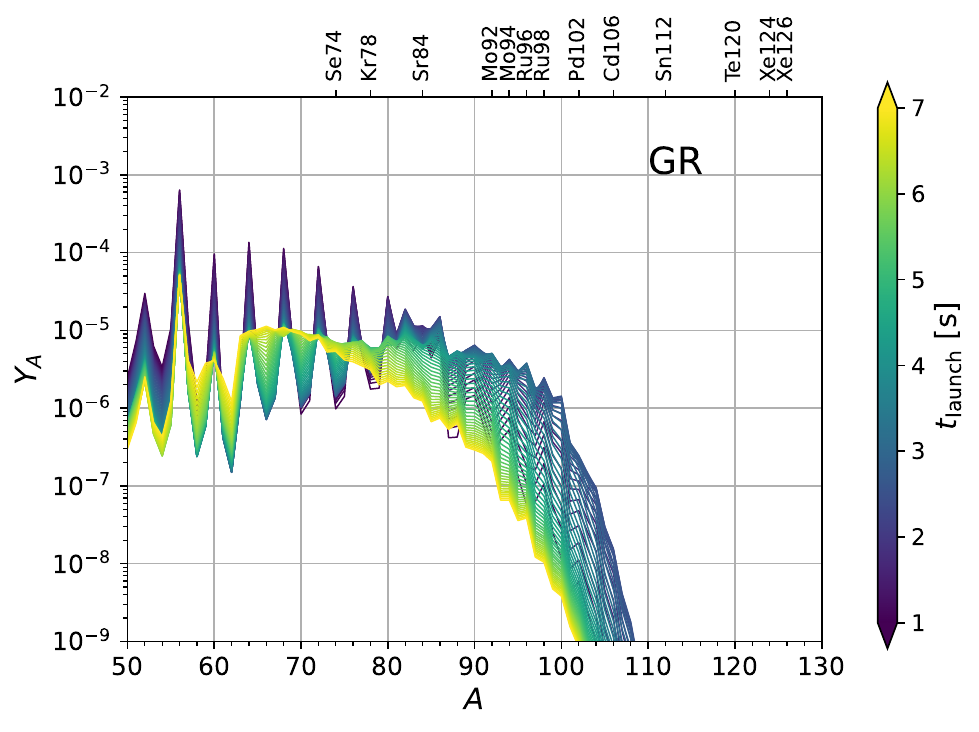}
  \caption{Instantaneous yields for NW (upper) and GR (lower) for different launch times $t_{\rm launch}$, in our $18~M_{\odot}$ progenitor model. Darker \textcolor{black}{purple} lines correspond to earlier launch times ($t_{\rm launch}\approx 1$~s), while \textcolor{black}{brighter yellow} curves are related to later ones.   
  }
  \label{fig:instyields}
\end{figure}

We model the SN explosion and the subsequent $\nu p$ yields through the methods described in Sec.~\ref{sec:nup-method}. Here we focus on the time evolution of the yields produced by subsonic NDOs in the cooling phase ($t_{\rm launch}\lesssim~8.5$~s in the NW case and $t_{\rm launch}\lesssim~7$~s for GR) {of the $18\,M_\odot$ progenitor model}. The instantaneous abundances $Y_A$ obtained with {\tt SkyNet} are shown in Fig.~\ref{fig:instyields} for the fully NW (upper panel) and the fully GR (lower panel) cases.
In both the NW and GR scenarios, the abundances of nuclei with mass numbers $A \gtrsim 90$ follow a clear trend: they rise steadily from $t_{\rm launch} = 1$~s (darker \textcolor{black}{purple} lines) to around $2.5$~s, after which they begin to decline, reaching their lowest values at the final time considered (\textcolor{black}{brighter yellow} lines). This overall behavior is present in both scenarios, but the decline is notably sharper in the fully NW case. As illustrated in the upper panel of Fig.~\ref{fig:instyields} (NW), the abundance of $^{94}{\rm Mo}$ increases from $Y_{\rm Mo94}(1~{\rm s}) \approx 10^{-7}$ to $Y_{\rm Mo94}(2.5~{\rm s}) \approx 10^{-6}$, before dropping by three orders of magnitude to $Y_{\rm Mo94}(8~{\rm s}) \approx 10^{-9}$. In contrast, the GR scenario shows a more gradual evolution: the abundance starts at $Y_{\rm Mo94}(1~{\rm s}) \approx 7 \times 10^{-7}$, peaks at $Y_{\rm Mo94}(2.5~{\rm s}) \approx 5 \times 10^{-6}$, and then decreases to $Y_{\rm Mo94}(7~{\rm s}) \approx 7 \times 10^{-8}$. This rise-and-fall pattern is not unique to $^{94}{\rm Mo}$, but it is representative of other nuclides in the $90 \lesssim A \lesssim 100$ mass range.

\begin{figure}
\centering
  \includegraphics[width=0.49\columnwidth]{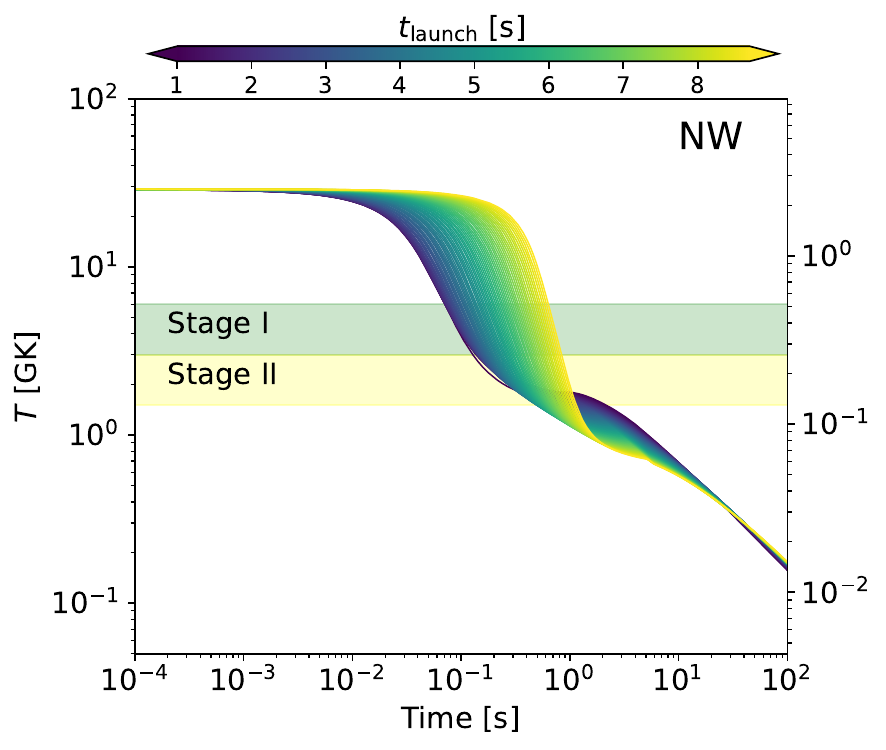}
  \includegraphics[width=0.49\columnwidth]{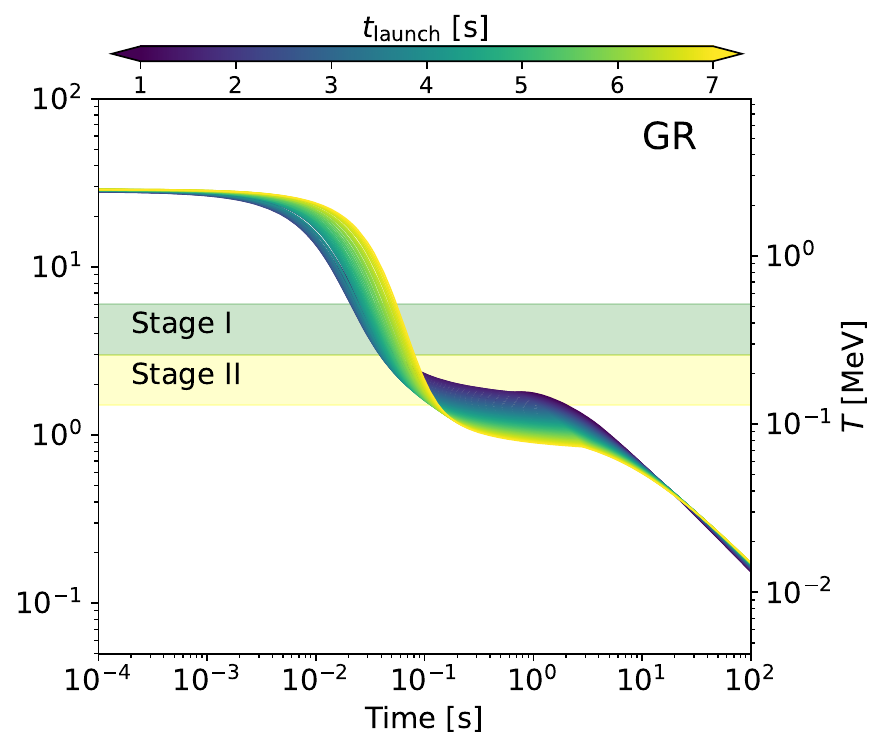}
  \includegraphics[width=0.49\columnwidth]{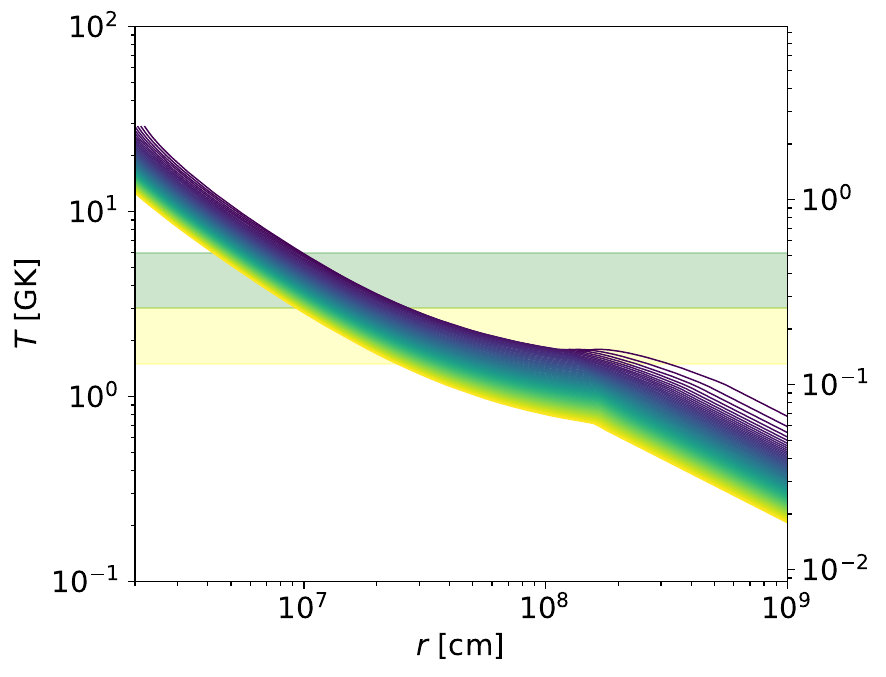}
  \includegraphics[width=0.49\columnwidth]{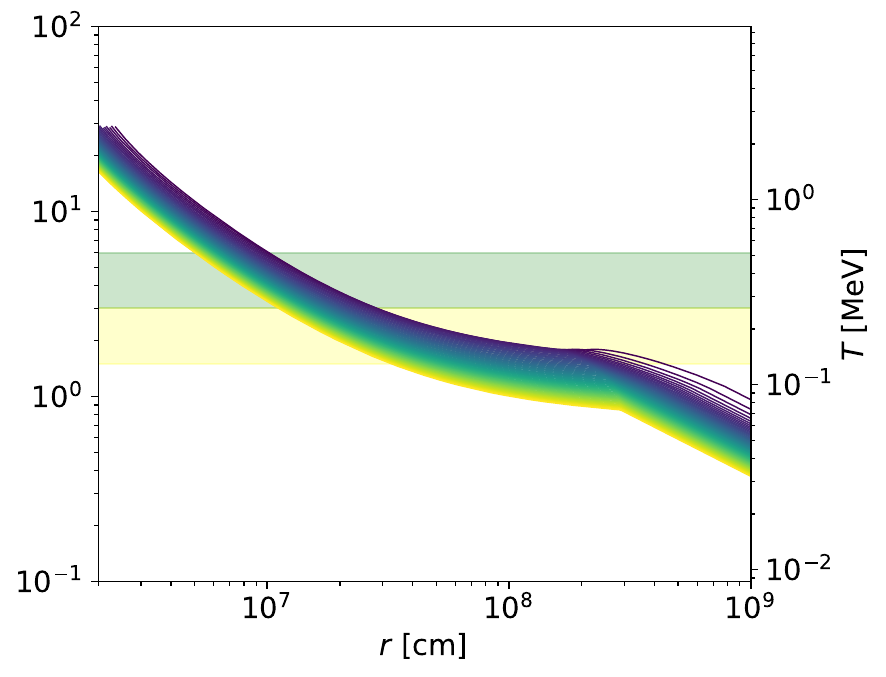}
  \caption{Evolution of the outflow temperature for NW (left panels) and GR (right panels) as a function of time (upper panels) and distance (lower panels) for different launch times, in our $18~M_{\odot}$ progenitor model. Darker \textcolor{black}{purple} lines correspond to earlier launch times ($t_{\rm launch}\approx 1$~s), while \textcolor{black}{brighter yellow} curves are related to later launch times. \textcolor{black}{The green band represents the $6\,\mathrm{GK} > T > 3\,\mathrm{GK}$ range, where the stage I of the $\nu p$-process occurs, while the yellow band indicates the $3\,\mathrm{GK} > T > 1.5\,\mathrm{GK}$ temperature window corresponding to the stage II.}
  }
  \label{fig:temp}
\end{figure}

The \textcolor{black}{isotopic abundances depend on} the evolution of the outflow temperature used by {\tt SkyNet} as input (see App.~\ref{app:setup-skynet} for more details) and reported in Fig.~\ref{fig:temp} for the NW (left panels) and GR (right panels) cases. 
In the upper panels of Fig.~\ref{fig:temp}, we show the evolution of the temperature for each snapshot as a function of the simulation time in {\tt SkyNet}, which starts at zero seconds and ends at $\sim$32 years. \textcolor{black}{The temperature of the outflow} steadily cools down as time progresses. \textcolor{black}{For each trajectory, we start} the reaction network at $T=2.5$~MeV, in conditions of \textcolor{black}{weak equilibrium and} NSE. The NW case is characterized by slower outflows compared to GR, almost by a factor of three in velocity for $t_{\rm launch}\approx 1$~s and even larger later. This makes the times \textcolor{black}{$\tau_1$ and $\tau_2$} spent in stages I and II, respectively (see the green and yellow bands in Fig.~\ref{fig:temp}), larger in the NW case---\textcolor{black}{especially for $\tau_1$}. Moreover, later snapshots (darker red lines) tend to go much faster through these bands \textcolor{black}{as the far pressure drops. At the same time,} the PNS contracts and the asymptotic value of $S$ increases. 

As GR trajectories move faster, \textcolor{black}{they begin to nearly match the far pressure} earlier, 
leading to a plateau in the temporal evolution of $T$ about $0.1\mbox{--}1$~s after their launch. The kink at the end of such a plateau corresponds to the merging to the homologously expanding material, modeled through the gluing prescription discussed in Sec.~\ref{sec:traj}. \textcolor{black}{Trajectories starting at $t_{\rm launch}\gtrsim 2.5$~s move through the $3\,\text{GK} > T > 1.5\,\text{GK}$ temperature band much more quickly, due to the reduced far pressure. As a consequence, the $\nu p$-process becomes less effective at these later times.}

The lower panels in Fig.~\ref{fig:temp} show the radial profiles of the temperature for the NW and GR cases. In both scenarios, the starting radius of the trajectories is shifted to smaller distances as the launch times increase, mainly because of the reduction of the PNS radius. Due to the larger speed in GR trajectories, these outflows undergo the different phases of the $\nu p$-process at larger radii than NW; an aspect which hurts the yield production as pointed out in Ref.~\cite{Friedland:2023kqp}. 

\begin{figure}
\centering
  \includegraphics[width=0.99\columnwidth]
  {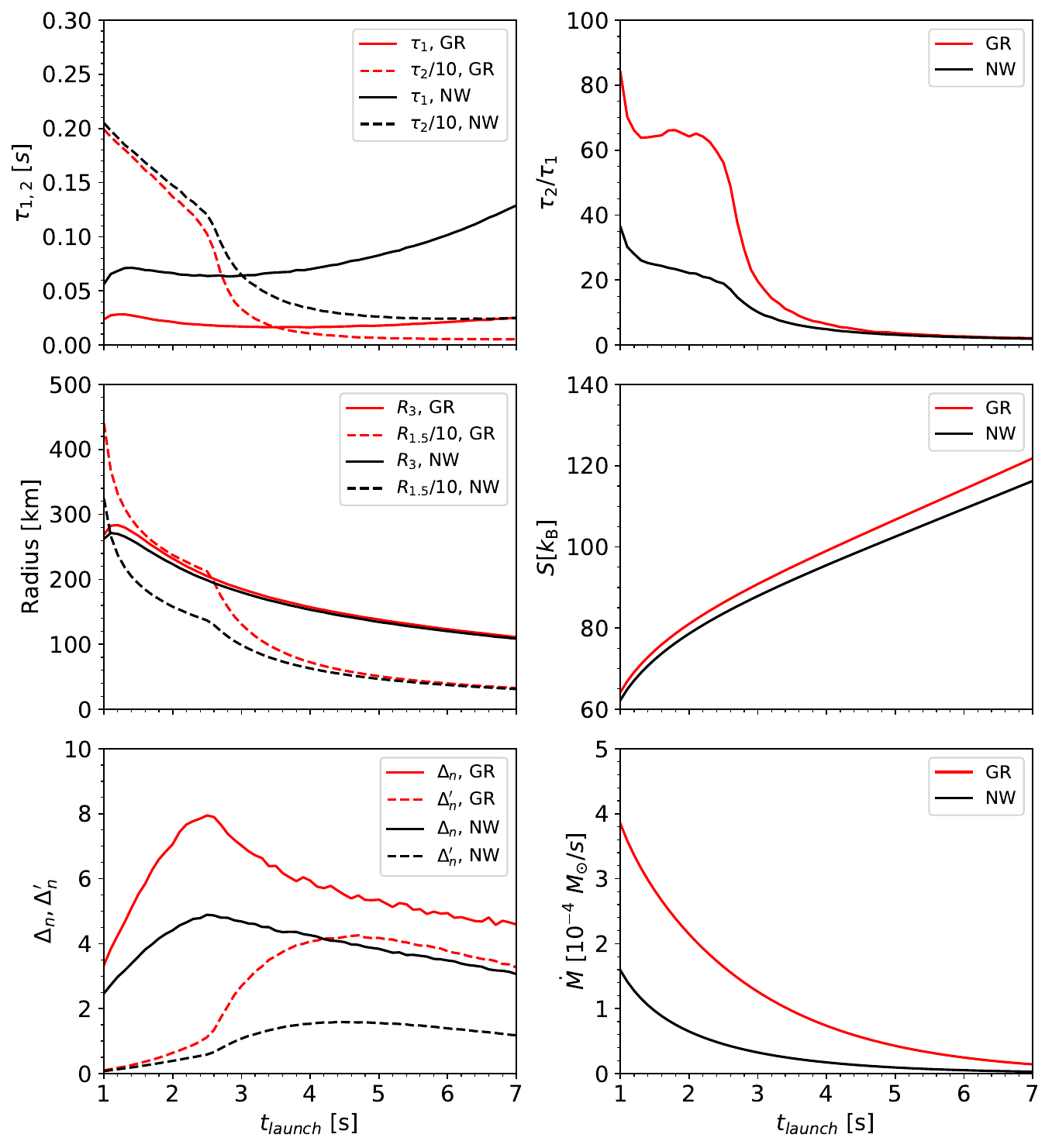}
  \caption{Upper panels: the duration of stage I ($\tau_1$) and stage II ($\tau_2$) on the left, and their ratio $\tau_2/\tau_1$ on the right panel. Central panels: the locations $R_{3}$ and $R_{1.5}$ (left) and the asymptotic value of the entropy $S$ (right). Lower panels: the number of neutrons during (after) the stage II, i.e. $\Delta_n$ ($\Delta^{\prime}_n$) on the left panel, and the mass outflow rate $\dot{M}$ on the right panel. We show these quantities for both the NW (black) and GR (red) cases as a function of launch time.
  }
  \label{fig:six-panels}
\end{figure}

The temperature evolution discussed above influences the time evolution of the parameters characterizing the efficiency of the $\nu p$-process, which are shown in Fig.~\ref{fig:six-panels} for launch times in the range $t_{\rm launch}\in[1,\,7]$~s for NW (black) and GR (red).

The upper panels in Fig.~\ref{fig:six-panels} display the change in the duration of the stages of the $\nu p$-process, with the values of $\tau_1$ (solid lines) and $\tau_2$ (dashed) in the left panel and their ratio in the right panel. Since NW  trajectories (black) are slower than GR (red), both $\tau_1$ and $\tau_2$ are larger in NW at a given $t_{\rm launch}$. Moreover, $\tau_1$ does not change drastically as the launch time advances, with $\tau_1\approx 20\mbox{--}30$~ms (GR) and $\tau_1\approx 50\mbox{--}130$~ms (NW), the largest values being achieved at late times. On the other hand, $\tau_2$ monotonically decreases with time, from $\tau_2 \approx 2$~s at $t_{\rm launch}=1$~s for both NW and GR down to $\tau_2 \approx 300$~ms (NW) and $\tau_2 \approx 50$~ms (GR). The kink observed in $\tau_2$ and its steep decrease at $t_{\rm launch}\approx 2.5$~s in both scenarios is due to the trajectories \textcolor{black}{exiting the $3\,\text{GK} > T > 1.5\,\text{GK}$ band quicker, as described above.}

The upper right panel of Fig.~\ref{fig:six-panels} shows the ratio of timescales $\tau_2/\tau_1$. The weak dependence of $\tau_1$ on the launch times results in a ratio $\tau_2/\tau_1$ (which can be used a measure of the effectiveness of the $\nu p$-process) that mostly follows the behavior of $\tau_2$. At $t_{\rm launch}\lesssim 1.5$~s, the ratio shows a sharp decline related to a reduction in $\tau_2$ and a small increase in $\tau_1$. For \textcolor{black}{the next 1 s}, the ratio remains approximately constant until $t_{\rm launch}\approx 2.5$~s, after which it \textcolor{black}{falls off sharply} due to the decline in $\tau_2$. 
At early times, before \textcolor{black}{this fall-off}, the ratio $\tau_2/\tau_1$ is almost three times larger for GR ($\tau_2/\tau_1\approx 65$) than NW ($\tau_2/\tau_1\approx 25$). On the other hand, at later times ($t_{\rm launch}\gtrsim 4$~s) the ratio becomes smaller, with similar values for both NW and GR.

The large ratio $\tau_2/\tau_1$ (which has a positive impact on the yields) in the GR case is partially balanced by the distance of the \textcolor{black}{$3\,\mathrm{GK} > T > 1.5\,\mathrm{GK}$ region} from the PNS. As shown in the central left panel of Fig.~\ref{fig:six-panels}, both the values of $R_3$ (solid line) and $R_{1.5}$ (dashed line) are larger for GR than NW.  \textcolor{black}{In general, the $\nu p$-process becomes less efficient at larger distances from the PNS due to a decrease in the antineutrino fluence.}

Qualitatively, there is no stark difference in the behavior of $R_3$ between the NW and GR cases. For early launch times ($t_{\rm launch} \lesssim 1.5$~s), $R_3$ is located slightly farther out in the GR scenario. As time progresses, both cases exhibit a steady decrease in $R_3$, dropping from approximately $250$--$300$~km at $t_{\rm launch} = 1.5$~s to about $100$~km by $t_{\rm launch} = 7$~s. On the other hand, $R_{1.5}$ displays a more pronounced difference between the two scenarios. At $t_{\rm launch} = 1$~s, $R_{1.5}\approx4000$~km in the GR case, approximately 1000~km larger than the NW one. As $t_{\rm launch}$ increases, $R_{1.5}$ decreases, and a noticeable kink appears near $t_{\rm launch} \approx 2.5$~s, \textcolor{black}{when the far temperature in the trajectories becomes lower than 1.5~GK due to the drop in far pressure,} marking a change in the conditions for effective nucleosynthesis. At later times, when the \textcolor{black}{stage-II duration} becomes too short to contribute meaningfully to yield production, we no longer observe significant differences in $R_{1.5}$ between NW and GR.

The third quantity that plays a key role in shaping the \textcolor{black} {yields} of the $\nu p$-process, alongside the location and duration of the different nucleosynthesis stages, is the entropy of the outflow, shown in the central right panel of Fig.~\ref{fig:six-panels}. As the neutrino luminosities and the PNS radius decrease over time, the entropy increases in both the GR and NW cases. This behavior is expected from the scaling relation $S \propto L_\nu(t)^{-1/6} R_{\rm PNS}(t)^{-2/3}$~\cite{Qian:1996xt}, yielding values of ${S(1{\rm ~s})\approx 65}$ and $S(7{\rm ~s}) \approx 120$ for both cases, with a slightly larger value for GR. Notably, the entropy difference between GR and NW also grows with the launch time. While the difference is modest at early times ($\Delta S (1{\rm ~s}) \approx 2$), it slightly increases to about $\Delta S (7{\rm ~s}) \approx 5$. \textcolor{black}{ A higher entropy implies a lower density at the same temperature, which deceases the triple-$\alpha$ reaction rate and lowers seed production, resulting in a higher proton-to-seed ratio~\cite{Wanajo:2010mc, Friedland:2023kqp}.}

All the quantities discussed so far influence the efficiency of the $\nu p$-process. A particularly informative diagnostic of this efficiency is the number of neutrons available per seed during stages II and III of the process, quantified by $\Delta_n$ and $\Delta_n^\prime$, defined in Eqs.~\eqref{eq:deltan} and~\eqref{eq:deltanprime}, respectively. As shown in the lower left panel of Fig.~\ref{fig:six-panels}, $\Delta_n$ peaks at $t_{\rm launch} \approx 2.5$~s in both scenarios, with $\Delta_n \approx 8$ in GR and $\Delta_n \approx 5$ in NW. This trend declines at later times, indicating that the \emph{optimal time window} for efficient $\nu p$-process is at $t_{\rm launch} \lesssim 3$~s. \textcolor{black}{ Ref.~\cite{Friedland:2023kqp} stressed the existence of this time window, occurring at 1-2 s after the core bounce. Here, the time dependence of the PNS radius, which sets the entropy of the outflow, extends the duration of this window up to about $3$~s.\footnote{\textcolor{black}{In Ref.~\cite{Friedland:2023kqp} a constant PNS radius was assumed, leading to slightly lower values of the entropy at later times since $S\propto L_\nu(t)^{-1/6}\,R_{\rm PNS}(t)^{-2/3}$.}}} The larger value of $\Delta_n$ in the GR case reflects a greater neutron availability {per seed nucleus} and, consequently, higher $\nu p$-process efficiency. We report in Tab.~\ref{tab:proton-seeds-deltas} the main quantities which determine the peak value of $\Delta_n$ at $t_{\rm launch}=2.5$~s. At this time and around $T=3$~GK, we find \textcolor{black}{proton-to-seed} ratios of $Y_p/Y_{A\geq12}\approx112$ (NW) and $Y_p/Y_{A\geq12}\approx471$ (GR). \textcolor{black}{On the contrary}, the number of captured antineutrinos during stage II \textcolor{black}{$n_{\bar{\nu}_e}$ is a factor $\sim 2.5$ smaller in GR than in NW.} The interplay between the increase in the \textcolor{black}{proton-to-seed} ratios and the decrease in the number of captured antineutrinos leads to a $\sim 65\%$ increase in $\Delta_n$ due to GR effects.

\begin{table}[t!]
\centering
\begin{tabular}{|c|c|c|c|c|c|c|c|c|c|l}
\hline\hline
Case & $t_{\rm launch}$& $Y_p$& $Y_{A\geq 12}$& $Y_p/Y_{A\geq 12}$& $n_{\bar{\nu}_e}$& $n^{\prime}_{\bar{\nu}_e}$&$\Delta_n$ & $\Delta_n^\prime$  &$v_{\rm peak}$\\
 & [s]& & [$\times 10^{-4}$]& & [$\times 10^{-3}$]& [$\times 10^{-3}$]& & &$[10^3~ 
\rm km/s]$\\
\hline
NW &  $2.5$& $0.1345$& $11.96$& $112.4$& $43.4$& $6.6$& $4.8$& $0.5$ &$2.09$\\
GR&  $2.5$& $0.1586$& $3.37$& $470.6$& $16.9$& $3.1$& $7.9$& $1.1$ &$7.47$\\
NW&  $5.0$& $0.1315$& $9.56$& $137.5$& $28.0$& $12.9$& $3.8$& $1.5$ &$1.33$\\
GR&  $5.0$& $0.1584$& $2.01$&  $786.8$& $6.8$& $5.8$& $5.3$& $4.1$ &$5.98$\\
\hline
\hline
\end{tabular}
\caption{Proton ($Y_p$) and seed abundances ($Y_{A\geq 12}$) at $T=3$~GK, the proton to seed ratio, the antineutrino numbers $n_{\bar{\nu}_e}$ and $n^{\prime}_{\bar{\nu}_e}$, the number of neutrons per seed $\Delta_n$ and $\Delta_n^\prime$, and the maximum velocity ($v_{\rm peak}$) of the NDO at the specified launch time. We compare these quantities in the NW and GR cases at early ($2.5$~s) and late ($5$~s) snapshots, when $\Delta_n$ and $\Delta_n^\prime$ are peaked, respectively.}
\label{tab:proton-seeds-deltas}
\end{table}

\textcolor{black}{On the other hand, the behavior of $\Delta_n^\prime$ differs from that of $\Delta_n$, with important consequences for the production of $\Nb$ (see Sec.~\ref{sec:Nb}). As illustrated by the dashed curves in the lower-left panel of Fig.~\ref{fig:six-panels}, $\Delta_n^\prime$ remains modest at early times, with $\Delta_n^\prime\simeq 1$ at $t_{\rm launch}\simeq 2.5$~s in the GR case, and rises sharply for $t_{\rm launch}\gtrsim 2.5$ s, coincident with the kink observed in $\tau_2$ and $R_{1.5}$ (upper and middle left panels). At later times, the NDOs reach the $T\lesssim1.5$~GK region earlier and at smaller radii, conditions that significantly increase the number of antineutrino captures during stage III ($n^\prime_{\bar{\nu}e}$) and thereby boost $\Delta_n^\prime$. This leads to a peak in $\Delta_n^\prime$ at $t_{\rm launch}\approx 5$~s, where $\Delta_n^\prime\approx 4$ in the GR case, almost three times the NW value. The origin of this enhancement lies in the GR-driven increase of the proton-to-seed ratio by a factor of $\sim 6$, which more than compensates for the $\sim 2.2$ reduction in $n^\prime_{\bar{\nu}_e}$ induced by GR (see Tab.~\ref{tab:proton-seeds-deltas}).}

\textcolor{black}{The different peak times of $\Delta_n$ and $\Delta_n^\prime$ govern how GR effects shape the nucleosynthesis outcome. As the PNS contracts (e.g., $R_{\rm PNS}(5~{\rm s})\approx 13~{\rm km} < R_{\rm PNS}(3~{\rm s})\approx 15~{\rm km}$, see Fig.~\ref{fig:pns-gr1d}), GR corrections become increasingly significant. This strengthening is reflected in the maximum velocity ($v_{\rm peak}$) of the NDO at a given $t_{\rm launch}$, which increases for larger GR effects (see the last column of Tab.~\ref{tab:proton-seeds-deltas}). The GR-to-NW ratio of $v_{\rm peak}$ grows from $\sim 3.5$ at $t_{\rm launch}=2.5$~s to $\sim 4.5$ at $t_{\rm launch}=5$~s and nearly 5 by $t_{\rm launch}=7$~s. A larger velocity implies a lower seed production, reducing the seed abundance $Y_{A \geq12}$. Combined with the stronger GR-induced reduction of $n_{\bar{\nu}e}$ relative to $n^\prime_{\bar{\nu}e}$, these effects imply that $\Delta_n^\prime$ receives a substantially larger boost than $\Delta_n$ for $t_{\rm launch}\gtrsim 2.5$ s. Because $\Delta_n^\prime$ peaks at later times, when GR effects are most pronounced, the net GR impact on the synthesis of late-time $\nu p$ nuclides such as $^{92}$Nb is particularly strong (see Sec.~\ref{sec:Nb}).}

Estimating the total abundances requires knowledge of the full time history of the mass outflow rate $\dot{M}$ from $t_{\rm launch} = 1$~s to $t_{\rm launch} = 7$~s. 
\textcolor{black}{Fortunately}, the optimal window \textcolor{black}{also generically} coincides with a period when the mass outflow rate is appreciable, resulting in more substantial nucleosynthesis. We show the time evolution of $\dot{M}$ in the lower right panel of Fig.~\ref{fig:six-panels}. In the GR case, we find $\dot{M}(1~{\rm s}) \approx 3.8 \times 10^{-4}~M_\odot/{\rm s}$, almost three times larger than in the NW case. By $t_{\rm launch} = 3$~s, this rate has dropped to $\dot{M} \approx 1.5 \times 10^{-4}~M_\odot/{\rm s}$, still about five times higher than in NW. These differences suggest that the GR case achieves significantly \textcolor{black}{greater} time-integrated yields also due to its consistently higher mass outflow rate.

\begin{figure} [t!]
\centering
  \includegraphics[width=0.7\columnwidth]{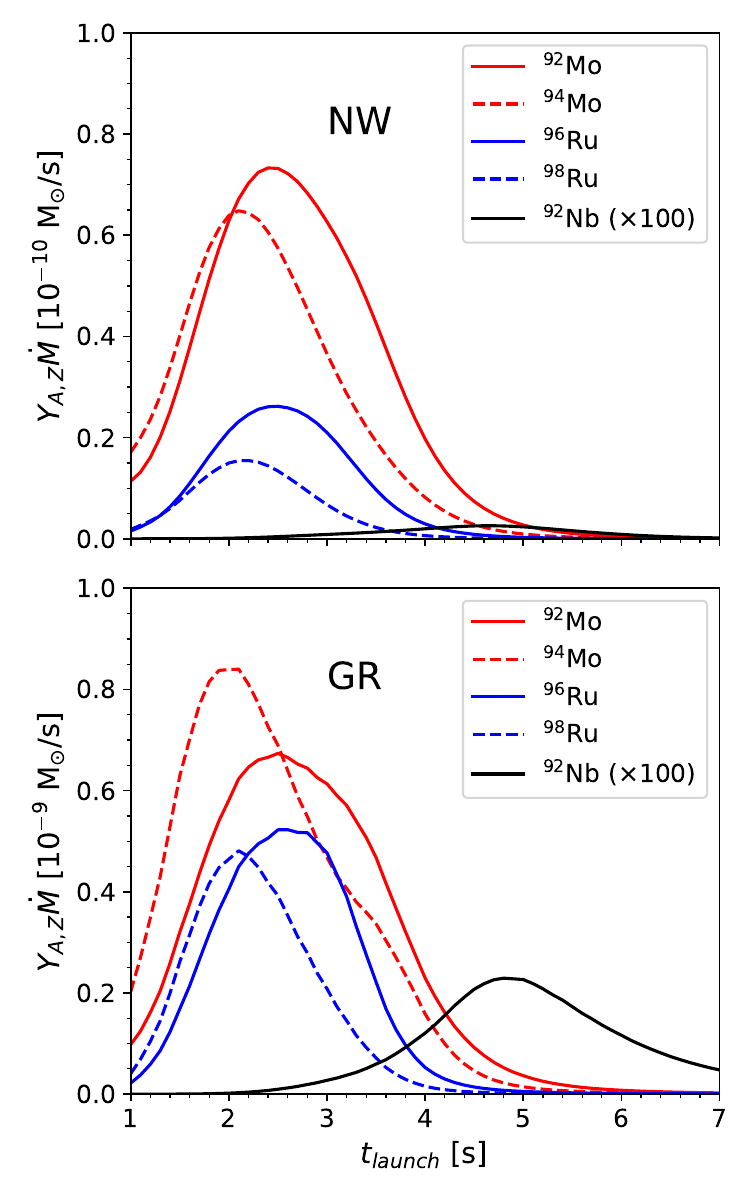}
  \caption{Time evolution of $\dot{M}$-weighted yields for NW (upper) and GR (lower) in our $18~M_\odot$ progenitor model. The selected nuclides are $\Mo$ (red), $\Ru$ (blue), and $\Nb$ (black). The production of $\Mo$ and $\Ru$ is optimal at $t_{\rm launch} \approx 2.5$~s, while $\Nb$ is mostly produced later at $t_{\rm launch} \approx 5$~s. Note that the $y$-axis scales differ between the two panels, with the NW scenario being an order of magnitude lower than the GR case.}
  \label{fig:timeyields}
\end{figure}

By multiplying the time-dependent abundances $Y_A$ by the corresponding mass outflow rates $\dot{M}$, we obtain the mass-outflow-rate-weighted abundances for the various nuclides. These weighted yields are shown in Fig.~\ref{fig:timeyields} for selected isotopes in the NW (upper panel) and GR (lower panel) cases. 
As expected, yields are larger in GR and in both cases the optimal time window for the production of $\Mo$ (red lines) and $\Ru$ (blue lines) is between 2~s and 3~s, when $\Delta_n$ is maximum. In contrast, the production of $\Nb$ \textcolor{black}{(black line) peaks} at later times, around $t_{\rm launch} \approx 5$~s, coinciding with the peak of $\Delta_n^\prime$, as indicated by the black line in Fig.~\ref{fig:timeyields}. At these times, the small value of $\Delta_n^\prime$ in the NW case results in almost negligible $\Nb$ synthesis, with the production rate remaining below $5 \times 10^{-14}~M_\odot/{\rm s}$. In the GR case, relativistic effects enhance the production of $\Nb$ by more than \textcolor{black}{one order} of magnitude, highlighting the critical role of GR in enabling the synthesis of this nuclide.

\subsection{Integrated yields and production factors}

\begin{figure}
\centering
\includegraphics[width=0.80\columnwidth]{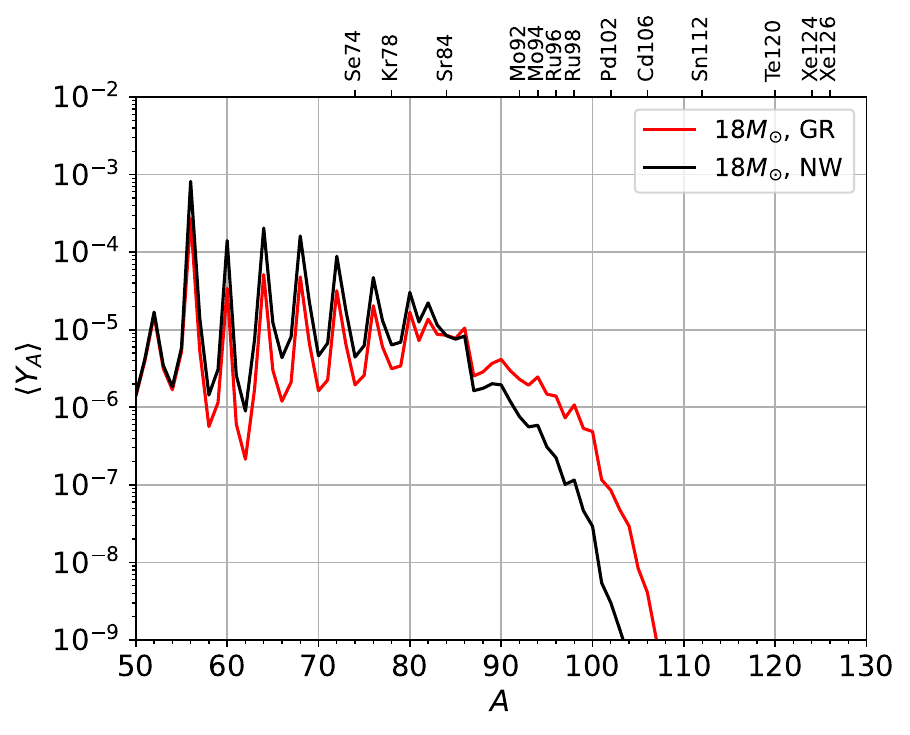}
  \caption{Time-averaged yields $\langle Y_A \rangle$ for NW (black) and GR (red) in our $18~M_\odot$ progenitor model, shown as a function of mass number $A$. Relativistic effects such as blueshift and  corrections to the hydrodynamics can enhance the time-integrated $p$-rich yields by 1--2 orders of magnitude in the $95 \lesssim A \lesssim 105$ mass range.}
  \label{fig:moneyplot}
\end{figure}

In order to estimate whether the abundances of nuclides produced by the $\nu p$-process can explain the observations in the solar system, one needs to compute time-integrated yields. To this end, for a nuclide $(A,\,Z)$ we define the time-averaged abundance $\langle Y_{A,\,Z}\rangle$ as~\cite{Friedland:2023kqp}
\begin{equation}
    \langle Y_{A,\,Z}\rangle = \frac{\int dt\, Y_{A,Z}(t)\dot{M}(t)}{\int dt\, \dot{M}(t)}\,,
\end{equation}
where the integral is defined over the launch time $t_{\rm launch}$, and $Y_{A,Z}$ and $\dot{M}$ are the instantaneous yields and the mass outflow rate as functions of $t_{\rm launch}$, as discussed in Sec.~\ref{sec:time-ev}. In Fig.~\ref{fig:moneyplot}, 
we show the time-averaged abundances $\langle Y_{A,Z}\rangle$ as a function of the mass number $A$ obtained with our benchmark $18~M_\odot$ SN model for NW (black) and GR (red). One can see that GR leads to larger yields than NW for $A\gtrsim 90$, with a factor of a few difference for $A=90$ and an enhancement of more than one order of magnitude for $A\gtrsim 100$.

More precisely, we report in Tab.~\ref{tab:ratioGRNWintegrated} the values of the time-averaged yields for a few selected isotopes in the mass window $92\leq A\leq 106$, showing that GR effects enhance the production of $\Mo$ by a factor $\sim 4$ and that of $\Ru$ by almost one order of magnitude. Furthermore, we observe an even larger increase in the production of heavier nuclides, such as $^{102}{\rm Pd}$, whose production is enhanced by a factor $\sim 30$. Nonetheless, other nucleosynthesis processes, such as the $\gamma$-process, may help explain the abundance of heavy nuclides with $A\gtrsim100$. For this reason, we will not comment on them any further. Moreover, as \textcolor{black}{mentioned earlier and to be} further discussed in Sec.~\ref{sec:Nb}, GR also plays a crucial role in enhancing the production of \textcolor{black}{long-lived radioactive isotopes} $\Nb$ and $\Tc$, with their abundances increased by factors of approximately $25$ and $350$, respectively.

\begin{table}[t!]
\centering
\begin{tabular}{|c|c|c|c|}
\hline\hline
Element & NW & GR &  GR/NW \\
\hline
$^{92}{\rm Mo}$ & $8.18\times 10^{-7}$ & $2.42\times 10^{-6}$ & 3.0 \\
$^{94}{\rm Mo}$ & $6.30\times 10^{-7}$ & $2.59\times 10^{-6}$ & 4.1 \\
$^{96}{\rm Ru}$ &  $2.42\times 10^{-7}$&  $1.47\times 10^{-6}$ & 6.1 \\
$^{98}{\rm Ru}$ &  $1.24\times 10^{-7}$& $1.13\times 10^{-6}$ & 9.1 \\
$^{102}{\rm Pd}$ &  $3.25\times 10^{-9}$ & $9.01\times 10^{-8}$ & 27.7 \\
$\Nb$ & $3.38\times 10^{-10}$ & $8.14\times 10^{-9}$ & 24.1 \\
$^{98}{\rm Tc}$ &  $1.56\times 10^{-13}$ & $5.52\times 10^{-11}$ & 354 \\
\hline\hline
\end{tabular}
\caption{Time-averaged yields $\langle Y_A\rangle$ for various nuclides in both the NW and GR cases, along with their corresponding ratios (GR/NW), are tabulated for the $18~M_\odot$ progenitor model. These yields represent the total amount of each nuclide synthesized and ejected in the outflow. \textcolor{black}{All the nuclides shown here are stable, except for $\Nb$ and $\Tc$, which are long-lived radioactive isotopes.}}
\label{tab:ratioGRNWintegrated}
\end{table}

\begin{figure}
\centering
  \includegraphics[width=0.80\columnwidth]{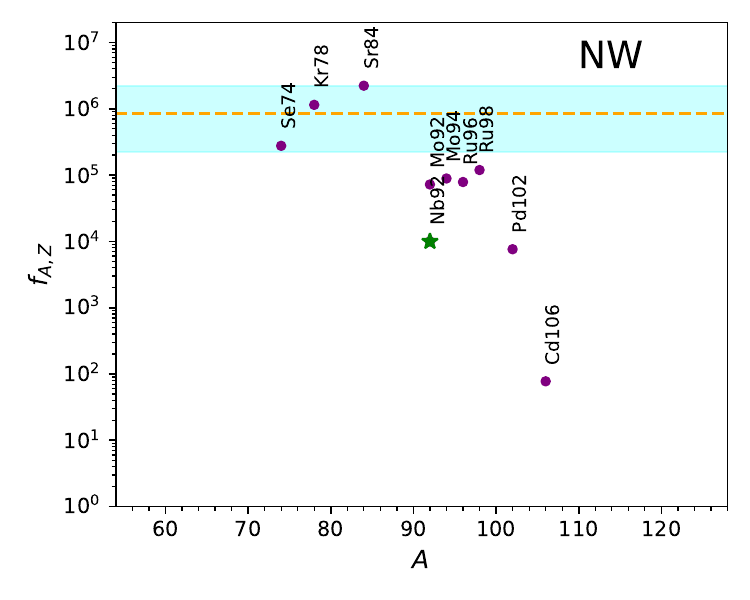}
    \includegraphics[width=0.80\columnwidth]{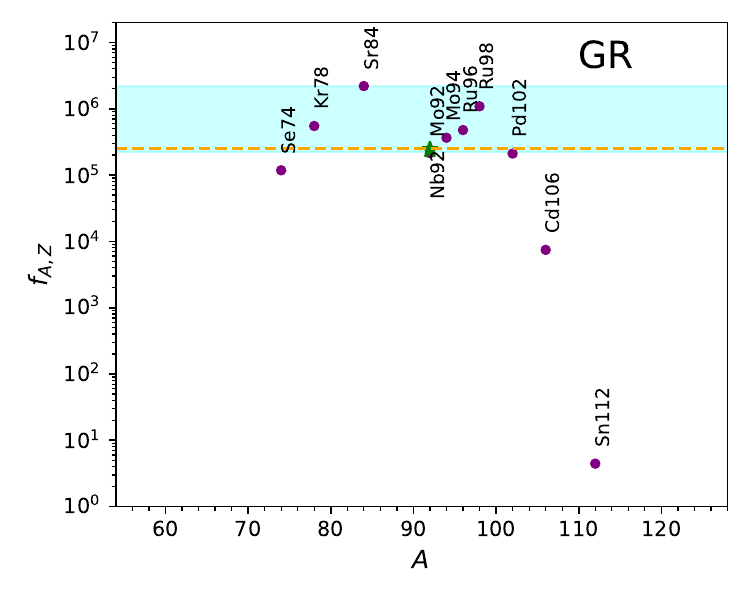}
  \caption{The production factor $f_{A,\,Z}$ in our $18~M_\odot$ progenitor model, shown as a function of the mass number $A$. Results show the NW (top) and GR (bottom) abundances of a view selected nuclides. The cyan band is the ``co-production'' region, i.e. the range $[f_{\rm max}/10,f_{\rm max}]$ (in this case $f_{\rm max}=f_{\rm Sr84}$). Data points above the dashed orange line have sufficiently high $f_{A,Z}$ to account for their solar abundances. See the main text for further discussion. 
  }
  \label{fig:prodfac}
\end{figure}

\textcolor{black}{Following standard practice in the nucleosynthesis literature (see, e.g., Refs.~\cite{Wanajo:2010mc,Friedland:2023kqp}), we compare the $\nu p$ nucleosynthesis yields obtained with {\tt SkyNet} to the observed solar-system abundances~\cite{Lodders:2003} by introducing the isotopic \emph{production factor}
\begin{equation}
    f_{A,Z} = \frac{\langle X_{A,Z} \rangle}{X_{A,Z}^\odot}\,,
\label{eq:fAZ}
\end{equation}
where $X_{A,Z}^\odot$ is the solar-system mass fraction of the isotope $(A,Z)$ and $\langle X_{A,Z} \rangle$ is its time-averaged mass fraction computed from the {\tt SkyNet} output, defined as $\langle X_{A,\,Z}\rangle = A \langle Y_{A,Z}\rangle$, provided $\sum_{A,Z}\, A\, Y_{A,\,Z} = 1$. The derivation of the solar mass fractions $X_{A,Z}^\odot$ from the tabulated measured isotopic abundances $N_{A,Z}^\odot$ of Ref.~\cite{Lodders:2003} is detailed in App.~\ref{app:solar-ab}.}
The production factors $f_{A,Z}$ are indicators of the \emph{relative abundance} of different isotopes produced in an environment. Nuclides with $f_{A,Z}$ larger than one-tenth of the largest production factor $f_{\rm max}$ are considered to be \emph{co-produced} in significant quantities, within an uncertainty of one order of magnitude~\cite{Wanajo:2010mc,Bliss:2018djg}. In Fig.~\ref{fig:prodfac}, we show the isotopic production factors of various $p$ nuclides for NW (upper panel) and GR (lower panel). In both  cases the largest production factor is the one for $^{84}{\rm Sr}$, i.e. $f_{\rm max}\approx2\times 10^{6}$. The shaded cyan band shows the ``co-production region'', which covers values of $f_{A,Z}$ between $f_{\rm max}/10$ and $f_{\rm max}$. For NW, all the nuclides in the mass window $90\leq A \leq 100$ lie below this band, whereas the GR case co-produces $p$ nuclides up to $^{102}{\rm Pd}$. 

The co-production region gives information about the \emph{relative} abundances of the synthesized yields. On the other hand, the \emph{absolute} isotopic yields and the efficiency of the $\nu p$-process are characterized by the so-called \emph{overproduction factor}~\cite{Woosley:1994ux,Wanajo:2010mc,Friedland:2023kqp}. It is defined as $f_{A,Z}\times (M_{\rm out}/M_{\rm ejec})$, where $f_{A,Z}$ is the production factor [Eq.~\eqref{eq:fAZ}] and $(M_{\rm out}/M_{\rm ejec})$ is the ratio between the total mass driven out in the NDO ($M_{\rm out} = \int dt\,\dot{M}(t)$) and the total mass ejected in the explosion ($M_{\rm ejec} \approx M_{\rm prog} - M_{\rm PNS}$). This ratio represents the astrophysical dilution factor. An overproduction factor larger than $\sim 10$ in a SN event is required to explain the solar-system abundance of a given nuclide~\cite{Wanajo:2010mc,Woosley:1994ux,Friedland:2023kqp}. 

For our benchmark $18~M_\odot$ model, $M_{\rm ejec}\approx 16.2~M_\odot$, independently of GR effects. However, due to a larger $\dot{M}$ in GR, the outflow mass is a factor of 4 larger in GR than NW, with $M_{\rm out}\approx 2\times 10^{-4}~M_\odot$ for NW and $M_{\rm out}\approx 8\times 10^{-4}~M_\odot$ for GR. Therefore, to explain the observed solar abundance, the production factor of each nuclide is required to be $\gtrsim 8 \times 10^{5}$ for NW and $\gtrsim 2 \times 10^{5}$ for GR, as shown by the horizontal dashed orange line in Fig.~\ref{fig:prodfac}. This means that in the NW case (upper panel of Fig.~\ref{fig:prodfac}), the production of nuclides with $A \gtrsim 90$ is not only suppressed relative to lighter $p$ nuclides but also absolutely insufficient, as indicated by their production factors falling below both the cyan band and the dashed orange line. By contrast, $p$ nuclides up to $^{102}{\rm Pd}$ are efficiently produced when including all relativistic effects. Additionally, as we will further discuss in Sec.~\ref{sec:Nb}, the production factor of $\Nb$ (the green star in Fig.~\ref{fig:prodfac}) is large enough to explain the observed solar abundance of this nuclide in GR, while it is strongly suppressed in the NW scenario. 

\textcolor{black}{ It is a nontrivial result that, in the GR case, the same distribution of yields that ensures co-production of $\Mo$ and $\Ru$ (and $\Nb$) with the lighter $p$ nuclides also ensures sufficient absolute production of these nuclides to satisfy the solar system observations. In Fig.~\ref{fig:prodfac}, this amounts to saying that the dashed orange line signifying the threshold for absolute production also falls within the blue co-production band, which is not guaranteed \textit{a priori}. This also ensures that none of the $p$ nuclides are over-produced relative to the solar abundances. On the other hand, in the NW case, even though the dashed orange line still overlaps with the blue co-production band, the production factors of $\Mo$, $\Ru$ and $\Nb$ fail to meet both the co-production and absolute thresholds.}

\textcolor{black}{Note that, these results were contingent on specific choices regarding neutrino luminosities and energy spectra, PNS radius, the EoS, and the far boundary condition (determined by shock velocity). Changes to any of these could, in principle, change the outcome; however, we expect our conclusions to remain robust over a range of physical conditions. As an example, adding baryonic gas to the EoS (for details, refer to App.~\ref{sec:baryonic-gas}) reduces the yields, if all else is the same. However, we also found that this reduction can be reversed by decreasing the PNS radius by about 20\%. The final outcome thus exhibits some degree of degeneracy across changes in the physical conditions.}

\subsection{Stage III nucleosynthesis and $\Nb$ production}
\label{sec:Nb}

\textcolor{black}{As discussed in the previous sections, relativistic effects have a quantifiable impact on the production of metastable isotopes, such as $\Nb$ and $\Tc$, which are produced in stage III of the $\nu p$-process, when the temperature is lower than 1.5~GK. Notably, the time-averaged yields of these nuclides are enhanced by a factor of $\sim 25$ and $\sim 350$, as shown in Tab.~\ref{tab:ratioGRNWintegrated}. This strong enhancement is related to the optimal production time window of $\Nb$, which occurs around $t_{\rm launch}\sim5$~s in our benchmark model; about $2.5$~s after the peak production of the other $p$ nuclides such as $\Mo$ and $\Ru$ (see Fig.~\ref{fig:timeyields}). These metastable nuclides are produced at later times due to an increase in the number of {neutrons per seed in stage III}, $\Delta_n^\prime$ defined in Eq.~\eqref{eq:deltanprime}. 
As discussed in Sec.~\ref{sec:time-ev}, GR effects have a large impact on $\Delta_n^\prime$, due to the interplay between the reduction in $n_{\bar{\nu}_e}^\prime$ and the enhancement of the proton-to-seed ratio $Y_p/Y_{A\geq 12}$. The combined result is a pronounced boost in $\Delta_n^\prime$ precisely within the optimal $\Nb$-production window, making the synthesis of these late-time nuclides exceptionally responsive to GR corrections.}

As shown by the green star in Fig.~\ref{fig:prodfac}, $\Nb$ is sufficiently produced in the GR case, while its production is extremely suppressed for NW, being two orders of magnitude lower than what suggested by observations. Here, we normalize the $\Nb$ production factor to $3\times 10^{-3}$ times the solar system meteoritic $^{92}{{\rm Mo}}$ abundance. We choose this normalization since $\Nb$ itself is unstable and cannot be observed in the present-day solar system. However, studies of meteoritic compositions, combined with galactic chemical evolution models, have shown that the ratio of $\Nb/^{92}{\rm Mo}$ at production has to be in the range $10^{-3}$--$10^{-2}$~\cite{Rauscher:2013,Lugaro:2016zuf,Iizuka:2016,Hibiya:2023}. This amount is sufficient to explain the solar system observations of $^{92}{\rm Zr}$, into which $\Nb$ eventually decays~\cite{Rauscher:2013,Lugaro:2016zuf,Iizuka:2016,Hibiya:2023}. Similar predictions can also be made for the other shielded isotopes produced by late-time neutrons. For $\Tc$, we predict an abundance of approximately $5 \times 10^{-5}$ relative to $^{98}\mathrm{Ru}$, which is consistent with the current upper limit observed in the solar system~\cite{DAUPHAS2003C287}, owing to its half-life of $4.2$ Myr~\cite{nndc-walletcard}.

\textcolor{black}{This discussion highlights that the most pronounced GR effects appear in the production of metastable isotopes synthesized through late-time neutron captures, such as $\Nb$ and $\Tc$, partly owing to the fact their production peaks at later launch times, when the GR effects on seed production are stronger.}
It is also noteworthy that $\Nb$, along with a fraction of $^{92}{\rm Mo}$, may alternatively be produced in mildly neutron-rich NDOs~\cite{Hoffman:1996}.

\section{Dependence on the Progenitor Mass and Shock Velocity}
\label{sec:prog-mass}

\begin{figure}
\centering
  \includegraphics[width=0.8\columnwidth]{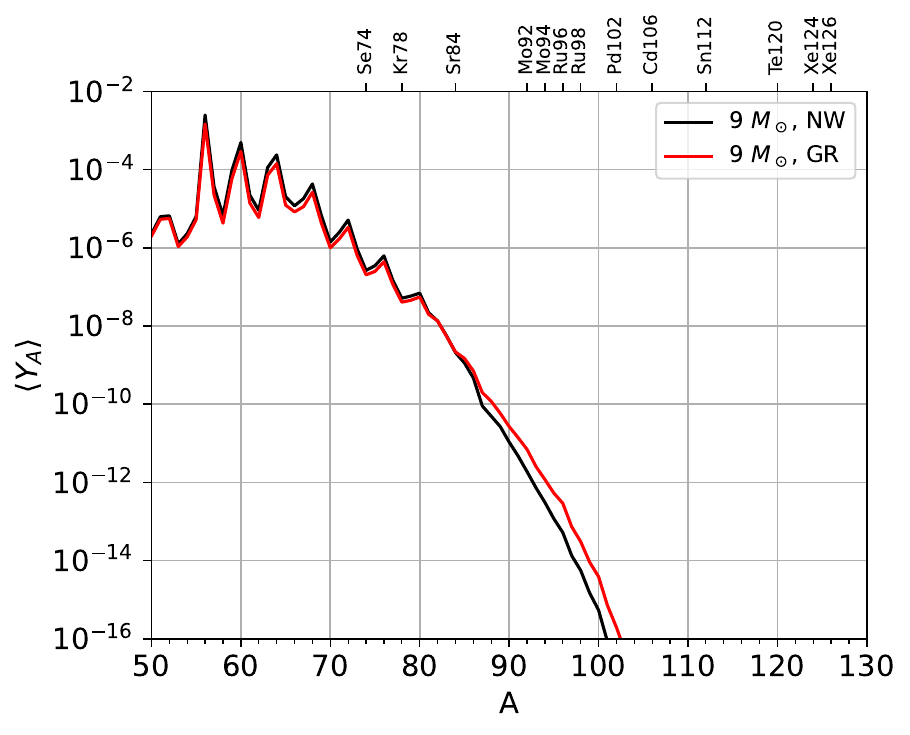}
  \caption{Time-averaged yields $\langle Y_A \rangle$ obtained with a $9~M_{\odot}$ progenitor model and $v_{\rm FS}=10^4$~km/s, with (red) and without (black) GR effects. Very energetic explosions generally result in poor yields even with the enhancement of relativistic corrections. Notice that abundances of nuclides in the relevant range $90 \lesssim A \lesssim 105$ are approximately 5--6 orders of magnitude lower than our benchmark $18~M_{\odot}$ progenitor model in Fig.~\ref{fig:moneyplot}.
  }
  \label{fig:nuclides-9}
\end{figure}

Numerical results on the nucleosynthetic yields presented in the previous sections are obtained using as benchmark case an $18~M_\odot$ progenitor with a front shock velocity of $v_{\rm FS} = 6000$~km/s and a $1.8~M_\odot$ PNS. It is, however, important to investigate how the results depend on the progenitor mass and other relevant physical factors, in particular the FS velocity. \textcolor{black}{We verified that increasing the FS velocity to $v_{\rm FS} = 8000$~km/s in the $18~M_\odot$ progenitor model does not alter the qualitative behavior of the outflows. They remain subsonic within the optimal time window, and the resulting nucleosynthetic yields are qualitatively consistent with those of our benchmark case, though reduced overall by a factor of a few. Therefore, here} we present results for lighter progenitors, specifically $9~M_\odot$ and $12.75~M_\odot$. 
In order to do so, we compute tracer trajectories and the $\nu p$ yields in the time window $t\in[1,\,8]$~s.  We find that both decreasing the progenitor mass and increasing $v_{\rm FS}$ generally \textcolor{black}{reduce} the $\nu p$ yields. This is particularly evident for the $9~M_\odot$ progenitor with $v_{\rm FS} = 10^4$~km/s, where we also use a smaller ($1.3~M_\odot$) PNS mass, because CCSN explosions of such progenitors do not involve an extended accretion phase that culminates in a heavy PNS~\cite{Huedepohl:2009wh, Fischer:2009, Burrows:2019rtd, Burrows:2019zce}. In this model, the lower confining pressure makes the outflows supersonic at early times, precisely at $t\approx 1$~s (GR) and at $t\approx1.5$~s (NW). It has been demonstrated that supersonic outflows \textcolor{black}{are not conducive to robust} $\nu p$-process yields~\cite{Friedland:2023kqp}. As shown in Fig.~\ref{fig:nuclides-9}, the time-averaged yields are extremely small in the mass window $90 \leq A \leq 100$, with $Y_{\rm Mo} \approx 10^{-12}$ and $Y_{\rm Ru} \approx 10^{-13}$. For comparison, observationally consistent yields  should be on the order of $10^{-6}$, as previously illustrated in Fig.~\ref{fig:moneyplot}. This makes $9~M_\odot$ progenitors very inefficient environments for the $\nu p$-process. In such extreme cases, GR corrections still enhance the abundances of the $p$ nuclides, but not enough to make them compatible with observations.

\begin{figure}
\centering
  \includegraphics[width=0.49\columnwidth]{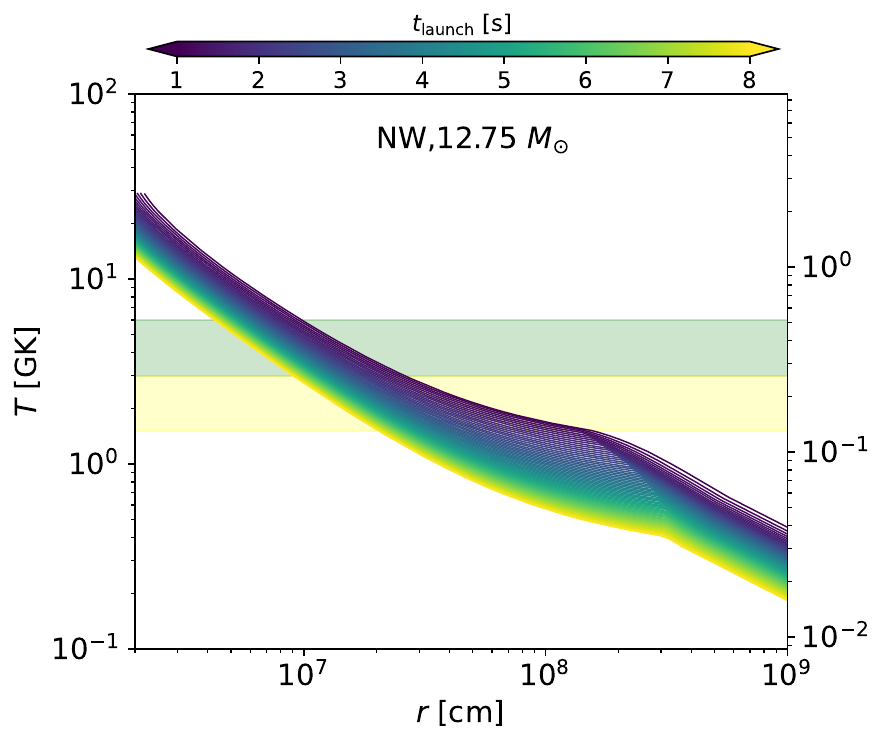}
  \includegraphics[width=0.49\columnwidth]{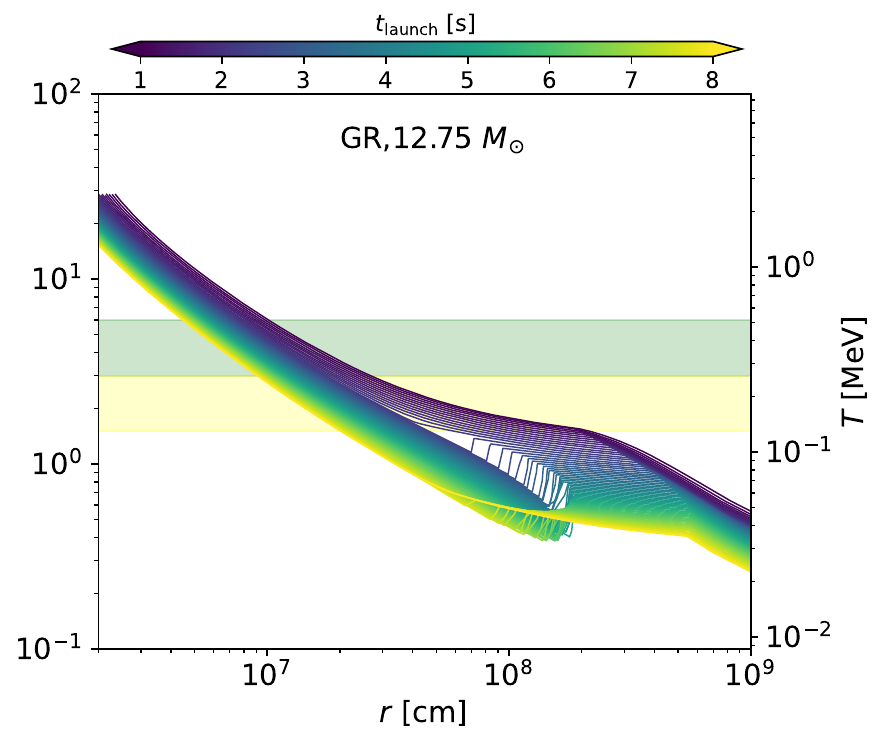}
  \caption{Evolution of the outflow temperature for NW (left panel) and GR (right panel) calculations for different launch times in our $12.75~M_{\odot}$ progenitor model. The front shock velocity of $v_{\rm FS}=8000 \ {\rm km/s}$ is assumed. Here, GR corrections qualitatively change the nature of the outflow, accelerating the material to supersonic speeds for $t_{\rm launch}\gtrsim 2.3$~s. The corresponding jump in temperature at the location of the reverse shock is evident. \textcolor{black}{In both panels, the green band represents the $6\,\mathrm{GK} > T > 3\,\mathrm{GK}$ range, where the stage I of the $\nu p$-process occurs, while the yellow band indicates the $3\,\mathrm{GK} > T > 1.5\,\mathrm{GK}$ temperature window corresponding to the stage II.}
  }
  \label{fig:temp-8k}
\end{figure}

On the other hand, GR effects \textcolor{black}{can} qualitatively change the outcome in intermediate cases \textcolor{black}{which may exhibit} transonic outflows for more energetic explosions, i.e., with higher FS velocity. For instance, for a $12.75~M_\odot$ progenitor, if $v_{\rm FS}=6000$~km/s, outflows are subsonic in all the considered time window for both GR and NW. For a slightly higher speed (${v_{\rm FS}=8000}$~km/s), NW outflows still remain subsonic, while GR outflows become transonic at $t_{\rm launch}\gtrsim2.3$~s, as shown in Fig.~\ref{fig:temp-8k}. \textcolor{black}{The appearance of a discontinuity in the temperature profiles at $t_{\rm launch}\gtrsim2.3$~s in the GR case (right panel) marks the formation of the reverse shock, which is absent in the NW case (left panel), where the temperature profiles remain smooth. This behavior is qualitatively different from our benchmark $18~M_\odot$ model, in which the outflows remain subsonic for both NW and GR cases, resulting always in smooth temperature profiles (see Fig.~\ref{fig:temp}).} A more detailed discussion on transonic outflows can be found in Sec.~\ref{sec:outflow:results:supersonic}. This example illustrates the near critical nature of the outflow~\cite{Friedland:2020ecy}. For even larger FS velocities, the transition to the supersonic phase in the GR scenario occurs earlier. 

\begin{figure}
\centering
\includegraphics[width=0.80\columnwidth]{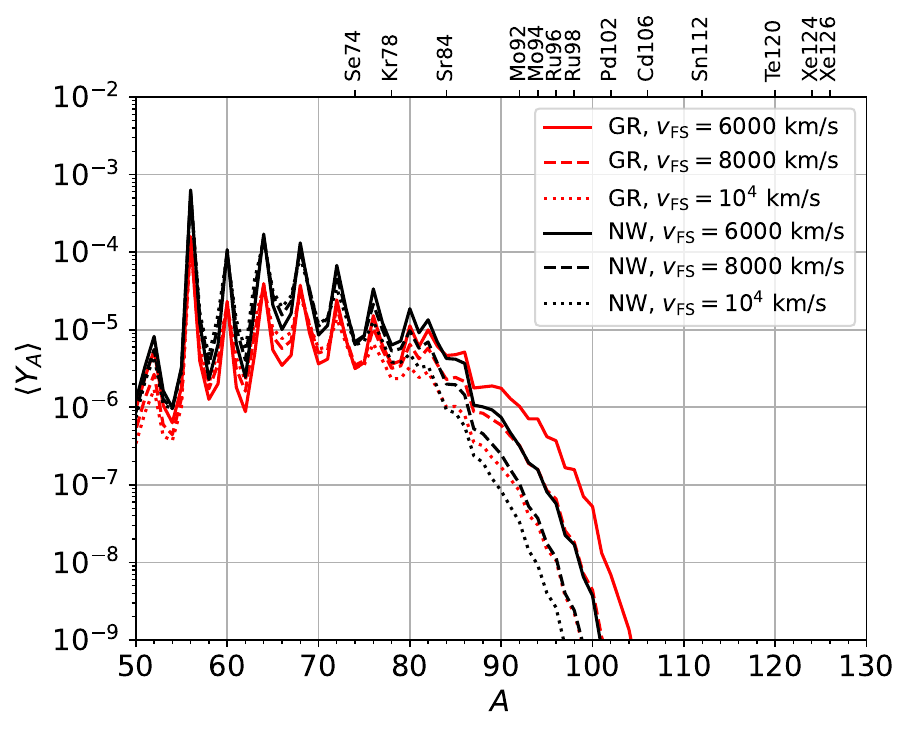}
  \caption{Time-averaged yields $\langle Y_A \rangle$ assuming a $12.75~M_{\odot}$ progenitor and the fully GR (red) and NW (black) cases, while varying the front shock velocity $v_{\rm FS}$. Higher yields are obtained for lower values of $v_{\rm FS}$, with abundances approximately one order of magnitude greater when ${v_{\rm FS}=6000}$~km/s than ${v_{\rm FS}=10^4}$~km/s. As before, relativistic corrections work in favor of larger yields in the ${90 \lesssim A \lesssim 105}$ mass range.}
  \label{fig:yields-vFS}
\end{figure}

In Fig.~\ref{fig:yields-vFS}, we show the time-averaged yields of the $12.75~M_{\odot}$ SN model, with FS velocities ranging from 6000~km/s (solid lines) to $10^4$~km/s (dotted) considering both the fully NW (black) and GR (red) cases. 
As a general trend, larger values of $v_{\rm FS}$ imply smaller yields. In particular, for $v_{\rm FS}=6000$~km/s, yields are a factor of a few smaller than those of our benchmark $18~M_\odot$ SN model (see Fig.~\ref{fig:moneyplot}), therefore marginally compatible with observations. \textcolor{black}{For $v_{\rm FS}=8000$~km/s (dashed lines), the outflow transitions to supersonic at $t_{\rm launch}=2.3$~s and we find a further reduction in the yields by a factor $\sim 3$ for $^{92}{\rm Mo}$ and about one order of magnitude for $^{98}{\rm Ru}$, relative to the $v_{\rm FS} = 6000$~km/s case. Consequently, the resulting yields are not compatible with the measured abundances, as the production factors of $\Mo$ and $\Ru$ fall short of those required by solar-system observations by at least one order of magnitude. For $v_{\rm FS} = 10^4$~km/s, the outflow becomes transonic at $t_{\rm launch}=1.7$~s. In this case, the production factors of $\Ru$ become nearly two orders of magnitude lower than in the $v_{\rm FS} = 6000$~km/s reference model. Regardless of the overall normalization of the integrated abundances, larger FS velocities prevent the co-production of key $p$ isotopes in the $90\leq A \leq 100$ range in the $12.75~M_{\odot}$ model}  (see, e.g., the dotted lines in Fig.~\ref{fig:yields-vFS} corresponding to $v_{\rm FS}=10^4$~km/s)\,\footnote{Such large value of $v_{\rm FS}$ allows us to show the yield suppression. However, this is much higher than the FS velocities in simulations, ranging from $6000$~km/s to $8000$~km/s for progenitor masses $\gtrsim 10~M_\odot$~\cite{Burrows:2019rtd,Bollig:2020phc,Burrows:2019zce}.}. 

\begin{figure}
\centering
  \includegraphics[width=0.7\columnwidth]{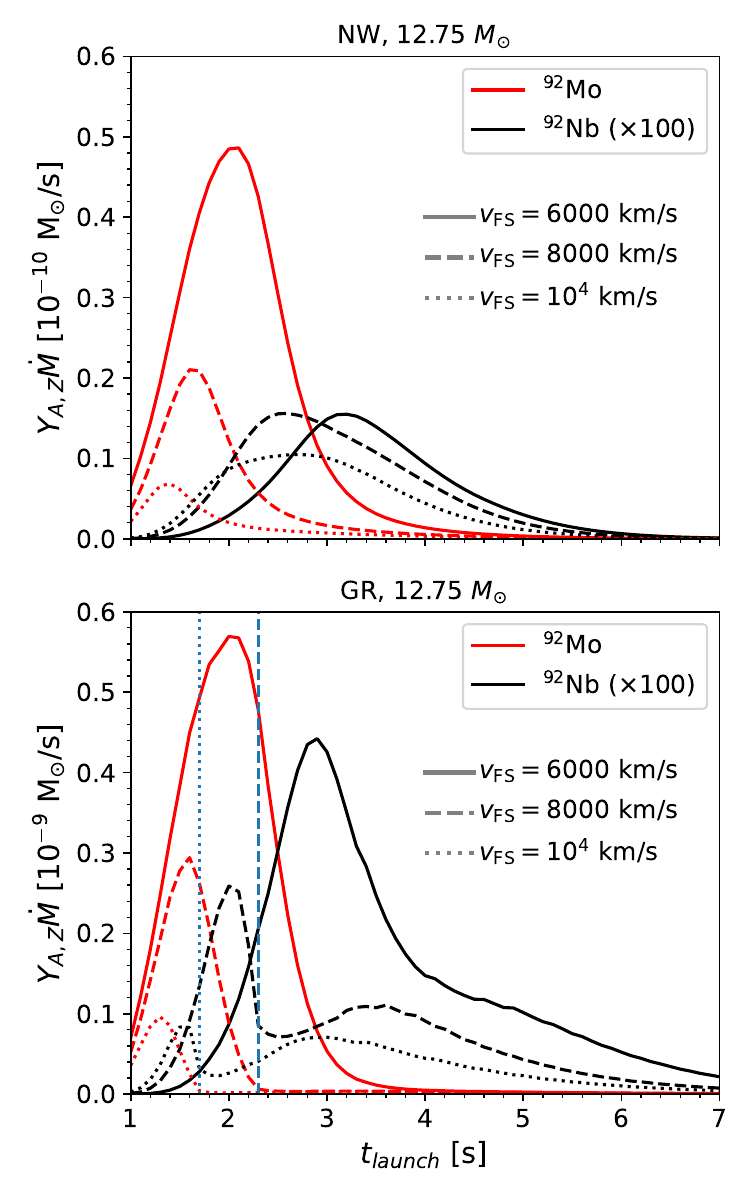}
  \caption{Time evolution of $^{92}$Mo (red) and $^{92}$Nb (black) assuming a $12.75~M_{\odot}$ progenitor for the fully NW (top) and GR (bottom) cases. Different front shock velocities $v_{\rm FS}$ are implemented: 6000~km (solid), 8000~km/s (dashed) and $10^4$~km/s (dotted). Vertical lines mark the time when the corresponding outflow becomes transonic; no vertical line means the outflow remains subsonic. One can see that as the shock velocity increases, the yields become progressively lower. As seen in the bottom panel, the presence of a termination shock affects severely the $^{92}$Nb production.}
  \label{fig:comparisons-vFS}
\end{figure}

As shown in Fig.~\ref{fig:nuclides-9} and Fig.~\ref{fig:yields-vFS}, relativistic effects do \textcolor{black}{enhance} the yields by almost one order of magnitude in the mass window $90\leq A \leq 100$ for all the different cases studied, confirming the robustness of the conclusions drawn from the analysis of the $18~M_\odot$ case. Notably, independently of the FS velocity, even for explosions developing termination shocks, there is always an optimal window for the production of $p$ nuclides, during which GR effects enhance $\nu p$ yields. This is clearly shown in Fig.~\ref{fig:comparisons-vFS}, where we plot the time evolution of the $^{92}$Mo (red lines) and $^{92}$Nb (black lines) production in the NW (upper panel) and GR (lower panel) scenarios for the $12.75~M_\odot$ SN model and different FS velocities. 
For $v_{\rm FS}=6000$~km/s (solid lines), in both NW and GR cases, we observe a temporal evolution analogous to what observed for the $18~M_\odot$ SN model in Fig.~\ref{fig:timeyields}, with an optimal production window for $^{92}{\rm Mo}$ at $t_{\rm launch}\sim 2$~s, and the $\Nb$ production peaked at later times. 

Increasing the FS velocity leads to faster outflows and reduces the production of $^{92}$Mo (as well as of $^{94}$Mo and $^{96,98}$Ru), as shown by the dashed and dotted red lines in Fig.~\ref{fig:comparisons-vFS}. For $v_{\rm FS}=8000$~km/s, the NW outflow solution remains subsonic and no termination shock is formed; the higher FS velocity moves the optimal window of $\Nb$ to earlier times but the overall height of the peak remains almost unaffected. On the other hand, in the GR case, both the $^{92}{\rm Mo}$ and the $\Nb$ productions are peaked at $t_{\rm launch}\lesssim 2$~s and, due to the transition to the supersonic regime, are shut off at $2.3$~s (see the dashed lines in the lower panel of Fig.~\ref{fig:comparisons-vFS}). Since in the supersonic regime outflows move faster and enter the region with $T\lesssim 1.5$~GK earlier and closer to the PNS, $\Nb$ production slightly increases again in this phase before being suppressed at $t_{\rm launch}\gtrsim 3.5$~s. A similar trend is observed for the extremely high FS velocity $v_{\rm FS}=10^{4}$~km/s (dotted lines), for which the optimal window is shifted at even earlier times---prior to the supersonic transition at $t\sim 1.7$~s---and the yields are further suppressed. 

In summary, we observe that transonic outflows lead to poorer yields compared to subsonic outflows, with two important implications. First, the $\nu p$-process is not efficient in very light progenitors, such as the $9~M_\odot$ model. Second, for the cases in which GR effects induce supersonic transitions, as in the $12.75~M_\odot$ SN model, the optimal production window of $p$ nuclides is shifted to earlier times, when the outflow is still subsonic.

\section{Summary and Conclusions} \label{sec:conclusion}

In this paper, we have investigated and quantified the effects of general relativity on the yields of the $\nu p$-process in a core-collapse SN. Our findings can be summarized as follows.
\begin{itemize}
    \item Although the main stage of the process---the one involving repeated captures of protons and neutrons on seed nuclei---occurs hundreds of kilometers from the surface of the PNS where GR effects are small, the GR corrections to the $\nu p$-process yields are nonetheless significant. The reason for this is that the GR effects impact the crucial region tens of kilometers above the PNS, which hosts the engine of the neutrino-driven outflow. Changes there affect the entire test particle trajectory, even qualitatively altering the nature of the outflow, under some circumstances. The inclusion of GR effects generally leads to improved $\nu p$ yields, but with several important caveats.
    \item  In the language of post-Newtonian corrections, the two main impacts of GR are due to the gravitational shift of the neutrino energies with radius and the deepening of the effective gravitational potential. This was shown by investigating ``partial GR'' implementations, which were designed to deconstruct the relevant physics. Since neutrino cross sections are energy-dependent, taking the gravitational shifts into account changes the total energy deposited into the outflow~\cite{Otsuki:1999kb}, thereby affecting the expansion speed, the mass loss rate, and the entropy per baryon in the outflow. The deepening of the effective gravitational potential also impacts the value of the entropy, but with an opposite sign. 
    \item In turn, these two effects can influence the $\nu p$ yields in several different ways. The increase in the expansion speed close to the PNS surface decreases the time the material spends in the seed-forming region, which suppresses seed production, thus leading to a more favorable neutron-to-seed ratio during the main $\nu p$ stage. The would-be negative impact on the entropy caused by the neutrino energy shift is more than compensated by the modification of the effective gravitational potential and the resulting modest entropy boost further assists the yields.
    \item On the other hand, the enhanced heating can, under certain circumstances, accelerate the material to supersonic speeds. As we had shown before~\cite{Friedland:2023kqp}, supersonic outflows are generally characterized by poor $\nu p$ yields and our studies here further confirm this.  Whether GR effects cause the outflow to become supersonic depends on the interplay of the neutrino luminosities, the progenitor mass and the front shock expansion rate. \textcolor{black}{In the cases considered here, the 18 $M_\odot$ model remained safely subsonic, while the $9\,M_\odot$ model was supersonic for the entire time window considered. A very instructive case is provided by the 12.75 $M_\odot$ model, which perfectly illustrates the near-critical nature of the outflow~\cite{Friedland:2020ecy} in SN conditions. For a front shock velocity of 6000 km/s, the model remained subsonic throughout its time-evolution, resulting in robust yields. On the other hand, the model with an 8000 km/s front shock velocity exhibited a supersonic transition 2.3 seconds into the explosion, accompanied by a drop in the $\nu p$ yields.}
    \item The calculations with the 18 $M_\odot$ model confirm that, in the fully-subsonic regime, most of the crucial $p$ nuclides, especially $^{92,94}$Mo and $^{96,98}$Ru, are produced in a well-defined time window, lasting 1-2 seconds. This optimal window is defined by the interplay of two factors: (i) the duration of time the material spends in the temperature region between 3 and 1.5 GK and (ii) the PNS radius, which sets the entropy of the outflow material. Thus, in modeling the $\nu p$-process yields, it is important to take the time evolution of the PNS radius into account.    
    \item The $\nu p$ yields in the 12.75 $M_\odot$ model exhibit a different time pattern \textcolor{black}{in the case when it becomes transonic. Before the transonic transition,  when the outflow is subsonic but near-critical, the conditions at stages I and II are fairly optimal, resulting in good instantaneous yields. However, after the transition the yields drop off sharply and the resulting integrated yields are suboptimal. }
    \item The analysis of the $^{92}$Nb yields deserves a special mention. This important nuclide serves as an cosmochronometer~\cite{Lugaro:2016zuf,Haba:2021PNAS} and its origins have been the focus of much research over the decades~\cite{DAUPHAS2003C287,Rauscher:2013}. It was shown in~\cite{Friedland:2023kqp} that, despite being shielded from beta decay paths by the stable $^{92}$Mo, it is nonetheless produced in the $\nu p$-process. This production is driven by late-time neutrons, produced after the mass element enters the temperature region $\lesssim 1.5$ GK. Here we show that the optimal time window for the production of $^{92}$Nb occurs later than that of the standard $\nu p$ nuclides. At these times, the radius of the PNS is considerably decreased~\cite{Fischer:2009, Roberts:2016rsf, Bollig:2020phc, Nagakura:2021lma}, therefore, the GR effects on the outflow and hence on the yields are very pronounced. We find that GR boosts the $^{92}$Nb yields by as much as a factor of 25.
\end{itemize}

\textcolor{black}{It should be emphasized that the $18\,M_\odot$ model is not fine-tuned in any way and in fact serves as an excellent illustration of what is generically expected for an explosion with a sufficiently massive progenitor star.
It is therefore remarkable that this model is able to reproduce the solar-system $p$-nuclide abundances in the \emph{entire} mass range $74 \le A  \le 102$, while also reproducing the observed abundance of the extinct nuclide $^{92}$Nb. This is the principal result of this paper.}

\textcolor{black}{We note that, apart from the predecessor of this work~\cite{Friedland:2023kqp}, this is the only study of the $\nu p$-process incorporating GR effects and self-consistently treating the hydrodynamics for each time snapshot. The yields in the present paper are different than in~\cite{Friedland:2023kqp}, owing to a 
number of physics improvements implemented here. Of these, the three major factors turn out to be: (i) a more accurate implementation of GR effects, (ii) the inclusion of the time dependence of the PNS radius, and (iii) the inclusion of the neutrino-electron heating mechanism. The latter two were completely omitted in~\cite{Friedland:2023kqp}, while GR effects were implemented there, but using the equations from Ref.~\cite{Cardall:1997bi}. As we showed in this paper, those equations were missing certain factors. Other relevant improvements in the present paper include a more accurate treatment of variable relativistic degrees of freedom and a refined procedure for merging the two stages of the trajectory. Together, all these  improvements significantly increase the maximum outflow speed, suppressing seed production and therefore enhancing the $\nu p$ yields.}

\textcolor{black}{On the subject of comparing our results to the earlier literature on neutrino-driven outflows, it is worth mentioning that the large GR-induced entropy increases reported in several earlier works~\cite{Cardall:1997bi,Otsuki:1999kb,Thompson:2001} are not present in our results here. The apparent discrepancy is not due to the differences in the underlying equations, but because of a different convention about the neutrino spectra. Namely, the earlier works compared the GR and Newtonian calculations assuming identical neutrino spectra at the PNS surface. We instead consider our spectra given at 500 km, which is how most modern numerical simulations report their spectra~\cite{Fischer:2009, Radice:2017ykv, Burrows:2019rtd, Burrows:2019zce, Vartanyan:2018xcd, Nagakura:2021lma, Bollig:2020phc,Fiorillo:2023frv}. It should also be noted that this is the relevant convention from the experimental point of view. 
}

\textcolor{black}{The influence of the neutrino spectra on the resulting neutrino-driven outflows underscores the importance of key external inputs in our analysis---such as neutrino luminosities and energy spectra, as well as the PNS mass and radius---which are informed by available core-collapse SN simulations.}
The resulting $\nu p$ yields are strongly sensitive to these inputs, via the value of the electron fraction $Y_e$, the value of entropy per baryon $S$, the outflow velocity profile, etc. We therefore hope that our study provides an additional physical motivation to the groups working on incorporating state-of-the-art physics in the explosion simulations, particularly improved neutrino transport, modern nuclear EoS, and the effects of the PNS convection~\cite{1988PhR...163...51B,Dessart:2005ck,Huedepohl:2009wh,Roberts:2011yw,Mirizzi:2015eza,Roberts:2016mwj,Roberts:2016rsf,Fischer:2018kdt,Nagakura:2019tmy,Oertel:2020pcg,Pascal:2022qeg,Suleiman:2023bdf,Fiorillo:2023frv,Lucente:2024ngp}.

Additionally, future studies should explore the impact of flavor oscillations, quantify the limitations of the steady-state approximation for the first stage of the outflow and assess the role of multi-dimensional hydrodynamics. The latter two can be pursued by incorporating into the nucleosynthesis study the outputs of detailed numerical simulations (see, e.g., Refs.~\cite{Sieverding:2020wxw, Wang:2023vkk, Wang:2023vml, Wang:2024fnv, Zha:2024fyo}, albeit not in a $\nu p$-process context). Our results in this paper should provide guidance for identifying optimal conditions for these studies and for ensuring the relevant GR effects are included.

\section*{Acknowledgements}
\textcolor{black}{We thank the anonymous referee for their helpful comments, which improved the clarity of the presentation.} The work of AF, DL, GL, and IPG at SLAC was supported by the U.S. Department of Energy under contract number DE-AC02-76SF00515. AVP acknowledges partial support from the U.S. Department of Energy under contract number DE-FG02-87ER40328 at the University of Minnesota, and would also like to thank SLAC for their hospitality and support during the period of completion of this project. IPG is supported by NSF Physics Frontier Center Award number 2020275. The authors would like to thank P.~Mukhopadhyay for useful discussions during our previous work on this topic, wherein seeds of many of the ideas pursued in the current work were sown.

\section*{Code and data availability}
\textcolor{black}{The open-source nucleosynthesis network code {\tt SkyNet}, and the wrapper implementing the in-medium triple-alpha reaction rate enhancement are publicly available~\cite{skynet,triplealpha}. To conduct the calculations presented in this work, these codes had to be suitably modified, as described in Sec.~\ref{sec:nup-method} and App.~\ref{app:setup-skynet}. The steady-state outflow solver and/or the tracer-particle trajectories for nucleosynthesis may be shared with the reader upon reasonable request to the authors.
} 

\bibliographystyle{bibi}
\bibliography{Nucleobib}

@article{Friedland:2023kqp,
    author = "Friedland, Alexander and Mukhopadhyay, Payel and Patwardhan, Amol V.",
    title = "{Successful \ensuremath{\nu}p-process in neutrino-driven outflows in core-collapse supernovae}",
    eprint = "2312.03208",
    archivePrefix = "arXiv",
    primaryClass = "astro-ph.HE",
    reportNumber = "SLAC-PUB-17746, INT-PUB-23-040",
    doi = "10.1088/1475-7516/2025/02/005",
    journal = "JCAP",
    volume = "02",
    pages = "005",
    year = "2025"
}

@ARTICLE{Anders:1989,
       author = {{Anders}, E. and {Grevesse}, N.},
        title = "{Abundances of the elements: Meteoritic and solar}",
      journal = {\gca},
         year = 1989,
        month = jan,
       volume = {53},
       number = {1},
        pages = {197-214},
          doi = {10.1016/0016-7037(89)90286-X},
       adsurl = {https://ui.adsabs.harvard.edu/abs/1989GeCoA..53..197A}
}

@ARTICLE{Meyer:1994,
       author = {{Meyer}, Bradley S.},
        title = "{The r-, s-, and p-Processes in Nucleosynthesis}",
      journal = {\araa},
         year = 1994,
        month = jan,
       volume = {32},
        pages = {153-190},
          doi = {10.1146/annurev.aa.32.090194.001101},
       adsurl = {https://ui.adsabs.harvard.edu/abs/1994ARA&A..32..153M}
}

@book{Shapiro:1983du,
    author = "Shapiro, S. L. and Teukolsky, S. A.",
    title = "{Black holes, white dwarfs, and neutron stars: The physics of compact objects}",
    doi = "10.1002/9783527617661",
    isbn = "978-0-471-87316-7, 978-3-527-61766-1",
    year = "1983"
}

@book{Kolb:1990vq,
    author = "Kolb, Edward W. and Turner, Michael S.",
    title = "{The Early Universe}",
    reportNumber = "FERMILAB-BOOK-1990-01",
    doi = "10.1201/9780429492860",
    isbn = "978-0-429-49286-0, 978-0-201-62674-2",
    publisher = "Taylor and Francis",
    volume = "69",
    month = "5",
    year = "2019"
}

@article{Husdal:2016haj,
    author = "Husdal, Lars",
    title = "{On Effective Degrees of Freedom in the Early Universe}",
    eprint = "1609.04979",
    archivePrefix = "arXiv",
    primaryClass = "astro-ph.CO",
    doi = "10.3390/galaxies4040078",
    journal = "Galaxies",
    volume = "4",
    number = "4",
    pages = "78",
    year = "2016"
}

@article{Roberts:2016rsf,
    author = "Roberts, Luke F. and Reddy, Sanjay",
    title = "{Neutrino Signatures From Young Neutron Stars}",
    eprint = "1612.03860",
    archivePrefix = "arXiv",
    primaryClass = "astro-ph.HE",
    doi = "10.1007/978-3-319-21846-5_5",
    month = "12",
    year = "2016"
}

@article{Friedland:2020ecy,
	archiveprefix = {arXiv},
	author = {Friedland, Alexander and Mukhopadhyay, Payel},
	doi = {10.1016/j.physletb.2022.137403},
	eprint = {2009.10059},
	journal = {Phys. Lett. B},
	pages = {137403},
	primaryclass = {astro-ph.HE},
	reportnumber = {SLAC-PUB-17562},
	title = {{Near-critical supernova outflows and their neutrino signatures}},
	volume = {834},
	year = {2022}}

@article{Schatz:1998zz,
	author = {Schatz, H. and others},
	doi = {10.1016/S0370-1573(97)00048-3},
	journal = {Phys. Rept.},
	pages = {167--263},
	title = {{rp-process nucleosynthesis at extreme temperature and density conditions}},
	volume = {294},
	year = {1998}}

@phdthesis{Mukhopadhyay:2022yrd,
    author = "Mukhopadhyay, Payel",
    title = "{Neutrino driven outflows in supernovae : from hydrodynamics to nucleosynthesis}",
    school = "Stanford U.",
    year = "2022"
}

@article{ArnouldGoriely2003,
	adsurl = {https://ui.adsabs.harvard.edu/abs/2003PhR...384....1A},
	author = {{Arnould}, M. and {Goriely}, S.},
	doi = {10.1016/S0370-1573(03)00242-4},
	journal = {Phys. Rept.},
	month = sep,
	number = {1-2},
	pages = {1-84},
	title = {{The p-process of stellar nucleosynthesis: astrophysics and nuclear physics status}},
	volume = {384},
	year = 2003}

@article{Rauscher:2014fea,
	archiveprefix = {arXiv},
	author = {Rauscher, T.},
	doi = {10.1063/1.4868239},
	eprint = {1403.2015},
	journal = {AIP Adv.},
	pages = {041012},
	primaryclass = {astro-ph.SR},
	title = {{Challenges in nucleosynthesis of trans-iron elements}},
	volume = {4},
	year = {2014}}

@article{Cameron:1958vx,
	author = {Cameron, A. G. W},
	doi = {10.1146/annurev.ns.08.120158.001503},
	journal = {Ann. Rev. Nucl. Part. Sci.},
	pages = {299--326},
	title = {{Nuclear astrophysics}},
	volume = {8},
	year = {1958}}

@article{Roberts:2016mwj,
	archiveprefix = {arXiv},
	author = {Roberts, Luke F. and Reddy, Sanjay},
	doi = {10.1103/PhysRevC.95.045807},
	eprint = {1612.02764},
	journal = {Phys. Rev. C},
	number = {4},
	pages = {045807},
	primaryclass = {astro-ph.HE},
	reportnumber = {INT-PUB-16-048},
	title = {{Charged current neutrino interactions in hot and dense matter}},
	volume = {95},
	year = {2017}}

@article{DAUPHAS2003C287,
	author = {N. Dauphas and T. Rauscher and B. Marty and L. Reisberg},
	doi = {https://doi.org/10.1016/S0375-9474(03)00934-5},
	issn = {0375-9474},
	journal = {Nuclear Physics A},
	pages = {C287-C295},
	title = {Short-lived p-nuclides in the early solar system and implications on the nucleosynthetic role of X-ray binaries},
	url = {https://www.sciencedirect.com/science/article/pii/S0375947403009345},
	volume = {719},
	year = {2003}}

@article{Bollig:2020phc,
	archiveprefix = {arXiv},
	author = {Bollig, Robert and Yadav, Naveen and Kresse, Daniel and Janka, H. -Th. and M\"uller, Bernhard and Heger, Alexander},
	doi = {10.3847/1538-4357/abf82e},
	eprint = {2010.10506},
	journal = {Astrophys. J.},
	number = {1},
	pages = {28},
	primaryclass = {astro-ph.HE},
	title = {{Self-consistent 3D Supernova Models From \ensuremath{-}7 Minutes to +7 s: A 1-bethe Explosion of a \ensuremath{\sim}19 $M_\odot$ Progenitor}},
	volume = {915},
	year = {2021}}

@article{Fujibayashi:2015rma,
	archiveprefix = {arXiv},
	author = {Fujibayashi, Sho and Yoshida, Takashi and Sekiguchi, Yuichiro},
	doi = {10.1088/0004-637X/810/2/115},
	eprint = {1507.05945},
	journal = {Astrophys. J.},
	number = {2},
	pages = {115},
	primaryclass = {astro-ph.HE},
	title = {{Nucleosynthesis in neutrino-driven winds in hypernovae}},
	volume = {810},
	year = {2015}}

@article{Xiong:2020ntn,
	archiveprefix = {arXiv},
	author = {Xiong, Zewei and Sieverding, Andre and Sen, Manibrata and Qian, Yong-Zhong},
	doi = {10.3847/1538-4357/abac5e},
	eprint = {2006.11414},
	journal = {Astrophys. J.},
	number = {2},
	pages = {144},
	primaryclass = {astro-ph.HE},
	title = {{Potential Impact of Fast Flavor Oscillations on Neutrino-driven Winds and Their Nucleosynthesis}},
	volume = {900},
	year = {2020}}

@article{OConnor:2018sti,
	archiveprefix = {arXiv},
	author = {O'Connor, Evan and others},
	doi = {10.1088/1361-6471/aadeae},
	eprint = {1806.04175},
	journal = {J. Phys. G},
	number = {10},
	pages = {104001},
	primaryclass = {astro-ph.HE},
	title = {{Global Comparison of Core-Collapse Supernova Simulations in Spherical Symmetry}},
	volume = {45},
	year = {2018}}

@article{Mirizzi:2015eza,
    author = "Mirizzi, Alessandro and Tamborra, Irene and Janka, Hans-Thomas and Saviano, Ninetta and Scholberg, Kate and Bollig, Robert and Hudepohl, Lorenz and Chakraborty, Sovan",
    title = "{Supernova Neutrinos: Production, Oscillations and Detection}",
    eprint = "1508.00785",
    archivePrefix = "arXiv",
    primaryClass = "astro-ph.HE",
    doi = "10.1393/ncr/i2016-10120-8",
    journal = "Riv. Nuovo Cim.",
    volume = "39",
    number = "1-2",
    pages = "1--112",
    year = "2016"
}

@article{Eichler:2017kvd,
	archiveprefix = {arXiv},
	author = {Eichler, M. and Nakamura, K. and Takiwaki, T. and Kuroda, T. and Kotake, K. and Hempel, M. and Cabez{\'o}n, R. and Liebend{\"o}rfer, M. and Thielemann, F-K.},
	doi = {10.1088/1361-6471/aa8891},
	eprint = {1708.08393},
	journal = {J. Phys.},
	number = {1},
	pages = {014001},
	primaryclass = {astro-ph.SR},
	slaccitation = {%%CITATION = ARXIV:1708.08393;%%},
	title = {{Nucleosynthesis in 2D Core-Collapse Supernovae of 11.2 and 17.0 M$_{\odot}$ Progenitors: Implications for Mo and Ru Production}},
	volume = {G45},
	year = {2018}}

@article{Cowan:2019pkx,
	archiveprefix = {arXiv},
	author = {Cowan, John J. and Sneden, Christopher and Lawler, James E. and Aprahamian, Ani and Wiescher, Michael and Langanke, Karlheinz and Mart{\'\i}nez-Pinedo, Gabriel and Thielemann, Friedrich-Karl},
	eprint = {1901.01410},
	primaryclass = {astro-ph.HE},
	slaccitation = {%%CITATION = ARXIV:1901.01410;%%},
	title = {{Making the Heaviest Elements in the Universe: A Review of the Rapid Neutron Capture Process}},
	year = {2019}}

@article{Woosley:1994ux,
	author = {Woosley, S. E. and Wilson, J. R. and Mathews, G. J. and Hoffman, R. D. and Meyer, B. S.},
	doi = {10.1086/174638},
	journal = {Astrophys. J.},
	pages = {229-246},
	slaccitation = {%%CITATION = ASJOA,433,229;%%},
	title = {{The r process and neutrino heated supernova ejecta}},
	volume = {433},
	year = {1994}}

@article{Vartanyan:2018iah,
	archiveprefix = {arXiv},
	author = {Vartanyan, David and Burrows, Adam and Radice, David and Skinner, Aaron M. and Dolence, Joshua},
	doi = {10.1093/mnras/sty2585},
	eprint = {1809.05106},
	journal = {Mon. Not. Roy. Astron. Soc.},
	number = {1},
	pages = {351-369},
	primaryclass = {astro-ph.HE},
	slaccitation = {%%CITATION = ARXIV:1809.05106;%%},
	title = {{A Successful 3D Core-Collapse Supernova Explosion Model}},
	volume = {482},
	year = {2019}}

@article{Fischer:2009,
	archiveprefix = {arXiv},
	author = {Fischer, T. and Whitehouse, S. C. and Mezzacappa, A. and Thielemann, F. -K. and Liebendorfer, M.},
	doi = {10.1051/0004-6361/200913106},
	eprint = {0908.1871},
	journal = {Astron. Astrophys.},
	pages = {A80},
	primaryclass = {astro-ph.HE},
	slaccitation = {%%CITATION = ARXIV:0908.1871;%%},
	title = {{Protoneutron star evolution and the neutrino driven wind in general relativistic neutrino radiation hydrodynamics simulations}},
	volume = {517},
	year = {2010}}

@article{Qian:1996xt,
	archiveprefix = {arXiv},
	author = {Qian, Y. Z. and Woosley, S. E.},
	doi = {10.1086/177973},
	eprint = {astro-ph/9611094},
	journal = {Astrophys. J.},
	pages = {331-351},
	slaccitation = {%%CITATION = ASTRO-PH/9611094;%%},
	title = {{Nucleosynthesis in neutrino driven winds: 1. The Physical conditions}},
	volume = {471},
	year = {1996}}

@article{Thompson:2001,
	archiveprefix = {arXiv},
	author = {Thompson, Todd A. and Burrows, Adam and Meyer, Bradley S.},
	doi = {10.1086/323861},
	eprint = {astro-ph/0105004},
	journal = {Astrophys. J.},
	pages = {887},
	primaryclass = {astro-ph},
	slaccitation = {%%CITATION = ASTRO-PH/0105004;%%},
	title = {{The Physics of protoneutron star winds: implications for r-process nucleosynthesis}},
	volume = {562},
	year = {2001}}

@Inbook{Thompson2004,
        author="Thompson, Todd A.",
        title="Protoneutron Star Winds",
        bookTitle="Stellar Collapse",
        year="2004",
        publisher="Springer Netherlands",
        address="Dordrecht",
        pages="175--202",
        isbn="978-0-306-48599-2",
        doi="10.1007/978-0-306-48599-2_6",
        url="https://doi.org/10.1007/978-0-306-48599-2_6"
}

@article{Duncan1986,
	adsurl = {http://adsabs.harvard.edu/abs/1986ApJ...309..141D},
	author = {{Duncan}, R.~C. and {Shapiro}, S.~L. and {Wasserman}, I.},
	doi = {10.1086/164587},
	journal = {Astrophys. J.},
	month = oct,
	pages = {141-160},
	title = {{Neutrino-driven winds from young, hot neutron stars}},
	volume = 309,
	year = 1986}

@article{Arcones:2006,
	archiveprefix = {arXiv},
	author = {Arcones, A. and Janka, Hans-Thomas and Scheck, L.},
	doi = {10.1051/0004-6361:20066983},
	eprint = {astro-ph/0612582},
	journal = {Astron. Astrophys.},
	pages = {1227},
	primaryclass = {astro-ph},
	slaccitation = {%%CITATION = ASTRO-PH/0612582;%%},
	title = {{Nucleosynthesis-relevant conditions in neutrino-driven supernova outflows. 1. Spherically symmetric hydrodynamic simulations}},
	volume = {467},
	year = {2007}}

@article{Huedepohl:2009wh,
	archiveprefix = {arXiv},
	author = {Hudepohl, L. and Muller, B. and Janka, H. -T. and Marek, A. and Raffelt, G. G.},
	doi = {10.1103/PhysRevLett.104.251101, 10.1103/PhysRevLett.105.249901},
	eprint = {0912.0260},
	journal = {Phys. Rev. Lett.},
	note = {[Erratum: Phys. Rev. Lett.105,249901(2010)]},
	pages = {251101},
	primaryclass = {astro-ph.SR},
	slaccitation = {%%CITATION = ARXIV:0912.0260;%%},
	title = {{Neutrino Signal of Electron-Capture Supernovae from Core Collapse to Cooling}},
	volume = {104},
	year = {2010}}

@article{Keil:2002in,
	archiveprefix = {arXiv},
	author = {Keil, Mathias Th. and Raffelt, Georg G. and Janka, Hans-Thomas},
	doi = {10.1086/375130},
	eprint = {astro-ph/0208035},
	journal = {Astrophys. J.},
	pages = {971-991},
	primaryclass = {astro-ph},
	slaccitation = {%%CITATION = ASTRO-PH/0208035;%%},
	title = {{Monte Carlo study of supernova neutrino spectra formation}},
	volume = {590},
	year = {2003}}

@article{Burbidge:1957vc,
	author = {Burbidge, M. E. and Burbidge, G. R. and Fowler, W. A. and Hoyle, F.},
	doi = {10.1103/RevModPhys.29.547},
	journal = {Rev. Mod. Phys.},
	pages = {547-650},
	slaccitation = {%%CITATION = RMPHA,29,547;%%},
	title = {{Synthesis of the elements in stars}},
	volume = {29},
	year = {1957}}

@article{Sukhbold:2015,
	archiveprefix = {arXiv},
	author = {Sukhbold, Tuguldur and Ertl, T. and Woosley, S. E. and Brown, Justin M. and Janka, H. -T.},
	doi = {10.3847/0004-637X/821/1/38},
	eprint = {1510.04643},
	journal = {Astrophys. J.},
	number = {1},
	pages = {38},
	primaryclass = {astro-ph.HE},
	slaccitation = {%%CITATION = ARXIV:1510.04643;%%},
	title = {{Core-Collapse Supernovae from 9 to 120 Solar Masses Based on Neutrino-powered Explosions}},
	volume = {821},
	year = {2016}}

@article{Vartanyan:2018xcd,
	archiveprefix = {arXiv},
	author = {Vartanyan, David and Burrows, Adam and Radice, David and Skinner, M. Aaron and Dolence, Joshua},
	doi = {10.1093/mnras/sty809},
	eprint = {1801.08148},
	journal = {Mon. Not. Roy. Astron. Soc.},
	number = {3},
	pages = {3091-3108},
	primaryclass = {astro-ph.HE},
	slaccitation = {%%CITATION = ARXIV:1801.08148;%%},
	title = {{Revival of the Fittest: Exploding Core-Collapse Supernovae from 12 to 25 M$_{\odot}$}},
	volume = {477},
	year = {2018}}

@article{Radice:2017ykv,
	archiveprefix = {arXiv},
	author = {Radice, David and Burrows, Adam and Vartanyan, David and Skinner, M. Aaron and Dolence, Joshua C.},
	doi = {10.3847/1538-4357/aa92c5},
	eprint = {1702.03927},
	journal = {Astrophys. J.},
	number = {1},
	pages = {43},
	primaryclass = {astro-ph.HE},
	reportnumber = {LA-UR-17-20973},
	slaccitation = {%%CITATION = ARXIV:1702.03927;%%},
	title = {{Electron-Capture and Low-Mass Iron-Core-Collapse Supernovae: New Neutrino-Radiation-Hydrodynamics Simulations}},
	volume = {850},
	year = {2017}}

@article{Burrows:2019rtd,
	archiveprefix = {arXiv},
	author = {Burrows, A. and Radice, D. and Vartanyan, D.},
	doi = {10.1093/mnras/stz543},
	eprint = {1902.00547},
	journal = {Mon. Not. Roy. Astron. Soc.},
	number = {3},
	pages = {3153-3168},
	primaryclass = {astro-ph.SR},
	slaccitation = {%%CITATION = ARXIV:1902.00547;%%},
	title = {{Three-dimensional supernova explosion simulations of 9-, 10-, 11-, 12-, and 13-$M_\odot$ stars}},
	volume = {485},
	year = {2019}}

@article{Burrows:2019zce,
    author = "Burrows, Adam and Radice, David and Vartanyan, David and Nagakura, Hiroki and Skinner, M. Aaron and Dolence, Joshua",
    title = "{The Overarching Framework of Core-Collapse Supernova Explosions as Revealed by 3D Fornax Simulations}",
    eprint = "1909.04152",
    archivePrefix = "arXiv",
    primaryClass = "astro-ph.HE",
    doi = "10.1093/mnras/stz3223",
    journal = "Mon. Not. Roy. Astron. Soc.",
    volume = "491",
    number = "2",
    pages = "2715--2735",
    year = "2020"
}

@article{Nagakura:2021lma,
	archiveprefix = {arXiv},
	author = {Nagakura, Hiroki and Burrows, Adam and Vartanyan, David},
	eprint = {2102.11283},
	month = {2},
	primaryclass = {astro-ph.HE},
	title = {{Supernova neutrino signals based on long-term axisymmetric simulations}},
	year = {2021}}

@article{Fischer:2018kdt,
	archiveprefix = {arXiv},
	author = {Fischer, Tobias and Guo, Gang and Dzhioev, Alan A. and Mart\'\i{}nez-Pinedo, Gabriel and Wu, Meng-Ru and Lohs, Andreas and Qian, Yong-Zhong},
	doi = {10.1103/PhysRevC.101.025804},
	eprint = {1804.10890},
	journal = {Phys. Rev. C},
	number = {2},
	pages = {025804},
	primaryclass = {astro-ph.HE},
	title = {{Neutrino signal from proto-neutron star evolution: Effects of opacities from charged-current\textendash{}neutrino interactions and inverse neutron decay}},
	volume = {101},
	year = {2020}}

@article{Cardall:1997bi,
	archiveprefix = {arXiv},
	author = {Cardall, Christian Y. and Fuller, George M.},
	doi = {10.1086/310838},
	eprint = {astro-ph/9701178},
	journal = {Astrophys. J. Lett.},
	pages = {L111},
	title = {{General relativistic effects in the neutrino driven wind and r process nucleosynthesis}},
	volume = {486},
	year = {1997}}

@article{Otsuki:1999kb,
    author = "Otsuki, Kaori and Tagoshi, Hideyuki and Kajino, Toshitaka and Wanajo, Shin-ya",
    title = "{General relativistic effects on neutrino driven wind from young, hot neutron star and the r process nucleosynthesis}",
    eprint = "astro-ph/9911164",
    archivePrefix = "arXiv",
    doi = "10.1086/308632",
    journal = "Astrophys. J.",
    volume = "533",
    pages = "424",
    year = "2000"
}

@article{Fisker:2009,
	adsurl = {https://ui.adsabs.harvard.edu/abs/2009ApJ...690L.135F},
	archiveprefix = {arXiv},
	author = {{Fisker}, J. L. and {Hoffman}, R. D. and {Pruet}, J.},
	doi = {10.1088/0004-637X/690/2/L135},
	eprint = {0711.1502},
	journal = {Astrophys. J. Lett.},
	month = jan,
	number = {2},
	pages = {L135-L139},
	primaryclass = {astro-ph},
	title = {{On the Origin of the Lightest Molybdenum Isotopes}},
	volume = {690},
	year = 2009}

@article{Rauscher:2013,
	adsurl = {https://ui.adsabs.harvard.edu/abs/2013RPPh...76f6201R},
	archiveprefix = {arXiv},
	author = {{Rauscher}, T. and {Dauphas}, N. and {Dillmann}, I. and {Fr{\"o}hlich}, C. and {F{\"u}l{\"o}p}, Zs and {Gy{\"u}rky}, Gy},
	doi = {10.1088/0034-4885/76/6/066201},
	eid = {066201},
	eprint = {1303.2666},
	journal = {Reports on Progress in Physics},
	month = jun,
	number = {6},
	pages = {066201},
	primaryclass = {astro-ph.SR},
	title = {{Constraining the astrophysical origin of the p-nuclei through nuclear physics and meteoritic data}},
	volume = {76},
	year = 2013}

@article{Bliss:2018djg,
	archiveprefix = {arXiv},
	author = {Bliss, J. and Arcones, A. and Qian, Y-Z.},
	doi = {10.3847/1538-4357/aade8d},
	eprint = {1804.03947},
	journal = {Astrophys. J.},
	number = {2},
	pages = {105},
	primaryclass = {astro-ph.HE},
	slaccitation = {%%CITATION = ARXIV:1804.03947;%%},
	title = {{Production of Mo and Ru isotopes in neutrino-driven winds: implications for solar abundances and presolar grains}},
	volume = {866},
	year = {2018}}

@article{Wanajo:2006rp,
	archiveprefix = {arXiv},
	author = {Wanajo, S.},
	doi = {10.1086/505483},
	eprint = {astro-ph/0602488},
	journal = {Astrophys. J.},
	pages = {1323-1340},
	slaccitation = {%%CITATION = ASTRO-PH/0602488;%%},
	title = {{The rp-process in neutrino-driven winds}},
	volume = {647},
	year = {2006}}

@ARTICLE{Woosley1992,
       author = {{Woosley}, S.~E. and {Baron}, E.},
        title = "{The Collapse of White Dwarfs to Neutron Stars}",
      journal = {\apj},
         year = 1992,
        month = may,
       volume = {391},
        pages = {228},
          doi = {10.1086/171338},
       adsurl = {https://ui.adsabs.harvard.edu/abs/1992ApJ...391..228W}
}

@article{Wanajo:2010mc,
	archiveprefix = {arXiv},
	author = {Wanajo, S. and Janka, H-T. and Kubono, S.},
	doi = {10.1088/0004-637X/729/1/46},
	eprint = {1004.4487},
	journal = {Astrophys. J.},
	pages = {46},
	primaryclass = {astro-ph.SR},
	slaccitation = {%%CITATION = ARXIV:1004.4487;%%},
	title = {{Uncertainties in the nu p-process: supernova dynamics versus nuclear physics}},
	volume = {729},
	year = {2011}}

@article{Janka:2025tvf,
    author = "Janka, H. -Thomas",
    title = "{Long-Term Multidimensional Models of Core-Collapse Supernovae: Progress and Challenges}",
    eprint = "2502.14836",
    archivePrefix = "arXiv",
    primaryClass = "astro-ph.HE",
    month = "2",
    year = "2025"
}

@article{Jin:2020,
	author = {Jin, S. and Roberts, L. F. and Austin, S. M. and Schatz, H.},
	doi = {10.1038/s41586-020-2948-7},
	journal = {Nature},
	number = {7836},
	pages = {57--60},
	title = {{Enhanced triple-\ensuremath{\alpha} reaction reduces proton-rich nucleosynthesis in supernovae}},
	volume = {588},
	year = {2020}}

@article{Lodders:2003,
	adsurl = {https://ui.adsabs.harvard.edu/abs/2003ApJ...591.1220L},
	author = {{Lodders}, K.},
	doi = {10.1086/375492},
	journal = {ApJ},
	month = jul,
	number = {2},
	pages = {1220-1247},
	title = {{Solar System Abundances and Condensation Temperatures of the Elements}},
	volume = {591},
	year = 2003}

@article{Travaglio:2011,
       author = {{Travaglio}, C. and {R{\"o}pke}, F.~K. and {Gallino}, R. and {Hillebrandt}, W.},
        title = "{Type Ia Supernovae as Sites of the p-process: Two-dimensional Models Coupled to Nucleosynthesis}",
      journal = {Astrophys. J},
         year = 2011,
        month = oct,
       volume = {739},
       number = {2},
          eid = {93},
        pages = {93},
          doi = {10.1088/0004-637X/739/2/93},
archivePrefix = {arXiv},
       eprint = {1106.0582},
 primaryClass = {astro-ph.SR},
       adsurl = {https://ui.adsabs.harvard.edu/abs/2011ApJ...739...93T}
}

@article{Travaglio:2014,
	author = {Travaglio, C. and Gallino, R. and Rauscher, T. and R{\"o}pke, F. K. and Hillebrandt, W.},
	doi = {10.1088/0004-637x/799/1/54},
	eprint = {astro-ph/1411.2399},
	issn = {1538-4357},
	journal = {The Astrophysical Journal},
	month = {Jan},
	number = {1},
	pages = {54},
	publisher = {American Astronomical Society},
	title = {TESTING THE ROLE OF SNe Ia FOR GALACTIC CHEMICAL EVOLUTION OFp-NUCLEI WITH TWO-DIMENSIONAL MODELS AND WITHs-PROCESS SEEDS AT DIFFERENT METALLICITIES},
	url = {http://dx.doi.org/10.1088/0004-637X/799/1/54},
	volume = {799},
	year = {2015}}

@article{Sasaki:2021ffa,
    author = "Sasaki, Hirokazu and Yamazaki, Yuta and Kajino, Toshitaka and Kusakabe, Motohiko and Hayakawa, Takehito and Cheoun, Myung-Ki and Ko, Heamin and Mathews, Grant J.",
    title = "{Impact of Hypernova \ensuremath{\nu}p-process Nucleosynthesis on the Galactic Chemical Evolution of Mo and Ru}",
    eprint = "2106.01679",
    archivePrefix = "arXiv",
    primaryClass = "astro-ph.GA",
    doi = "10.3847/1538-4357/ac34f8",
    journal = "Astrophys. J.",
    volume = "924",
    number = "1",
    pages = "29",
    year = "2022"
}

@article{Sasaki:2023ysp,
    author = "Sasaki, Hirokazu and Yamazaki, Yuta and Kajino, Toshitaka and Mathews, Grant J.",
    title = "{Effects of Hoyle state de-excitation on $\nu p$-process nucleosynthesis and Galactic chemical evolution}",
    eprint = "2307.02785",
    archivePrefix = "arXiv",
    primaryClass = "astro-ph.HE",
    reportNumber = "LA-UR-23-27158",
    journal = "",
    month = "7",
    year = "2023"
}

@article{Nevins:2024dkr,
    author = "Nevins, Brian and Roberts, Luke F.",
    title = "{Proto-neutron star convection and the neutrino-driven wind: implications for the {\ensuremath{\nu}}p-process}",
    eprint = "2404.07324",
    archivePrefix = "arXiv",
    primaryClass = "astro-ph.HE",
    doi = "10.1093/mnras/stae1005",
    journal = "Mon. Not. Roy. Astron. Soc.",
    volume = "530",
    number = "2",
    pages = "2001--2011",
    year = "2024"
}

@article{Frohlich:2005ys,
	archiveprefix = {arXiv},
	author = {Frohlich, C. and Martinez-Pinedo, Gabriel and Liebendorfer, M. and Thielemann, F. -K. and Bravo, E. and Hix, W. R. and Langanke, K. and Zinner, N. T.},
	doi = {10.1103/PhysRevLett.96.142502},
	eprint = {astro-ph/0511376},
	journal = {Phys. Rev. Lett.},
	pages = {142502},
	title = {{Neutrino-induced nucleosynthesis of a\ensuremath{>}64 nuclei: the nu p-process}},
	volume = {96},
	year = {2006}}

@article{Nishimura:2019jlh,
	archiveprefix = {arXiv},
	author = {Nishimura, N. and Rauscher, T. and Hirschi, R. and Cescutti, G. and Murphy, A. St. J. and Fr{\"o}hlich, C.},
	doi = {10.1093/mnras/stz2104},
	eprint = {1907.13129},
	journal = {Mon. Not. Roy. Astron. Soc.},
	pages = {1379},
	primaryclass = {astro-ph.SR},
	slaccitation = {%%CITATION = ARXIV:1907.13129;%%},
	title = {{Uncertainties in $\nu$p-process nucleosynthesis from Monte Carlo variation of reaction rates}},
	volume = {489},
	year = {2019}}

@article{Arcones:2011zj,
	archiveprefix = {arXiv},
	author = {Arcones, A. and Frohlich, C. and Martinez-Pinedo, G.},
	doi = {10.1088/0004-637X/750/1/18},
	eprint = {1112.4651},
	journal = {Astrophys. J.},
	pages = {18},
	primaryclass = {astro-ph.SR},
	title = {{Impact of supernova dynamics on the $\nu p$-process}},
	volume = {750},
	year = {2012}}

@article{Bliss:2014qiz,
	author = {Bliss, J. and Arcones, A.},
	doi = {10.22323/1.204.0073},
	editor = {Elekes, Zolt\'an and F\"ul\"op, Zsolt},
	journal = {PoS},
	pages = {073},
	title = {{Nucleosynthesis of Mo in neutrino-driven winds}},
	volume = {NICXIII},
	year = {2015}}

@article{Beard:2017jpg,
	archiveprefix = {arXiv},
	author = {Beard, M. and Austin, S. M. and Cyburt, R.},
	doi = {10.1103/PhysRevLett.119.112701},
	eprint = {1708.07204},
	journal = {Phys. Rev. Lett.},
	number = {11},
	pages = {112701},
	primaryclass = {nucl-th},
	title = {{Enhancement of the Triple Alpha Rate in a Hot Dense Medium}},
	volume = {119},
	year = {2017}}

@article{Pruet:2005qd,
	archiveprefix = {arXiv},
	author = {Pruet, J. and Hoffman, R. D. and Woosley, S. E. and Janka, H. -T. and Buras, R.},
	doi = {10.1086/503891},
	eprint = {astro-ph/0511194},
	journal = {Astrophys. J.},
	pages = {1028-1039},
	primaryclass = {astro-ph},
	slaccitation = {%%CITATION = ASTRO-PH/0511194;%%},
	title = {{Nucleosynthesis in early supernova winds. 2. the role of neutrinos}},
	volume = {644},
	year = {2006}}

@article{Arnould:2007gh,
	archiveprefix = {arXiv},
	author = {Arnould, M. and Goriely, S. and Takahashi, K.},
	doi = {10.1016/j.physrep.2007.06.002},
	eprint = {0705.4512},
	journal = {Phys. Rept.},
	pages = {97--213},
	primaryclass = {astro-ph},
	title = {{The r-process of stellar nucleosynthesis: Astrophysics and nuclear physics achievements and mysteries}},
	volume = {450},
	year = {2007}}

@article{Kaeppeler:2010kk,
	archiveprefix = {arXiv},
	author = {Kaeppeler, F. and Gallino, R. and Bisterzo, S. and Aoki, W.},
	doi = {10.1103/RevModPhys.83.157},
	eprint = {1012.5218},
	journal = {Rev. Mod. Phys.},
	pages = {157},
	primaryclass = {astro-ph.SR},
	title = {{The s Process: Nuclear Physics, Stellar Models, Observations}},
	volume = {83},
	year = {2011}}

@article{Woosley:1978,
	adsurl = {https://ui.adsabs.harvard.edu/abs/1978ApJS...36..285W},
	author = {{Woosley}, S.~E. and {Howard}, W.~M.},
	doi = {10.1086/190501},
	journal = {Astrophys. J. Suppl.},
	month = feb,
	pages = {285-304},
	title = {{The p-processes in supernovae.}},
	volume = {36},
	year = 1978}

@ARTICLE{Howard:1991,
       author = {{Howard}, W. Michael and {Meyer}, Bradley S. and {Woosley}, S.~E.},
        title = "{A New Site for the Astrophysical Gamma-Process}",
      journal = {Astrophys. J. Letters},
         year = 1991,
        month = may,
       volume = {373},
        pages = {L5},
          doi = {10.1086/186038},
       adsurl = {https://ui.adsabs.harvard.edu/abs/1991ApJ...373L...5H}
}

@article{Burrows:2020qrp,
	archiveprefix = {arXiv},
	author = {Burrows, Adam and Vartanyan, David},
	doi = {10.1038/s41586-020-03059-w},
	eprint = {2009.14157},
	journal = {Nature},
	number = {7840},
	pages = {29--39},
	primaryclass = {astro-ph.SR},
	title = {{Core-Collapse Supernova Explosion Theory}},
	volume = {589},
	year = {2021}}

@misc{triplealpha,
	howpublished = {{\tt TripleAlphaInMediumEnhancement} source code available at: \url{https://bitbucket.org/lroberts/triplealphainmediumenhancement/src/master/}}
}

@misc{nndc-walletcard,
  author       = {Tuli, Jagdish K.},
  title        = {Nuclear Wallet Cards},
  year         = {2023},
  howpublished = {\url{https://www.nndc.bnl.gov/walletcards/}},
  note         = {National Nuclear Data Center, Brookhaven National Laboratory},
  institution  = {Brookhaven National Laboratory},
  url          = {https://www.nndc.bnl.gov/walletcards/}
}

@article{Burrows:2002jv,
    author = "Burrows, Adam and Thompson, Todd A.",
    title = "{Neutrino - matter interaction rates in supernovae: The Essential microphysics of core collapse}",
    eprint = "astro-ph/0211404",
    archivePrefix = "arXiv",
    pages = "133--174",
    month = "11",
    year = "2002"
}

@article{Lippuner:2017tyn,
	archiveprefix = {arXiv},
	author = {Lippuner, Jonas and Roberts, Luke F.},
	doi = {10.3847/1538-4365/aa94cb},
	eprint = {1706.06198},
	journal = {Astrophys. J. Suppl.},
	number = {2},
	pages = {18},
	primaryclass = {astro-ph.HE},
	title = {{SkyNet: A modular nuclear reaction network library}},
	volume = {233},
	year = {2017}}

@misc{skynet,
	howpublished = {{\tt SkyNet} source code available at: \url{https://bitbucket.org/jlippuner/skynet/src/master/}}}

@article{Sieverding:2020wxw,
	archiveprefix = {arXiv},
	author = {Sieverding, Andre and M\"uller, Bernhard and Qian, Yong-Zhong},
	doi = {10.3847/1538-4357/abc61b},
	eprint = {2008.12831},
	journal = {Astrophys. J.},
	number = {2},
	pages = {163},
	primaryclass = {astro-ph.HE},
	title = {{Nucleosynthesis of an $11.8\,M_\odot$ Supernova with 3D Simulation of the Inner Ejecta: Overall Yields and Implications for Short-lived Radionuclides in the Early Solar System}},
	volume = {904},
	year = {2020}}

@ARTICLE{reaclib2010,
       author = {{Cyburt}, Richard H. and {Amthor}, A. Matthew and {Ferguson}, Ryan and {Meisel}, Zach and {Smith}, Karl and {Warren}, Scott and {Heger}, Alexander and {Hoffman}, R.~D. and {Rauscher}, Thomas and {Sakharuk}, Alexander and {Schatz}, Hendrik and {Thielemann}, F.~K. and {Wiescher}, Michael},
        title = "{The JINA REACLIB Database: Its Recent Updates and Impact on Type-I X-ray Bursts}",
      journal = {Astrophys. J. Supplement Series},
         year = 2010,
        month = jul,
       volume = {189},
       number = {1},
        pages = {240-252},
          doi = {10.1088/0067-0049/189/1/240},
       adsurl = {https://ui.adsabs.harvard.edu/abs/2010ApJS..189..240C}
}

@misc{reaclibweb,
	howpublished = {JINA Reaclib Database, available at: \url{https://reaclib.jinaweb.org/}}}

@article{Lugaro:2016zuf,
    author = "Lugaro, Maria and Pignatari, Marco and Ott, Ulrich and Zuber, Kai and Travaglio, Claudia and Gyurky, Gyorgy and Fulop, Zsolt",
    title = "{Origin of the p-process radionuclides 92Nb and 146Sm in the early Solar System and inferences on the birth of the Sun}",
    eprint = "1601.05986",
    archivePrefix = "arXiv",
    primaryClass = "astro-ph.SR",
    doi = "10.1073/pnas.1519344113",
    journal = "Proc. Nat. Acad. Sci.",
    volume = "113",
    pages = "907",
    year = "2016"
}

@ARTICLE{Haba:2021PNAS,
       author = {{Haba}, Makiko K. and {Lai}, Yi-Jen and {Wotzlaw}, J{\"o}rn-Frederik and {Yamaguchi}, Akira and {Lugaro}, Maria and {Sch{\"o}nb{\"a}chler}, Maria},
        title = "{Precise initial abundance of Niobium-92 in the Solar System and implications for p-process nucleosynthesis}",
      journal = "Proc. Nat. Acad. Sci.",
         year = 2021,
        month = feb,
       volume = {118},
       number = {8},
          eid = {2017750118},
        pages = {2017750118},
          doi = {10.1073/pnas.2017750118},
       adsurl = {https://ui.adsabs.harvard.edu/abs/2021PNAS..11820177H}
}

@article{Sasaki:2017jry,
    author = "Sasaki, H. and Kajino, T. and Takiwaki, T. and Hayakawa, T. and Balantekin, A. B. and Pehlivan, Y.",
    title = "{Possible effects of collective neutrino oscillations in three-flavor multiangle simulations of supernova $\nu p$ processes}",
    eprint = "1707.09111",
    archivePrefix = "arXiv",
    primaryClass = "astro-ph.HE",
    doi = "10.1103/PhysRevD.96.043013",
    journal = "Phys. Rev. D",
    volume = "96",
    number = "4",
    pages = "043013",
    year = "2017"
}

@article{Rauscher:2019mcn,
    author = {Rauscher, T. and Nishimura, N. and Cescutti, G. and Hirschi, R. and Murphy, A. St. J. and Fr\"ohlich, C.},
    editor = "Kawabata, T. and others",
    title = "{Impact of Uncertainties in Astrophysical Reaction Rates on Nucleosynthesis in the $\nu p$ Process}",
    eprint = "1909.03235",
    archivePrefix = "arXiv",
    primaryClass = "astro-ph.HE",
    doi = "10.7566/JPSCP.31.011026",
    journal = "JPS Conf. Proc.",
    volume = "31",
    pages = "011026",
    year = "2020"
}

@ARTICLE{Iizuka:2016,
       author = {{Iizuka}, Tsuyoshi and {Lai}, Yi-Jen and {Akram}, Waheed and {Amelin}, Yuri and {Sch{\"o}nb{\"a}chler}, Maria},
        title = "{The initial abundance and distribution of $^{92}$Nb in the Solar System}",
      journal = {Earth and Planetary Science Letters},
         year = 2016,
        month = apr,
       volume = {439},
        pages = {172-181},
          doi = {10.1016/j.epsl.2016.02.005},
archivePrefix = {arXiv},
       eprint = {1602.00966},
 primaryClass = {astro-ph.SR},
       adsurl = {https://ui.adsabs.harvard.edu/abs/2016E&PSL.439..172I}
}

@ARTICLE{Hibiya:2023,
       author = {{Hibiya}, Yuki and {Iizuka}, Tsuyoshi and {Enomoto}, Hatsuki and {Hayakawa}, Takehito},
        title = "{Evidence for Enrichment of Niobium-92 in the Outer Protosolar Disk}",
      journal = {Astrophys. J. Lett.},
         year = 2023,
        month = jan,
       volume = {942},
       number = {1},
          eid = {L15},
        pages = {L15},
          doi = {10.3847/2041-8213/acab5d},
       adsurl = {https://ui.adsabs.harvard.edu/abs/2023ApJ...942L..15H}
}

@ARTICLE{Hoffman:1996,
       author = {{Hoffman}, R.~D. and {Woosley}, S.~E. and {Fuller}, G.~M. and {Meyer}, B.~S.},
        title = "{Production of the Light p-Process Nuclei in Neutrino-driven Winds}",
      journal = {Astrophys. J},
         year = 1996,
        month = mar,
       volume = {460},
        pages = {478},
          doi = {10.1086/176986},
       adsurl = {https://ui.adsabs.harvard.edu/abs/1996ApJ...460..478H}
}

@article{OConnor:2014sgn,
    author = "O'Connor, Evan",
    title = "{An Open-Source Neutrino Radiation Hydrodynamics Code for Core-Collapse Supernovae}",
    eprint = "1411.7058",
    archivePrefix = "arXiv",
    primaryClass = "astro-ph.HE",
    doi = "10.1088/0067-0049/219/2/24",
    journal = "Astrophys. J. Suppl.",
    volume = "219",
    number = "2",
    pages = "24",
    year = "2015"
}

@article{Horowitz:1999fe,
    author = "Horowitz, C. J. and Li, Gang",
    title = "{Charge conjugation violating interactions in supernovae and nucleosynthesis}",
    eprint = "astro-ph/9904171",
    archivePrefix = "arXiv",
    doi = "10.1103/PhysRevLett.82.5198",
    journal = "Phys. Rev. Lett.",
    volume = "82",
    pages = "5198",
    year = "1999"
}

@article{Steiner:2012rk,
    author = "Steiner, Andrew W. and Hempel, Matthias and Fischer, Tobias",
    title = "{Core-collapse supernova equations of state based on neutron star observations}",
    eprint = "1207.2184",
    archivePrefix = "arXiv",
    primaryClass = "astro-ph.SR",
    reportNumber = "INT-PUB-12-033",
    doi = "10.1088/0004-637X/774/1/17",
    journal = "Astrophys. J.",
    volume = "774",
    pages = "17",
    year = "2013"
}

@article{Fiorillo:2023frv,
    author = "Fiorillo, Damiano F. G. and Heinlein, Malte and Janka, Hans-Thomas and Raffelt, Georg and Vitagliano, Edoardo and Bollig, Robert",
    title = "{Supernova simulations confront SN 1987A neutrinos}",
    eprint = "2308.01403",
    archivePrefix = "arXiv",
    primaryClass = "astro-ph.HE",
    doi = "10.1103/PhysRevD.108.083040",
    journal = "Phys. Rev. D",
    volume = "108",
    number = "8",
    pages = "083040",
    year = "2023"
}

@article{Hempel:2009mc,
    author = "Hempel, Matthias and Schaffner-Bielich, Jurgen",
    title = "{Statistical Model for a Complete Supernova Equation of State}",
    eprint = "0911.4073",
    archivePrefix = "arXiv",
    primaryClass = "nucl-th",
    doi = "10.1016/j.nuclphysa.2010.02.010",
    journal = "Nucl. Phys. A",
    volume = "837",
    pages = "210--254",
    year = "2010"
}

@article{Oertel:2020pcg,
    author = "Oertel, Micaela and Pascal, Aur{\'e}lien and Mancini, Marco and Novak, Jerome",
    title = "{Improved neutrino-nucleon interactions in dense and hot matter for numerical simulations}",
    eprint = "2003.02152",
    archivePrefix = "arXiv",
    primaryClass = "astro-ph.HE",
    doi = "10.1103/PhysRevC.102.035802",
    journal = "Phys. Rev. C",
    volume = "102",
    number = "3",
    pages = "035802",
    year = "2020"
}

@article{Pascal:2022qeg,
    author = "Pascal, A. and Novak, J. and Oertel, M.",
    title = "{Proto-neutron star evolution with improved charged-current neutrino{\textendash}nucleon interactions}",
    eprint = "2201.01955",
    archivePrefix = "arXiv",
    primaryClass = "nucl-th",
    doi = "10.1093/mnras/stac016",
    journal = "Mon. Not. Roy. Astron. Soc.",
    volume = "511",
    number = "1",
    pages = "356--370",
    year = "2022"
}

@article{Suleiman:2023bdf,
    author = "Suleiman, Lami and Oertel, Micaela and Mancini, Marco",
    title = "{Modified Urca neutrino emissivity at finite temperature}",
    eprint = "2308.09819",
    archivePrefix = "arXiv",
    primaryClass = "nucl-th",
    doi = "10.1103/PhysRevC.108.035803",
    journal = "Phys. Rev. C",
    volume = "108",
    number = "3",
    pages = "035803",
    year = "2023"
}

@article{Lucente:2024ngp,
    author = "Lucente, Giuseppe and Heinlein, Malte and Janka, Hans-Thomas and Mirizzi, Alessandro",
    title = "{Simple fits for the neutrino luminosities from protoneutron star cooling}",
    eprint = "2405.00769",
    archivePrefix = "arXiv",
    primaryClass = "astro-ph.HE",
    doi = "10.1103/PhysRevD.110.063023",
    journal = "Phys. Rev. D",
    volume = "110",
    number = "6",
    pages = "063023",
    year = "2024"
}

@article{Roberts:2011yw,
    author = "Roberts, L. F. and Shen, G. and Cirigliano, V. and Pons, J. A. and Reddy, S. and Woosley, S. E.",
    title = "{Proto-Neutron Star Cooling with Convection: The Effect of the Symmetry Energy}",
    eprint = "1112.0335",
    archivePrefix = "arXiv",
    primaryClass = "astro-ph.HE",
    reportNumber = "INT-PUB-11-053",
    doi = "10.1103/PhysRevLett.108.061103",
    journal = "Phys. Rev. Lett.",
    volume = "108",
    pages = "061103",
    year = "2012"
}

@ARTICLE{1988PhR...163...51B,
       author = {{Burrows}, A. and {Lattimer}, J.~M.},
        title = "{Convection, Type II supernovae, and the early evolution of neutron stars.}",
      journal = {\physrep},
         year = 1988,
        month = jan,
       volume = {163},
       number = {1},
        pages = {51-62},
          doi = {10.1016/0370-1573(88)90035-X},
       adsurl = {https://ui.adsabs.harvard.edu/abs/1988PhR...163...51B}
}

@article{Dessart:2005ck,
    author = "Dessart, Luc and Burrows, A. and Livne, E. and Ott, C. D.",
    title = "{Multi-dimensional radiation/hydrodynamic simulations of protoneutron star convection}",
    eprint = "astro-ph/0510229",
    archivePrefix = "arXiv",
    doi = "10.1086/504068",
    journal = "Astrophys. J.",
    volume = "645",
    pages = "534--550",
    year = "2006"
}

@article{Nagakura:2019tmy,
    author = "Nagakura, Hiroki and Burrows, Adam and Radice, David and Vartanyan, David",
    title = "{A systematic study of proto-neutron star convection in three-dimensional core-collapse supernova simulations}",
    eprint = "1912.07615",
    archivePrefix = "arXiv",
    primaryClass = "astro-ph.HE",
    doi = "10.1093/mnras/staa261",
    journal = "Mon. Not. Roy. Astron. Soc.",
    volume = "492",
    number = "4",
    pages = "5764--5779",
    year = "2020"
}

@article{Kajino:2019abv,
    author = "Kajino, T. and Aoki, W. and Balantekin, A. B. and Diehl, R. and Famiano, M. A. and Mathews, G. J.",
    title = "{Current status of r -process nucleosynthesis}",
    eprint = "1906.05002",
    archivePrefix = "arXiv",
    primaryClass = "astro-ph.HE",
    doi = "10.1016/j.ppnp.2019.02.008",
    journal = "Prog. Part. Nucl. Phys.",
    volume = "107",
    pages = "109--166",
    year = "2019"
}

@article{Lugaro:2023qli,
    author = "Lugaro, Maria and Pignatari, Marco and Reifarth, Ren{\'e} and Wiescher, Michael",
    title = "{The s Process and Beyond}",
    doi = "10.1146/annurev-nucl-102422-080857",
    journal = "Ann. Rev. Nucl. Part. Sci.",
    volume = "73",
    pages = "315--340",
    year = "2023"
}

@ARTICLE{Nomoto:2013,
       author = {{Nomoto}, Ken'ichi and {Kobayashi}, Chiaki and {Tominaga}, Nozomu},
        title = "{Nucleosynthesis in Stars and the Chemical Enrichment of Galaxies}",
      journal = {\araa},
         year = 2013,
        month = aug,
       volume = {51},
       number = {1},
        pages = {457-509},
          doi = {10.1146/annurev-astro-082812-140956},
       adsurl = {https://ui.adsabs.harvard.edu/abs/2013ARA&A..51..457N}
}

@ARTICLE{Thielemann:2007,
       author = {{Thielemann}, F. -K. and {Fr{\"o}hlich}, C. and {Hirschi}, R. and {Liebend{\"o}rfer}, M. and {Dillmann}, I. and {Mocelj}, D. and {Rauscher}, T. and {Martinez-Pinedo}, G. and {Langanke}, K. and {Farouqi}, K. and {Kratz}, K. -L. and {Pfeiffer}, B. and {Panov}, I. and {Nadyozhin}, D.~K. and {Blinnikov}, S. and {Bravo}, E. and {Hix}, W.~R. and {H{\"o}flich}, P. and {Zinner}, N.~T.},
        title = "{Production of intermediate-mass and heavy nuclei}",
      journal = {Progress in Particle and Nuclear Physics},
         year = 2007,
        month = jul,
       volume = {59},
       number = {1},
        pages = {74-93},
          doi = {10.1016/j.ppnp.2006.12.019},
       adsurl = {https://ui.adsabs.harvard.edu/abs/2007PrPNP..59...74T}
}

@ARTICLE{Sneden:2008,
       author = {{Sneden}, C. and {Cowan}, J.~J. and {Gallino}, R.},
        title = "{Neutron-capture elements in the early galaxy.}",
      journal = {\araa},
         year = 2008,
        month = sep,
       volume = {46},
        pages = {241-288},
          doi = {10.1146/annurev.astro.46.060407.145207},
       adsurl = {https://ui.adsabs.harvard.edu/abs/2008ARA&A..46..241S}
}

@ARTICLE{Fischer:2024,
       author = {{Fischer}, Tobias and {Guo}, Gang and {Langanke}, Karlheinz and {Mart{\'\i}nez-Pinedo}, Gabriel and {Qian}, Yong-Zhong and {Wu}, Meng-Ru},
        title = "{Neutrinos and nucleosynthesis of elements}",
      journal = {Progress in Particle and Nuclear Physics},
         year = 2024,
        month = may,
       volume = {137},
          eid = {104107},
        pages = {104107},
          doi = {10.1016/j.ppnp.2024.104107},
archivePrefix = {arXiv},
       eprint = {2308.03962},
 primaryClass = {astro-ph.HE},
       adsurl = {https://ui.adsabs.harvard.edu/abs/2024PrPNP.13704107F}
}

@ARTICLE{Arcones:2023,
       author = {{Arcones}, Almudena and {Thielemann}, Friedrich-Karl},
        title = "{Origin of the elements}",
      journal = {\aapr},
         year = 2023,
        month = dec,
       volume = {31},
       number = {1},
          eid = {1},
        pages = {1},
          doi = {10.1007/s00159-022-00146-x},
       adsurl = {https://ui.adsabs.harvard.edu/abs/2023A&ARv..31....1A}
}

@article{Wang:2024fnv,
    author = "Wang, Tianshu and Burrows, Adam",
    title = "{Insights into the Production of $^{44}$Ti and Nickel Isotopes in Core-collapse Supernovae}",
    eprint = "2406.13746",
    archivePrefix = "arXiv",
    primaryClass = "astro-ph.HE",
    doi = "10.3847/1538-4357/ad6983",
    journal = "Astrophys. J.",
    volume = "974",
    number = "1",
    pages = "39",
    year = "2024"
}

@article{Wang:2023vml,
    author = "Wang, Tianshu and Burrows, Adam",
    title = "{Nucleosynthetic Analysis of Three-dimensional Core-collapse Supernova Simulations}",
    eprint = "2311.03446",
    archivePrefix = "arXiv",
    primaryClass = "astro-ph.HE",
    doi = "10.3847/1538-4357/ad12b8",
    journal = "Astrophys. J.",
    volume = "962",
    number = "1",
    pages = "71",
    year = "2024"
}

@article{Wang:2023vkk,
    author = "Wang, Tianshu and Burrows, Adam",
    title = "{Neutrino-driven Winds in Three-dimensional Core-collapse Supernova Simulations}",
    eprint = "2306.13712",
    archivePrefix = "arXiv",
    primaryClass = "astro-ph.SR",
    doi = "10.3847/1538-4357/ace7b2",
    journal = "Astrophys. J.",
    volume = "954",
    number = "2",
    pages = "114",
    year = "2023"
}

@article{Zha:2024fyo,
    author = {Zha, Shuai and M{\"u}ller, Bernhard and Powell, Jade},
    title = "{Nucleosynthesis in the Innermost Ejecta of Magnetorotational Supernova Explosions in Three Dimensions}",
    eprint = "2403.02072",
    archivePrefix = "arXiv",
    primaryClass = "astro-ph.HE",
    doi = "10.3847/1538-4357/ad4ae7",
    journal = "Astrophys. J.",
    volume = "969",
    number = "2",
    pages = "141",
    year = "2024"
}

\clearpage

\appendix

\section{\textcolor{black}{Hydrodynamic Equations in Spherical Symmetry}}
\label{app:outflow:equations}

Here we derive, in some detail, the evolution of the outflow exterior to the PNS, adopting as starting point the general approach of Ref.~\cite{Shapiro:1983du}. Given the formal differences in the relativistic fluid equations of Refs.~\cite{Cardall:1997bi, Otsuki:1999kb, Thompson:2001}, we hope to hereby clarify the origin of the equations we apply in our study. 

As discussed in Sec.~\ref{sec:outflow:equations}, the evolution of the stellar outflow derives from three physical principles, namely baryon-number conservation expressed by Eq.~\eqref{eq:fluid:continuity} (continuity equation), momentum conservation of the plasma given by Eq.~\eqref{eq:fluid:energy-momentum-conservation} (Euler equation), and the first law of thermodynamics in Eq.~\eqref{eq:thermo:id}.

As the dominant source of gravitation is the PNS, we assume the vacuum Schwarzschild spacetime, which determines Eq.~\eqref{eq:fluid:continuity} and Eq.~\eqref{eq:fluid:energy-momentum-conservation} for $n$ and $u^\mu$. In Eq.~(\ref{eq:fluid:energy-momentum-conservation}) we manifestly neglect direct energy deposition and momentum transfer from neutrino heating, a standard assumption across Newtonian and GR studies (e.g., Refs.~\cite{Duncan1986, Qian:1996xt, Otsuki:1999kb, Thompson:2001}). This is justified for the body of the neutrino-driven outflow, where the baryon mass and pressure gradients dominate over the external heating rate~\cite{Thompson2004}, except perhaps immediate to the PNS surface. In Eqs.~\eqref{eq:fluid:energy-momentum-conservation} and \eqref{eq:thermo:id}, we model the baryon-radiation plasma as an ensemble of non-relativistic baryon gas and radiation consisting of photons, electrons, and positrons in thermal equilibrium, described with a variable number of effective RDF. Heating of this plasma contributes to the temperature and density gradients which in turn propel the outflow outward, as described by Eq.~\eqref{eq:fluid:energy-momentum-conservation}. We adopt the four-velocity $u^\mu$, the baryon number density $n$, and the temperature $T$ as fundamental state variables, in which Eqs.~\eqref{eq:fluid:continuity}--\eqref{eq:thermo:id} are easily expressed.

We model SN outflows by applying the simplifying assumptions of \textit{spherical symmetry} and \textit{steady state}.  Modern simulations have certainly progressed beyond these simplifications to understand the origin, mechanism, and rich physics of CCSNe (Ref.~\cite{Janka:2025tvf} provides a recent review). For our nucleosynthesis study, nevertheless, these assumptions still capture essential physics and afford us a physical system in which GR corrections can be readily isolated and studied. Such an analysis of various corrections can be difficult to perform in actual simulations. Thus, the results we have obtained here can inform and motivate more rigorous studies based on state-of-the-art simulations. 

The conditions of spherical symmetry and steady-state outflow are imposed by setting the Lie derivatives of the fluid fields in the angular and temporal directions to zero. In the coordinate basis, these Lie derivatives coincide with the partial derivatives $\partial_\theta$, $\partial_\phi$, $\partial_t$, and the conditions are $\partial_\theta=\partial_\phi=\partial_t=0$ for all fields, which we impose throughout below. This reduces $u^\mu$ to only a radial component, denoted as $u\equiv u^r$.

The three principles can be reformulated to make explicit the evolution of $u$, $T$, and $n$ with radius. For the baryon-radiation ensemble, the thermodynamic identity \eqref{eq:thermo:id} has an equivalent form (in steady-state) for the entropy \textit{per baryon} $S$, given by Eq.~\eqref{eq:thermo:dS}. Here, $S$ is the total entropy with the components of radiation and monatomic baryon gas $S=S_{\rm r}+S_b$:
\begin{equation}
    S_{\rm r}=\frac{2\pi^2}{45} g_*^S(T) \frac{T^3}{n}, \quad S_b = \ln\frac{\left(m_N T\right)^{3/2}}{n}+\frac{5}{2}-\frac{3}{2}\ln 2\pi\,,
\end{equation}
where $g_*^S$ is the number of effective RDF for entropy (see App.~\ref{app:gstar} for a dedicated discussion about RDF). In terms of temperature and density, Eq.~\eqref{eq:thermo:dS} can be expressed as 
\begin{equation} \label{eq:app:std:dT}
    \left(4\beta_*P_{\rm r}+\frac{3}{2}P_b\right)\frac{dT}{dr} = \left(\frac{4g_*^S}{g_*^P}P_{\rm r}+P_b\right)\frac{T}{n}\frac{dn}{dr} + nT\frac{\dot{q}}{u}\,,
\end{equation}
which corresponds to Eq.~\eqref{eq:std:dT} and serves as the evolution equation  for temperature, once the density derivative is expressed as a function of $(u,T,n)$. In the equation above, the radiation and baryonic components of pressure are
\begin{equation}
P_{\rm r} = \frac{\pi^2g_*^P(T)}{90} T^4\,, \quad P_b = nT.
\end{equation}
We model the non-relativistic baryon gas as a \textit{monatomic} Maxwell-Boltzmann gas.  This is not appropriate after $\alpha$-particle formation, though in regions where this has significantly altered the composition, radiation dominates and the baryon gas becomes a small correction (typically $P_b < P_{\rm r}/10$).
Also, the entropy transfer from baryons to radiation due to $\alpha$-particle formation is  typically lower than 5 and occurs when $S_{\rm r}\gtrsim 60\mbox{--}70$ already for the cases of interest. Thus, this approximation, which exaggerates the contribution of the baryon gas component away from the PNS, is not expected to be impactful. Practically, this allows us to separate the computation of the outflow from the change in the composition of the nuclides.

The continuity equation \eqref{eq:fluid:continuity} under the steady-state condition reduces to Eq.~\eqref{eq:std:dn}, i.e.
\begin{equation} \label{eq:app:std:dn}
    \frac{1}{n}\frac{dn}{dr} = -\frac{1}{u}\frac{du}{dr} - \frac{2}{r}\,,
\end{equation}
which is a differential equation to evolve the density, once $du/dr$ is specified. 

The steady-state Euler equation \eqref{eq:fluid:energy-momentum-conservation} in Schwarzschild spacetime is
\begin{equation} \label{eq:app:fluid:Euler}
    u\frac{du}{dr} = -\frac{GM}{r^2} - \frac{\mu}{\rho+P} \frac{dP}{dr}\,,
\end{equation}
where the components of the total energy density $\rho\equiv \rho_b + \rho_{\rm r} + \bar{\rho}_b$ are
\begin{equation}
    \rho_b\equiv m_N n, \quad\rho_{\rm r} = \frac{\pi^2 g_*^\rho(T)}{30} T^4, \quad \bar{\rho}_b = \frac{3}{2} n T\,,
\end{equation}
the energy densities of baryon mass, thermal radiation, and thermal non-relativistic baryon gas, respectively. We note that the identification of relativistic corrections of $\mu$ and $\rho+P$, as discussed in Sec.~\ref{sec:outflow:Newtonian}, is transparent in this form of the Euler equation. We recover the Newtonian equation with $\mu\to 1$ and $\rho+P\to\rho_b$. Now, it is standard to formulate the Euler equation to reveal the singularity identified with the adiabatic sound speed $v_s$ \cite{Duncan1986,Qian:1996xt}. We introduce the sound speed via the pressure gradient
\begin{equation}
    \frac{dP}{dr} = \left(\frac{dP_{\rm r}}{dn}+\frac{dP_b}{dn}\right)\frac{dn}{dr} = \left[\frac{dP_{\rm r}}{dT}\frac{dT}{dn} + \left(\frac{\partial P_b}{\partial T}\frac{dT}{dn}+\frac{\partial P_b}{\partial n}\right)\right]\frac{dn}{dr} \,,\label{eq:app:fluid:pressure-gradient}
\end{equation}
where factors common to the sound speed appear:
\begin{equation} \label{eq:app:state:vs2}
    v_s^2 \equiv \left.\frac{dP}{d\rho}\right|_{q=0}= \left.\left(\frac{dP}{dT} \frac{dT}{dn}\right)\right|_{q=0} \frac{n}{\rho+P} \, .
\end{equation}
Here $\left.{dT}/{dn}\right|_{q=0}$ is obtained from Eq.~(\ref{eq:app:std:dT}) by annulling the heating term, while the remainder of ${dT}/{dn}$, corresponding to heating, contains only the state variables and the derivative ${dn}/{dr}$. The two components of pressure (baryon gas and radiation) contribute two components to the square of sound speed, appearing in Eq.~(\ref{eq:app:fluid:pressure-gradient}). Next, substituting the continuity equation \eqref{eq:app:std:dn} to eliminate the density derivative, we obtain the evolution equation for the coordinate velocity in the standard form
\begin{equation} \label{eq:app:std:du}
    \left(\frac{\mu v_s^2}{u}-u\right)\frac{du}{dr} = \frac{GM}{r^2} - \frac{2\mu v_s^2}{r} + \frac{\mu n}{\rho+P}\Pi_1^{{\rm r} b}\frac{\dot{q}}{u}\,,
\end{equation}
which is identical to Eq.~\eqref{eq:std:du}.

\section{Outflow Equations in Terms of Velocity, Temperature, and Entropy} \label{app:outflowDES}

Here, we seek to ascertain whether our outflow equations in the fundamental form---equivalent to Eqs.~(1)--(3) from \Otsuki~\cite{Otsuki:1999kb} or \Thompson~\cite{Thompson:2001}---transform to the corresponding equations in \CF~\cite{Cardall:1997bi} when expressed in terms of the entropy, physical velocity, and temperature as the dynamical quantities\footnote{Note that Ref.~\cite{Qian:1996xt}, while working in the Newtonian limit, presents the outflow equations in both these forms [see their equations (1)--(3), and (24)--(26), respectively]. We have explicitly checked that the latter set of equations follows from the former. For the relativistic case, it's a different story, as we show here.}. The motivation behind this exercise stemmed from our discovery that a numerical implementation of Eqs.~(1)--(3) from \CF~was found not to conserve the mass outflow rate $\dot{M} = 4\pi r^2 \rho_b u$. 
We start with the outflow Eqs.~(1)--(3) from \Otsuki:
\begin{align}
    \dot{M} &= 4 \pi r^2 \rho_b u \label{eq:app:MdotGR} = \text{constant} \implies \frac{2}{r} + \frac{1}{\rho_b} \frac{d\rho_b }{dr} + \frac{1}{u} \frac{du}{dr} = 0, \\
    u \frac{du}{dr} &= -\frac{1}{\rho + P} \frac{dP}{dr} \left(1 + u^2 - \frac{2GM}{r}\right) - \frac{GM}{r^2} \label{eq:app:dudrGR}, \\
    \dot Q &= u \left(\frac{d\epsilon}{dr} - \frac{P}{\rho_b^2} \frac{d\rho_b}{dr} \right) \label{eq:app:qdotGR}.
\end{align}
where $u = dr/d\tau = vy$ is the radial component of the four-velocity, with $y = [(1 - 2GM/r)/(1- v^2)]^{1/2}$. \textcolor{black}{In Sec.~\ref{sec:outflow:equations}, we defined the specific heating rate per baryon, $\dot{q} \equiv dq/d\tau$. It is more standard to use the heating rate per baryon mass, denoted above as $\dot{Q}$, defined as:
\begin{equation} \label{eq:Qdot}
    \dot{Q} \equiv \frac{\dot{q}}{m_N}.
\end{equation}}

In the relativistic case, an aspect that requires careful consideration is the distinction between the baryon rest-mass density $\rho_b = m_N n$ and the total density $\rho = \rho_b + \rho_b \epsilon$, where $\epsilon$ is the specific internal energy. As a result,
the sound speed $v_s$ acquires a relativistic correction when compared to the Newtonian expression. In what follows in this section, we neglect baryonic gas contributions to the pressure, internal energy, and entropy.

Here, for simplicity, we adopt the same approximations of \CF: a constant number of RDF ($g_\star = 11/2$) and omission of the baryonic gas in the thermal ensemble. The pressure, specific internal energy, and entropy per baryon then assume the forms:
\begin{align}
    P = P_r &= \frac{11\pi^2}{180} T^4, \label{eq:app:P} \\
    \epsilon = \frac{\rho_r}{\rho_b} &= \frac{11 \pi^2}{60} \frac{T^4}{\rho_b}  = \frac{3P}{\rho_b}, \label{eq:app:eps} \\
    S = S_r &= \frac{11\pi^2}{45} \frac{T^3}{\rho_b/m_N}, \label{eq:app:S}
\end{align}
and the sound speed can be calculated as:
\begin{equation} \label{eq:app:vsGR}
    v_s^2 = \frac{\partial P}{\partial \rho} \, \bigg\vert_S = \frac{1}{\partial (\rho_b + \rho_b \epsilon)/\partial P \, \big\vert_S} = \frac{1}{3 + \partial \rho_b/\partial P \, \big\vert_S} = \frac{TS}{3m_N} \left(1 + \frac{TS}{m_N}\right)^{-1},
\end{equation}
where the last equality above follows from 
\begin{equation}
\frac{\partial P}{\partial \rho_b} \bigg\vert_S = \frac{11\pi^2}{180} \frac{\partial}{\partial \rho_b} \left( \frac{45}{11\pi^2} \frac{\rho_b S}{m_N} \right)^{4/3} \bigg\vert_S = \frac{4P}{3\rho_b} = \frac{TS}{3m_N}, \label{eq:app:dPdrhob}    
\end{equation}
which incidentally is the expression for $v_s^2$ in the Newtonian limit.
 Next, from Eqs.~\eqref{eq:app:P} and \eqref{eq:app:S}, we have 
 \begin{equation} \label{eq:app:rhoplusP}
    \rho + P =  \rho_b + 4P = \rho_b \left(1 + \frac{TS}{m_N}\right).
\end{equation} 
Moreover, from the definition $u = v\,y$,
\begin{equation} \label{eq:app:ysq}
    1 + u^2 - \frac{2GM}{r} =  1 - \frac{2GM}{r} + \frac{v^2}{1-v^2} \left( 1 - \frac{2GM}{r} \right) = \frac{1 - 2GM/r}{1-v^2} = y^2,
\end{equation}
which shows that $y^2$ corresponds to the GR factor $\mu$ in Eq.~\eqref{eq:mu}, appearing in our outflow equations. For later use, we also calculate the derivative of $y$:
\begin{equation} \label{eq:app:dydr}
\begin{split}
    \frac{dy}{dr} &= \frac{1}{1-v^2} \left[ - \frac{(1-v^2)^{1/2}}{(1-2GM/r)^{1/2} } \, \frac{d}{dr} \left(\frac{GM}{r}\right) + \frac{(1-2GM/r)^{1/2}}{(1-v^2)^{1/2}} \, v \, \frac{dv}{dr} \right]  \\[5pt]
    &= \frac{1}{1-v^2} \left[-\frac{1}{y} \, \frac{d}{dr} \left(\frac{GM}{r}\right) + yv \, \frac{dv}{dr} \right].
\end{split}
\end{equation}

Let's begin our exercise with the \textit{heat equation}: Eq.~\eqref{eq:app:qdotGR}. Since, $dq = TdS$, and with $dQ = dq/m_N$ (per unit mass vs per baryon), one can write
\begin{gather}
    \dot Q = \frac{dQ}{d\tau} = \frac{1}{m_N} \frac{dq}{dr} \frac{dr}{d\tau} = \frac{uT}{m_N} \frac{dS}{dr}  \ \implies \ \boxed{vy\frac{dS}{dr} = \frac{\dot Q m_N}{T}} \label{eq:app:dSdrGR}
\end{gather}
which is precisely Eq.~(3) from \CF~and our Eq.~\eqref{eq:thermo:dS}, given that $y = \sqrt{\mu}$.

From Eq.~\eqref{eq:app:S}, one can then write
\begin{equation} \label{eq:app:qdotvy}
    \frac{\dot Q}{vy} = \frac{3S}{m_N} \frac{dT}{dr} - \frac{TS}{m_N} \frac{1}{\rho_b} \frac{d\rho_b}{dr},
\end{equation}
where we have used Eqs.~\eqref{eq:app:S} and \eqref{eq:app:P} to obtain the last equality. 

Next, we can deal with the \textit{Euler equation}: Eq.~\eqref{eq:app:dudrGR}. This has the same form of our Eq.~\eqref{eq:app:fluid:Euler}. Using Eqs.~\eqref{eq:app:eps}, \eqref{eq:app:rhoplusP}, and \eqref{eq:app:ysq}, we can write
\begin{equation} \label{eq:app:dudr2}
    u \frac{du}{dr} = -\frac{1}{\rho_b(1+TS/m_N)} \frac{11\pi^2}{45} T^3 \, \frac{dT}{dr} \, y^2 - \frac{GM}{r^2} = -\frac{S/m_N}{1+TS/m_N} \, y^2 \, \frac{dT}{dr} - \frac{GM}{r^2},
\end{equation}
which, using the product rule for derivatives, can be further rewritten as
\begin{equation}
    u \frac{du}{dr} = -\frac{1}{1 + TS/m_N} \, y^2 \left[ \frac{d}{dr} \left(\frac{TS}{m_N}\right) - \frac{T}{m_N} \frac{dS}{dr} \right] + \frac{d}{dr} \left(\frac{GM}{r}\right),
\end{equation}
where we have assumed that the mass $M$ supplying the gravitational potential is overwhelmingly concentrated within the PNS, allowing us to neglect the $dM/dr$ term. The $dS/dr$ term on the right-hand side (RHS) can be replaced using Eq.~\eqref{eq:app:dSdrGR}. Dividing through by $y^2$ and using $u = vy$ on the left-hand side (LHS), this becomes
\begin{equation}
    v \frac{dv}{dr} + \frac{v^2}{y} \frac{dy}{dr} =  -\frac{d}{dr} \log\left(1 + \frac{TS}{m_N}\right) + \frac{1}{1+TS/m_N} \frac{\dot Q}{vy} + \frac{1}{y^2} \, \frac{d}{dr} \left(\frac{GM}{r}\right)
\end{equation}

Plugging in Eq.~\eqref{eq:app:dydr} on the LHS and using Eq.~\eqref{eq:app:ysq} we obtain
\begin{equation}
\begin{split}
   & \frac{d}{dr} \left[\frac{v^2}{2}\right] + \frac{v^2}{1-v^2} \frac{d}{dr} \left[\frac{v^2}{2}\right] - \frac{v^2}{1-2GM/r} \,  \frac{d}{dr} \left(\frac{GM}{r}\right)   \\&= -\frac{d}{dr} \log\left(1 + \frac{TS}{m_N}\right)  + \frac{1}{1+TS/m_N} \frac{\dot Q}{vy} + \frac{1}{y^2} \, \frac{d}{dr} \left(\frac{GM}{r}\right).
\end{split}
\end{equation}

Gathering similar terms together, simplifying, and rearranging yields the result
\begin{equation}
    \boxed{vy \, \frac{d}{dr} \left[ -\frac12 \log(1-v^2) + \log\left(1 + \frac{TS}{m_N}\right)  + \frac12 \log \left(1 - \frac{2GM}{r} \right) \right] =  \frac{\dot Q}{1+TS/m_N}\,,}
\end{equation}
which is Eq.~(2) from \CF.

Finally, let us invoke the \textit{continuity equation}: Eq.~\eqref{eq:app:MdotGR}, corresponding to our Eq.~\eqref{eq:std:dn} since $\rho_b = m_N\,n$. Starting with Eq.~\eqref{eq:app:dudr2}, we can replace $S/m_N \, dT/dr$ using Eq.~\eqref{eq:app:qdotvy}, and 
$d\rho_b/dr$ using Eq.~\eqref{eq:app:MdotGR}, yielding:
\begin{equation}
\begin{split}
    u \frac{du}{dr} &= -\frac{y^2}{1+TS/m_N} \, \left[\frac{\dot Q}{3vy} + \frac{TS}{3m_N} \frac{1}{\rho_b} \frac{d\rho_b}{dr} \right] - \frac{GM}{r^2} \\
    &= -\frac{y^2}{1+TS/m_N} \, \left[\frac{\dot Q}{3vy} - \frac{TS}{3m_N} \left(\frac{2}{r} + \frac{1}{u} \frac{du}{dr} \right) \right] - \frac{GM}{r^2}\,.
\end{split}
\end{equation}

Using $u = vy$, dividing through by $y^2$, rearranging the terms, and using the expression for sound speed from Eq.~\eqref{eq:app:vsGR} gives
\begin{equation}
     \left(v - \frac{v_s^2}{v} \right) \frac{1}{y}\,\frac{d(vy)}{dr}
    =\frac{1}{r}\left[2v_s^2 - \frac{GM}{ry^2} \right] - \frac{1}{1+TS/m_N} \, \frac{\dot Q}{3vy}\,.
\end{equation}
Expanding the LHS and using Eq.~\eqref{eq:app:dydr} for $dy/dr$, one obtains:

\begin{equation}
    \left(v - \frac{v_s^2}{v} \right) \left[ \frac{dv}{dr}  + \frac{v^2}{1-v^2} \frac{dv}{dr} - \frac{v}{1-v^2} \, \frac{1}{y^2} \frac{d}{dr} \left(\frac{GM}{r}\right) \right]
    = \frac{2v_s^2}{r} - \frac{GM}{r^2 y^2} - \frac{\dot Q}{3vy\,(1+TS/m_N)}\,.
\end{equation}

Rearranging and grouping similar terms results in
\begin{equation}
\boxed{\frac{v}{1-v^2} \left(1 - \frac{v_s^2}{v^2} \right) \frac{dv}{dr}  =  \frac{1}{r} \left[
2v_s^2 - \frac{(1 - v_s^2)}{(1-2GM/r)} \frac{GM}{r} \right] - \frac{\dot Q}{3vy\,(1+TS/m_N)}\,.}
\label{eq:app:Eq3CF}
\end{equation}
where we have used Eq.~\eqref{eq:app:ysq} and again neglected the $dM/dr$ terms. {This result is similar to Eq.~(1) of \CF, \textit{but not exactly the same}}. \CF~have $TS/(3m_N)$ in place of $v_s^2$ inside the square brackets on the RHS, but as we know from Eq.~\eqref{eq:app:vsGR}, that is not entirely correct in the relativistic treatment. Since the continuity equation was explicitly imposed in the derivation of this equation, the omission of relativistic corrections to the sound speed in its numerical implementation evidently results in non-conservation of $\dot M$, as we discovered. Thus, we are able to recover Eqs.~(2) and (3) in \CF~and obtained a corrected form of their Eq.~(1), including the relativistic sound speed correction. We have verified that $\dot{M}$ is conserved when Eq.~\eqref{eq:app:Eq3CF} is employed.

\section{\textcolor{black}{Matching of Far Boundary Condition}}
\label{app:outflow:bvp}

In this Appendix, we elaborate on how we impose the far boundary condition of our BVP, namely the far pressure $P_f$. As described in Sec.~\ref{sec:outflow:bvp}, we impose $P_f$ at the merging radius $r_g$, where the NDO joins the expanding material behind the FS. This radius is defined as the point where the NDO velocity matches the Hubble-like velocity $v_h$ of the homologously expanding material, as prescribed by Eq.~\eqref{eq:FS:rg}. Consequently, unlike approaches that impose $P_f$ at a fixed outer radius, $r_g$ is not known a priori. Instead, the determination of $r_g$ and the matching of $P_f$ are carried out self-consistently within the same iterative routine. This represents an improvement over the treatments in Refs.~\cite{Friedland:2020ecy,Mukhopadhyay:2022yrd,Friedland:2023kqp}, wherein the far pressure was imposed on the steady-state solution at a fixed far radius of 10000 km.

Despite these differences, the existence of solutions to our BVP follows from the existence of solutions to the traditional BVP in which $P_f$ is imposed at a fixed radius. This is apparent from Fig.~\ref{fig:gbvp}, which shows velocity and pressure profiles for solutions of the GR steady-state outflow equations in Sec.~\ref{sec:outflow:equations}. All these outflow solutions have the same temperature and baryon density at the gain (starting) radius.

\begin{figure}
\includegraphics[width=\textwidth]{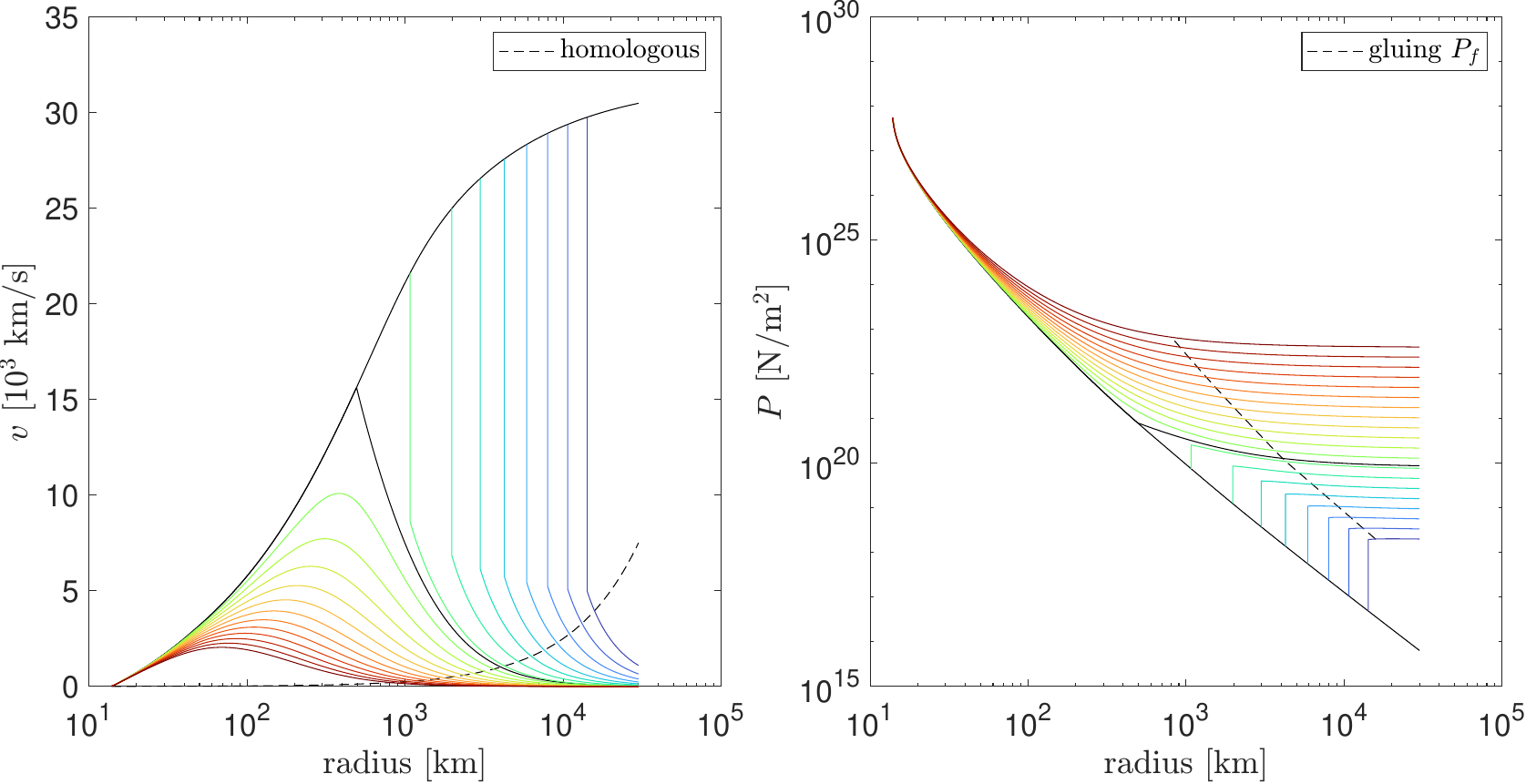}
\centering
\caption{An illustration of our BVP: velocity (left) and pressure (right). Each outflow, identified by a unique color, corresponds to a far pressure $P_f$ that is specified at the gluing radius $r_g$ where the velocity of homologous expansion matches that of the outflow. The critical (subsonic) and wind solutions are plotted in solid black. With the pressure profiles, the gluing positions $r_g$ are also shown (dashed line), from which it is clear that $P_f$ and $P_\infty$ are monotonic. The solutions in this plot were generated using neutrino parameters at $t=4$~s.}
\label{fig:gbvp}
\end{figure}

To find a solution, we assume that the intersection of $v_h$ with any physical subsonic outflow occurs at radii larger than the point where the subsonic velocity peaks. In addition, we consider only those transonic outflows  whose starting velocity of the post-shock subsonic segment is greater than $v_h$ at that radius.
These conditions are warranted by physical stellar and neutrino parameters, such as used in Fig.~\ref{fig:gbvp}. The right panel of Fig.~\ref{fig:gbvp} demonstrates that the gluing pressure $P_f$ varies monotonically with the pressure imposed at a fixed far radius, $P_\infty$. This bijection guarantees that for a given $P_f$, there exists a unique corresponding $P_\infty$, whose solution is known to exist uniquely. In practice, the shooting method of Ref.~\cite{Friedland:2020ecy} is directly available here and can be applied as already discussed in Sec.~\ref{sec:outflow:bvp}. This allows us to find a solution of the BVP for any value of $r_g$. In particular, for sufficiently far gluing positions, the pressure approaches its asymptotic value, and the outflow solution of our BVP becomes nearly indistinguishable from that obtained in the traditional case.

\section{\textcolor{black}{Relativistic Rankine-Hugoniot Conditions}}
\label{app:outflow:RH}

In this Appendix we describe how the state variables $n_2$, $u_2$ and $T_2$ behind the shock can be inferred given their values $n_1$, $u_1$ and $T_1$ upstream through the relativistic Rankine-Hugoniot (RH) conditions:
\begin{equation}
    \label{eq:app:RH:continuity}
    n_1 u_1 = n_2 u_2 \,,
\end{equation}
\begin{equation}
\label{eq:app:RH:momentum}
    (\rho_1+P_1)u_1^2 + \left(1-\frac{2GM}{r}\right)P_1 = (\rho_2+P_2)u_2^2 + \left(1-\frac{2GM}{r}\right)P_2\,,
\end{equation}
\begin{equation}
\label{eq:app:RH:energy}
    \left(\frac{\rho_1+P_1}{n_1}\right)^2\left(1+u_1^2-\frac{2GM}{r}\right) = \left(\frac{\rho_2+P_2}{n_2}\right)^2\left(1+u_2^2-\frac{2GM}{r}\right).
\end{equation}

Assuming radiation domination over the non-relativistic baryon gas (which usually holds by an order of magnitude in pressure), an appropriate approximation at such radii, Eqs.~\eqref{eq:app:RH:continuity} and \eqref{eq:app:RH:momentum} yield a quadratic equation of $n_2(T_2)$. Adopting the positive root, the energy equation (\ref{eq:app:RH:energy}) can then be applied to numerically solve for $T_2$.   

One may consider these relativistic revisions of the Newtonian RH conditions a rather tedious exercise, as the outflow is expected to become non-relativistic at such large radii. It is true that the gravitational correction $2GM/r$ at the sonic point is small, usually a few percents. The domination of mass energy density over radiation is also usually by about two orders of magnitude. Yet, for the RH conditions, the difference between Newtonian and relativistic sound speeds can be significant close to the sonic point. For simplicity, assuming $g_*$ is constant across the shock, the velocity discontinuity from the Newtonian conditions is \cite{Friedland:2020ecy}
\begin{equation} \label{eq:app:RH:Newton}
    \frac{v_2}{v_1} = \frac{\rho_1}{\rho_2} = \frac{{2T_1S_1}/{m_N}+v_1^2}{7v_1^2} = \frac{6v_s^2+v_1^2}{7v_1^2}\,,
\end{equation}
assuming Newtonian sound speed (with constant $g_*$) in radiation domination. Since before the shock, $v_1>v_s$, velocity is always diminished (density always increased) across the shock. Yet, the relativistic sound speed (c.f. Eq.~(\ref{eq:state:vs2:form}), with $P_b\to 0$ in radiation domination) is smaller than the Newtonian due to the incorporation of radiation mass. In this case, $2T_1S_1/m_N > 6v_s^2$, and one may thus obtain with Eq.~(\ref{eq:app:RH:Newton}) an unphysical positive jump in velocity and negative jump in baryon density, for termination shocks sufficiently close to the sonic point. This occurs in the near-critical regime where $v_1$ is only slightly larger than $v_s$. In fact, in computing outflows with termination shocks discussed in Sec.~\ref{sec:outflow:results:supersonic}, this issue of a positive velocity discontinuity was not infrequently encountered close to the sonic point, persisting outward for about 1\% of the sonic radius, which is still more than 20\% of the PNS radius. As Ref.~\cite{Friedland:2020ecy} argued, SN outflows can often be near-critical, so the regime close to the sonic point is not entirely an idle concern. This issue is resolved when using the relativistic conditions above.

At this juncture, we also observe that the implementation of three species of effective RDF, as described in App.~\ref{app:gstar}, is necessary to address the same issue. Had one implemented only one species of $g_*(T)$, the condition in Eq.~\eqref{eq:app:RH:Newton} still follows, and a positive velocity jump can also occur close to the sonic point and persist over an even larger radius than the GR effect. The impact of using $g_*(T)$, contra $g_*={\rm const}$, for transonic solutions with termination shocks is itself considerable (c.f. App.~\ref{sec:corr:gstar}), as the vacuum solution already differs significantly.

\section{Variable Relativistic Degrees of Freedom} \label{app:gstar}
\subsection{Formalism}

In Ref.~\cite{Friedland:2020ecy}, variable RDF, generically referred to as $g_*(T)$, was introduced in modeling the evolution of steady-state outflows. Given the temperature variation from a few to a fraction of MeV, this is a rather natural approach compared with the manual cutoff of entropy evolution when temperature decreases below the electron mass~\cite{Qian:1996xt}. The latter approximation also does not incorporate the effect of $g_*$'s variation itself on determining the outflow, as for instance enters into the value of sound speed.

The treatment in Ref.~\cite{Friedland:2020ecy}, nevertheless, was not fully consistent. There are different species of $g_*(T)$'s corresponding to different state functions. In our hydrodynamic equations, the three that enter are those for internal energy, pressure, and entropy, designated respectively as $g_*^\rho$, $g_*^P$, and $g_*^S$. That these are indeed different, as shown in the left panel of Fig.~\ref{fig:gstar}, is necessary for the thermodynamic identity $dU=TdS-PdV$, the conservation of $\dot{M}$, and the consistency of the Rankine-Hugoniot conditions across termination shocks, as important examples. As apparent from the same figure, $g_*^\rho>g_*^S>g_*^P$ at all temperatures.

\begin{figure}
\includegraphics[width=\textwidth]{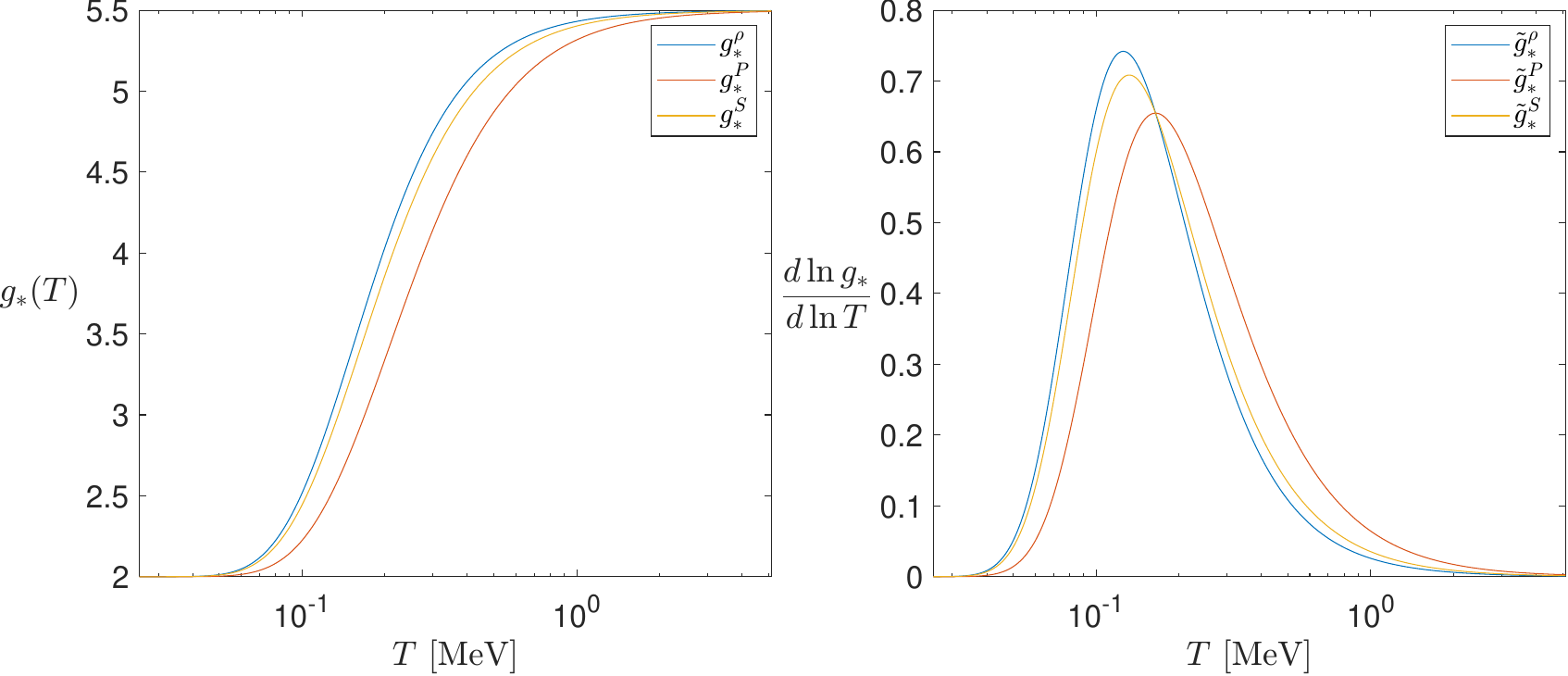}
\centering
\caption{Effective RDF (left) for different state functions and their logarithmic derivatives (right). }
\label{fig:gstar}
\end{figure}

In this Appendix, we specify the definition of the different $g_*$'s and key relations between them, which were implemented to produce our results in the paper. We assume $\mu_{e^\pm}=0$. Such an assumption is acceptable at the PNS surface where $S_{\rm r}\approx 6$ and it is excellent for the body of the NDO. After bounce, the post-shock region rapidly becomes low-density and non-degenerate. In this regime, $e^\pm$ pairs are created thermally via $\gamma + \gamma \leftrightarrow e^- + e^+$, leading to $n_{e^-} \approx n_{e^+}$. Thus, the net electron number is small, implying $\mu_{e^\pm} \approx 0$. Practically, this allows us to separate the computations of outflows from nucleosynthesis \cite{Qian:1996xt}. From the Fermi-Dirac distributions of $e^+$, $e^-$ populations, it follows that the state functions of total radiation, ($\gamma, e^+,e^-$), are \cite{Kolb:1990vq, Husdal:2016haj}:
\begin{equation}
    \rho_{\rm r} = \frac{\pi^2}{30}g_*^\rho T^4, \quad P_{\rm r} = \frac{\pi^2}{90}g_*^P T^4, \quad S_{\rm r}=\frac{2\pi^2}{45} g_*^S \frac{T^3}{n}\,,
\end{equation}
where
\begin{align}
    g_*^\rho(T) &\equiv 2 + \frac{60}{\pi^4}\int_{z_e}^\infty \frac{u^2\sqrt{u^2-z_e^2}}{1+e^u} du\,, \\
    g_*^P(T) &\equiv 2 + \frac{60}{\pi^4}\int_{z_e}^\infty \frac{\left(u^2-z_e^2\right)^{3/2}}{1+e^u} du\,,
\end{align}
using a dimensionless variable $z_e\equiv m_e/T$. The Euler relation $U = TS-PV$ for radiation then yields
\begin{equation}
    g_*^S(T) = \frac{1}{4} (3g_*^\rho+g_*^P) = 2 + \frac{15}{\pi^4}\int_{z_e}^\infty \frac{(4u^2-z_e^2)\sqrt{u^2-z_e^2}}{1+e^u} du.
\end{equation}
There exists another independent derivative relation (Gibbs-Duhem) between these species of $g_*$'s:
\begin{equation}
    \frac{dg_*^P}{dT} = \frac{3}{T}\left(g_*^\rho-g_*^P\right) \, ,
\end{equation}
which ensures the validity of the first law of thermodynamics. This can be directly verified from the integral expressions above. Thus, of the three $g_*$'s and their three derivatives, only three integrals need to be independently evaluated, e.g. $g_*^\rho$, $g_*^P$, $dg_*^\rho/dT$.

A logarithm derivative recurs in the hydrodynamic equations due to the differentiation of state functions:
\begin{equation}
    \tilde{g}_* \equiv \frac{d\ln g_*}{d\ln T} = \frac{T}{g_*} \frac{dg_*}{dT}
\end{equation}
for the $g_*$'s of different thermal state functions. For reference, these logarithmic derivatives are plotted in the right panel of Fig.~\ref{fig:gstar}.

\subsection{\textcolor{black}{Impact on hydrodynamics}} \label{sec:corr:gstar}

Here, we examine in greater detail the impact of variable $g_*(T)$ presented in Sec.~\ref{sec:outflow:gstar}.
When incorporating $g_*(T)$, the zeroth-order correction is to employ $g_*^P(T)$ in the matching of the far pressure $P_f$, without implementing $g_*(T)$ in the evolution equations. Otherwise, assuming $g_*=5.5$ leads to an underestimated terminal temperature. Thus, we compare outflows by enforcing the same far temperature $T_f$, rather than the same far pressure $P_f$. In addition, to avoid small higher-order shifts in the gluing radius $r_g$ caused by modest velocity differences we impose $T_f$ at a fixed far radius of $10^4$~km. Finally, to highlight the effect of $g_*(T)$ alone, we assume radiation domination.

As shown in the upper panels of Fig.~\ref{fig:gstarbvp-both}, for deeply subsonic outflows (e.g., our benchmark $18~M_\odot$ model), the influence of $g_*(T)$ is small in temperature and entropy (usually sub-percent level for the latter) while more noticeable in velocity and baryon density. The negligible change in entropy is reasonable, since the entropy development in the heating region is effectively concluded before the temperature decreases below the electron mass. In turn, the terminal baryon density is reduced by the fraction $g_*^S(T_f)/5.5$, since the far temperature $T_f$ is unchanged. An interesting consequence of the temperature dependence of $g_*(T)$ is a modest acceleration of the outflow once the temperature drops below the electron mass, where $g_*$ and its logarithmic derivative deviate from their constant values of 5.5 and 0, respectively. As shown in the upper-right panel of Fig.~\ref{fig:gstarbvp-both}, this effect increases the peak velocity of deeply subsonic outflows by several tens of percent---roughly an order of magnitude smaller than the enhancement produced by GR effects (cf. Fig.~\ref{fig:outflows:gr}). As the velocity at the gain radius approaches the critical value for the wind solution, the outflow with variable $g_*$ can have a peak subsonic velocity that is a few times higher than that in the constant-$g_*$ scenario. This is due to the correction of sound speed as given in Eq.~\eqref{eq:state:vs2:form}. Assuming radiation domination,
\begin{equation}
    v_s^2 = \frac{P_{\rm r}}{\rho+P_{\rm r}} \Bigg( \frac{4+\tilde{g}_*^P}{3+\tilde{g}_*^S} \Bigg) \,.
\end{equation}
For constant $g_*$, the fraction in parenthesis is ${4}/{3}$. In proper variation with temperature, it is usually less than 4/3 at lower temperatures (see Fig.~\ref{fig:gstar}). This reduction of sound speed produces a larger velocity gradient in Eq.~\eqref{eq:std:du} for subsonic outflows. Despite this considerable boost in velocity, the entropy, even in the case of near-criticality, is not appreciably altered for subsonic outflows, since the acceleration occurs after the outflow has exited the heating region for entropy growth.

The impact of $g_*(T)$ is more significant for the wind and outflows with termination shocks. As an example, the lower panels of Fig.~\ref{fig:gstarbvp-both} show GR outflows for the 12.75 $M_\odot$ progenitor at $t=4$~s, where $T_f$ is imposed at $10^4$~km. This system is near critical. The difference between the velocity of the outflow with $g_*(T)$ (solid blue) and that of $g_*=5.5$ (solid red) is prominent, with the former undergoing a supersonic transition. As shown in the lower right panel, the temperature with proper evolution of the effective RDF is also decreased at large radii due to the sonic transition, with minimal modification at smaller radii. The entropy profile is not significantly affected by incorporating a variable $g_*$, apart from the jump at the termination shock, since again entropy growth ceases as the outflow leaves the heating region, prior to the divergence in velocities.
These qualitative features of the sonic transition due to $g_*(T)$ and the percent-level difference in entropy hold also if we imposed $P_f$ at the radius $r_g$ where the velocity matches that of the homologous expansion, as described in Sec.~\ref{sec:outflow:bvp}.

Of more general interest are the critical and wind solutions, which are independent of $P_f$. The velocity of the critical solution (dotted) is not much affected by the incorporation of $g_*(T)$, as for entropy. In contrast, temperature is significantly increased, $\sim 50\%$ asymptotically. This translates into a doubling of the asymptotic (radiation) pressure, compared to assuming $g_*=5.5$ throughout. Thus, an outflow that is subsonic under constant $g_*=5.5$ can become transonic when the variation of RDF with temperature is properly taken into account. Baryon density of the outflow with $g_*(T)$ is increased at large radii to mostly compensate for the rise of temperature to yield approximately the same entropy.

For the winds, there is a notable developing velocity gain from incorporating $g_*(T)$. Temperature is also increased, though by a more modest extent ($\sim 30\%$ here). The entropies of the winds are, of course, effectively identical with the critical solutions. Thereby, in this case, the baryon density of the wind with variable RDF is actually lower, given the smaller temperature boost, than that of the wind assuming $g_*=5.5$. It is worth observing that the difference in temperature between the critical solutions as well as the divergence in velocity between the winds become prominent around the sonic point. At this radius, the temperature is about 0.1 MeV, significantly below the electron mass where $g_*(T)$ differs from 5.5 by a sizable amount. Also, between 0.1 and 0.2 MeV, the logarithmic derivatives $\tilde{g}_*$'s, which enter into the sound speed and fluid equations, attain their peak values (see Fig.~\ref{fig:gstar}), while they are zero when assuming a constant $g_*$. Thus it is sensible that $g_*(T)$ impacts the hydrodynamics in the near-critical regime of these outflows.

As noted in Sec.~\ref{sec:outflow:RH} and further discussed in App.~\ref{app:outflow:RH}, properly implementing different species of $g_*$'s for corresponding state functions $(\rho_{\rm r}, S_{\rm r}, P_{\rm r})$ is essential for the consistency of the Rankine-Hugoniot conditions across termination shocks.

\section{\textcolor{black}{Processes of Neutrino Heating and Cooling}}\label{app:heatcool}

Here we report the full expression for processes of neutrino heating and cooling. Following Refs.~\cite{Qian:1996xt, Otsuki:1999kb}, we consider free nucleon heating and cooling $\nu_e+n\leftrightarrow p + e^-$ and $\bar{\nu}_e+p\leftrightarrow n + e^+$, heating from elastic neutrino-electron scattering $e+\nu\rightarrow e+\nu$, and cooling through $e^+e^-$ annihilation $e^-+e^+\rightarrow \nu+\bar{\nu}$. These rates are computed making the following approximations~\cite{Qian:1996xt, Otsuki:1999kb}: (i) ignoring initial state chemical potentials as well as final state Pauli blocking, (ii) assuming that momentum transfer in these processes is small compared to the nucleon mass (so that the thermal motion of nucleons can be ignored in computing the rates), and (iii) assuming relativistic electrons and positrons (e.g., neglect electron mass\footnote{Strictly considered, if we retain the electron mass in our computations of effective RDF \textcolor{black}{as it appears} in the hydrodynamic equations in Sec.~\ref{sec:outflow:equations}, for consistency, we should evaluate the rates of processes involving $e^+e^-$ without making the ultra-relativistic assumption. However, we did not undertake this but verified that scaling these heating rates with $g_*(T)$ only altered the outflow at the percent level.} as well as the neutron-proton mass difference). These approximations are reasonable in the Kelvin-Helmholtz cooling phase, in the regions close to the PNS where most of the heating occurs~\cite{Qian:1996xt}.

\textcolor{black}{Here, we provide explicit expressions for the different contributions to the specific heating rate per unit mass, $\dot{Q}$, defined in Eq.~\eqref{eq:Qdot}}. The largest contribution to heating is given by the (anti)neutrino absorption by free nucleons $\nu_e+n\to p + e^-$ and $\bar{\nu}_e+p\to n + e^+$, described by 
\begin{equation}
    \dot{Q}_{\nu N}\approx 9.65\,X_N\,N_A\left[\left(1-Y_e)\right)L_{\nu_e,51}\varepsilon_{\nu_e}^2 + Y_e L_{\bar{\nu}_e,51}\varepsilon_{\bar{\nu}_e}^2\right]\,\frac{1-g_{1}(r)}{R_{\nu6}^2} \Phi(r)^6~{\rm MeV}~{\rm s}^{-1}~{\rm g}^{-1}\,,
\label{eq:app:nuN}
\end{equation}
where the first and second terms in parentheses refer to processes involving $\nu_e$ and $\bar{\nu}_e$, respectively. Additionally, $R_{\nu 6}$ is the neutrinosphere radius in units of $10^{6}$~cm, assumed to be equal to the PNS radius, $\varepsilon_i$ is the neutrino energy in MeV defined as $\varepsilon_i=(\langle E_i^3\rangle/\langle E_i\rangle)^{1/2}$, with $\langle E_i^n\rangle$ being the $n$-th energy moment of the neutrino ($i=\nu_e$) and antineutrino ($i=\bar{\nu}_e$) energy distribution, $N_A$ is the Avogadro number, $Y_e$ is the electron fraction, $L_{i,51}$ is the (anti)neutrino luminosity in units of $10^{51}$~erg~s$^{-1}$. We use values of $L_i$ and $\varepsilon_i$ at a reference radius $R_{\rm ref} = 500$~km~\cite{Huedepohl:2009wh}. Finally, $X_N$ is the free
nucleon mass fraction, assumed to be~\cite{Qian:1996xt, Woosley1992}
\begin{equation}
    X_N = \min\left[1,~ 828~\frac{T^{9/8}_{\rm MeV}}{ \rho^{3/4}_8}~\exp{\Big(- \frac{7.074}{T_{\rm MeV}} \Big)}\right]\,,
\label{eq:app:XN}
\end{equation}
where $T_{\rm MeV}$ is the temperature in MeV, $\rho_8$ in units of $10^8 \ {\rm g/cm^3}$. This allows us to take into account the transition from nucleons to $\alpha$-particles, which effectively suppresses the neutrino-nucleon interactions as the targets are depleted.

Two relativistic correction factors are included in Eq.~\eqref{eq:app:nuN}. The first one, $1-g_{1}(r)$, is the geometrical factor representing the effect of  geometric dilution, altered by the bending of null trajectories in a Schwarzschild geometry, with $g_{1}(r)$ given by
\begin{equation}
    g_1(r) = \left[1-\left(\frac{R_\nu}{r}\right)^2\frac{1-2GM/r}{1-2GM/R_\nu}\right]^{1/2}\,,
\end{equation}
where the ratio $(1-2GM/r)/(1-2GM/R_\nu)$ quantifies the effect of bending. Due to the massive PNS, neutrinos from behind it can also reach a mass element located at $r$ by virtue of the curved metric. As expected, $g_1=1$ in Newtonian geometry. 
Moreover, we define the blueshift factor in the Schwarzschild geometry as
\begin{equation}\label{eq:app:Phi}
    \Phi(r) = \sqrt{\frac{1-2GM/R_{\rm ref}}{1-2GM/r}}\,,
\end{equation}
which becomes unity in the Newtonian geometry. The second contribution to heating comes from neutrino and antineutrino scattering by electrons and positrons $e+\nu\rightarrow e+\nu$. Its rate is given by
\begin{equation}
    \dot{Q}_{\nu e} \approx 2.17 N_A \frac{T_{\rm MeV}}{\rho_8}\left(L_{\nu_e,51}\,\epsilon_{\nu_e} + L_{\bar{\nu}_e,51}\,\epsilon_{\bar{\nu}_e}+\frac{6}{7}\,L_{\nu_x,51}\,\epsilon_{\nu_x} \right)\frac{1-g_1(r)}{R_{\nu 6}^2} \Phi(r)^5~{\rm MeV}~{\rm s}^{-1}~{\rm g}^{-1}\,,
\end{equation}
where $\epsilon_i = \langle E_i^2 \rangle / \langle E_i \rangle$, ($i=\nu_e,\,\bar{\nu}_e,\,\nu_x$) is in MeV and we assume the same contribution from non-electron (anti)neutrinos $\nu_x=\nu_\mu,\,\bar{\nu}_\mu,\,\nu_\tau,\,\bar{\nu}_\tau$. Note that the redshift/blueshift correction to the heating terms $\dot{Q}_{\nu N}$ and $\dot{Q}_{\nu e}$ can significantly affect the heating rates, as they scale with the sixth and fifth power of the factor $\Phi$, respectively. For instance, assuming $M = 1.8~M_{\odot}$ and an $R_{\rm ref}$ of hundreds of kilometers, the value of $\Phi$ at $r = 30$~km is approximately $1.1$, leading to an enhancement of $78\%$ for $\dot{Q}_{\nu N}$ and $62\%$ for $\dot{Q}_{\nu e}$ due to blueshift.

The dominant cooling process is due to electron and positron captures by free nucleons, i.e. the inverse of the reactions in Eq.~\eqref{eq:app:nuN}. Its rate is given by
\begin{equation}
    \dot{Q}_{eN} \approx 2.27\,X_N\,N_A T_{\rm MeV}^6~{\rm MeV}~{\rm s}^{-1}~{\rm g}^{-1}\,,
\end{equation}
where $X_N$ is given by Eq.~\eqref{eq:app:XN}. Finally, the second contribution to cooling is represented by electron-positron pair annihilation into neutrino-antineutrino pairs of all flavors, approximately described by
\begin{equation}
    \dot{Q}_{e^+ e^-} \approx 0.144~N_A\,\frac{T_{\rm MeV}^9}{\rho_8} ~{\rm MeV}~{\rm s}^{-1}~{\rm g}^{-1}\,.
\end{equation}
Taking the aforementioned heating and cooling processes the net heating rate $\dot{Q}$ is given by
\begin{equation}
    \dot{Q}=\dot{Q}_{\nu N} + \dot{Q}_{\nu e} - \dot{Q}_{e N} - \dot{Q}_{e^+ e^-}\,. 
\end{equation}
\textcolor{black}{In the equation for the total $\dot{Q}$, we neglect the heating contribution stemming from neutrino-antineutrino pair annihilation into electron-positron pairs given its subdominant role. For the conditions relevant to our study, we checked that this process contributes only a few percent of the total heating rate at the PNS surface and declines very rapidly with radius thereafter, consistent with Ref.~\cite{Otsuki:1999kb}.}

\section{Methods: System Setup with {\tt SkyNet}}\label{app:setup-skynet}
 
In this Appendix, we provide a more detailed description of the set up of our nucleosynthesis calculation with the open source reaction network {\tt SkyNet}~\cite{Lippuner:2017tyn,skynet}. We adopt the majority of nuclear reaction rates from the {\tt Reaclib} library, version 2.2~\cite{reaclib2010,reaclibweb}, with two exceptions: (1) modern, medium-enhanced triple-$\alpha$ reaction rates from Ref.~\cite{Beard:2017jpg} are incorporated, as per Refs.~\cite{triplealpha, Jin:2020, Friedland:2023kqp}, and (2) the correct rate coefficient for $^{92}$Nb $\to$ $^{92}$Mo beta decay, consistent with the measured upper limit on this branch, is used, as discussed in Ref.~\cite{Friedland:2023kqp}.

Enabling the \lq self-heating\rq\ and \lq neutrino heating\rq\ options in {\tt SkyNet} allows it to recompute the temperature (or equivalently, the entropy) at each step, given the density profile and an initial temperature as inputs. The \lq self-heating\rq\ option takes into account the nuclear binding energy released as the reaction network is evolved. However, since we are already providing $T$ and $\rho$ profiles as inputs from the outflow calculations, 
we have chosen to turn off the self-heating and neutrino heating options in {\tt SkyNet} for the results shown here.
We have checked that the input entropy $S$ from the outflow solutions are in good agreement with the entropy per baryon computed by {\tt SkyNet}, and consequently, the nucleosynthetic yields are also not qualitatively different. 

To be consistent with the steady-state outflow solutions found in Sec.~\ref{sec:outflow}, we have modified {\tt SkyNet} to account for the blueshift/redshift of neutrino temperatures and luminosities, relative to those at a reference radius $R_\text{ref}$. For the {\tt SkyNet} runs requiring a blueshift (see Sec.~\ref{sec:8cases} for details), we implement the blueshift factor
\begin{equation}
    \Phi(r)=\sqrt{\frac{1-2GM/R_{\rm ref}}{1-2GM/r}} \approx \sqrt{\frac{1}{1-2GM/r}}\,,
\end{equation}
where we have taken $R_{\rm ref}=500$~km (where the neutrino luminosities are extracted) in Eq.~\eqref{eq:app:Phi}. Since the luminosities go as four powers of the neutrino energy, the input luminosities near the PNS are enhanced by $\Phi^4$ in {\tt SkyNet} whenever there is blueshift implemented in the models. Analogously, the neutrino temperatures $T_{\nu_i}$ are enhanced by one power of $\Phi$. 

We start each nucleosynthesis run at $T\approx 2.5$~MeV ($\sim$ 27 GK), in conditions of NSE. The abundance $Y_{A,Z}$ of each species is computed using an NSE calculation for $T>9$~GK, while at lower temperature {\tt SkyNet} switches to a full network calculation. At times beyond those provided in the input trajectory files, {\tt SkyNet} extrapolates the time evolution of the outgoing ejecta by assuming $T(t) \propto t^{-2/3}$ and $\rho(t) \propto t^{-2}$ which follow from mass flow conservation ($r^2\rho u = {\rm constant}$) and entropy conservation ($T^3/\rho = {\rm constant}$) in the homologous expansion phase ($r\propto t$, $u={\rm constant}$) after the outflow solution~\cite{Arcones:2006,Wanajo:2010mc}. Finally, the nucleosynthesis calculation ends $10^{9}$~s after $t_{\rm launch}$ when most beta decays have occurred and (meta)stable isotopes have been produced.

\section{\textcolor{black}{Baryonic Gas Component}}\label{sec:baryonic-gas}

In the numerical results concerning GR effects presented in the main text, we assumed radiation dominates in the thermal ensemble. Though a pertinent approximation for the body of the NDO, the baryon gas constitutes a coequal thermal component in the plasma around the gain radius. With the radiation entropy $S_{\rm r}=6$ at the PNS as we have employed, radiation pressure and internal energy somewhat exceed those of the baryon gas. Yet, since here is also the region of intense heating that determines entropy development, it is plausible that the incorporation of the baryonic gas has a formative effect on the character of the outflow.
The other place in our computations we assumed radiation domination is in computing the far pressure from the progenitor profile assuming homologous expansion. In this cold material, radiation pressure is comparable to the baryonic component, so $P_f$ will be increased. This latter correction of $P_f$, which renders outflows more subsonic, has a subdued effect for the outflows and nucleosynthesis of the 18 $M_\odot$ model, though for outflows with near-critical far pressures, such as for the 12.75 $M_\odot$ progenitor, this inclusion can proscribe or delay a supersonic transition. More influential and interesting is the incorporation of the baryon gas in the evolution equations themselves. 

Here, we study the impact of incorporating the baryon gas on subsonic outflows, based on our benchmark 18 $M_\odot$ progenitor model, which has comparatively optimal yields. As noted in Sec.~\ref{sec:outflow:equations}, we assume that the baryon gas is \textit{monatomic}, without tracking the composition of nuclides. This exaggerates its contribution at later stages of the outflow and thus tends to slightly overstate the impact of the baryon gas as presented here. The GR outflows at $t=2.5$~s, with and without including the monatomic non-relativistic gas (both in evolution and in $P_f$) are presented in Fig.~\ref{fig:profiles-idgas}, respectively in solid black and red. 
\begin{figure} 
\includegraphics[width=\textwidth]{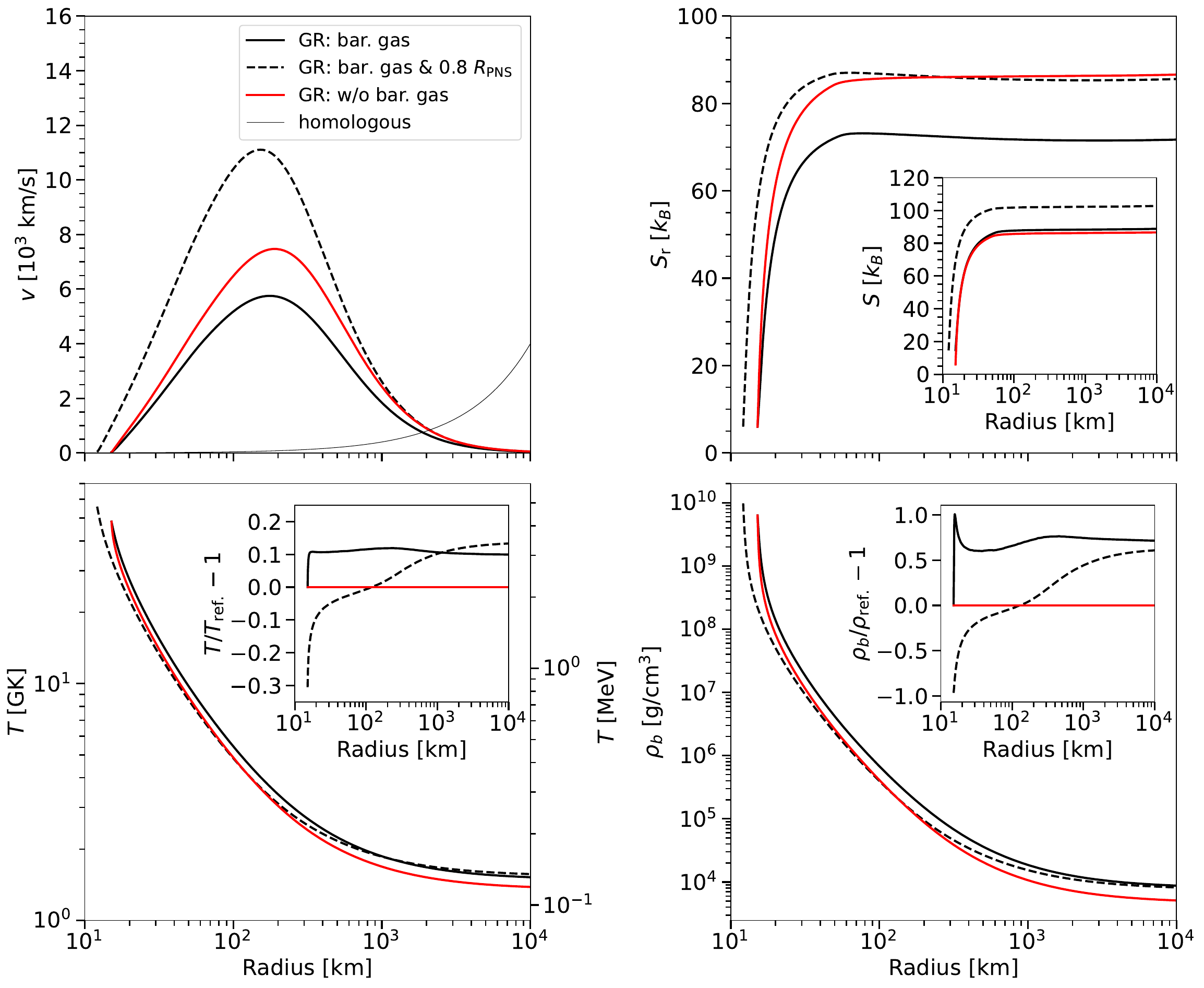}
\centering
\caption{Steady-state outflow for our benchmark $18~M_\odot$ progenitor model at $t=2.5~{\rm s}$ incorporating the baryon gas (in black) in the evolution equations and in the far pressure $P_f$, compared with our result assuming radiation domination (in red). Radiation entropy (top-right panel) is decreased with the introduction of the baryon gas, while total entropy (top-right inset) is not appreciable altered. Heuristically, we show that a $20\%$ reduction of the PNS radius (dashed black) can approximately restore $S_{\rm r}$ to its former value assuming radiation-domination. For both temperature and density profiles, the insets show the relative differences using the radiation-dominated outflow as the reference. The thin black line in the upper left panel represents the homologous expansion velocity $v_h$ defined in Eq.~\eqref{eq:FS:homologous}.} 
\label{fig:profiles-idgas}
\end{figure}
The most interesting result is for entropy. Whereas one would have naively expected an addition of $\sim$ 10 from the baryon $S_b$, the total entropy $S$ is only minimally increased ($\approx 2$ here). This means that the total entropy deposition is less when the baryon gas is included. From the figure, temperature is increased while velocity is decreased, with the latter changing by a larger factor. Yet, the total heating rate is also considerably reduced for the outflow with baryon gas, from more than $100\%$ immediate to the PNS to $\sim 30\%$ around the peak of heating. This reduction can be attributed to stronger nucleon cooling via electron and positron captures due to a higher temperature and weaker neutrino-electron scattering from a higher baryon density. Collected, $\dot{q}/uT$ is decreased for a curtailed growth of entropy, as in Eq.~\eqref{eq:thermo:dS}.

Still, what precisely causes this set of significant modifications when the baryon gas is involved? Sure, the radiation-dominated outflow has a lower $P_f$ at gluing since the baryonic contribution is ignored. Including this in $P_f$ leaves intact all the qualitative features observed above and the quantitative results are not really different. One might expect that the boost of sound speed from the baryonic component would reduce velocity (c.f. Eq.~\eqref{eq:std:du}). This would certainly be so for an initial value problem and still has some effect at smaller radii as a BVP. Yet, it turns out that the dominant factor that reduces velocity is $\Pi_2^{{\rm r} b}$ which appears in Eq.~\eqref{eq:std:dT}, controlling temperature evolution. Artificially taking $P_b\to 0$ in the $\Pi_2^{{\rm r} b}$ that appears in that equation (with no further modification) restores the velocity to the radiation-dominated values (with a slight shift to larger radii). Radiation entropy, however, does not significantly recover. On the other hand, demoting the sound speed in Eq.~\eqref{eq:std:du} to its radiation-dominated value, which also involves changing $\Pi_2^{{\rm r} b}$ (here we modify only in Eq.~\eqref{eq:std:du} but not in Eq.~\eqref{eq:std:dT}), radiation entropy is restored but the velocity deficit remains. If both of these substitutions are adopted simultaneously, then velocity becomes almost indistinguishable from that of the radiation-dominated outflow, and the temperature and density profiles also essentially agree. It follows that the radiation entropy is no longer reduced, and the total entropy is simply enhanced by $S_b$.

These identifications, obtained from rather mechanical manipulations, do not offer a qualitative explanation, part of the difficulty being due to the complex structure of our system as a BVP. Nonetheless, we want to observe that the corrections from incorporating the baryon gas can be attributed to the factors of sound speed and the ratio $\Pi_2^{{\rm r} b}$ related to the adiabatic derivative ${dT}/{dn}$.

\begin{figure}
\centering
  \includegraphics[width=0.80\columnwidth]{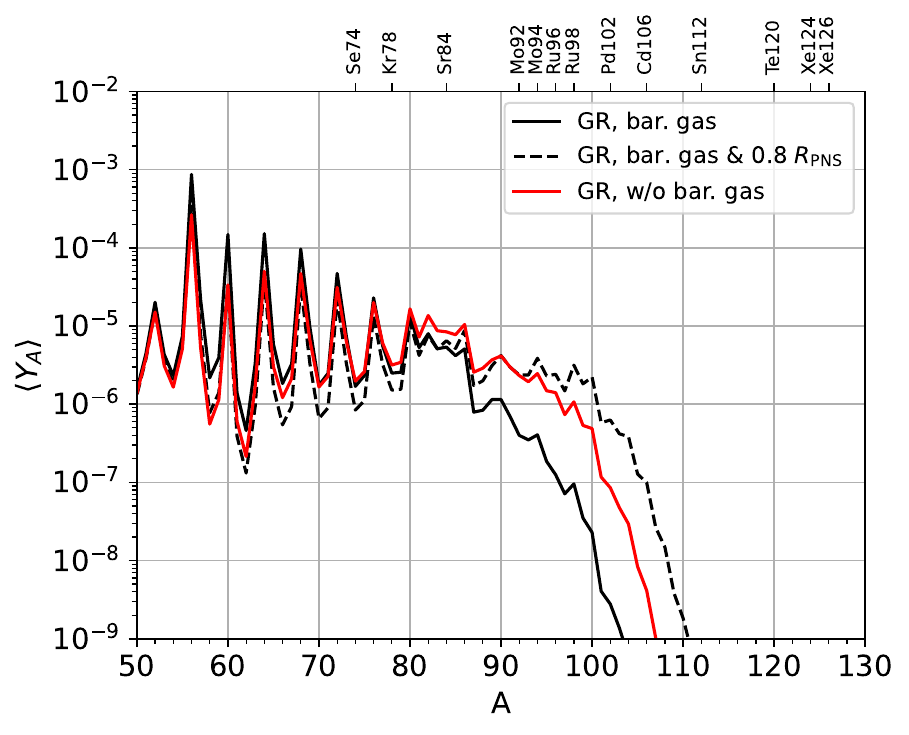}
  \caption{Time-averaged yields $\langle Y_A \rangle$ as a function of the mass number $A$ for our benchmark $18~M_\odot$ progenitor model when the baryon-gas component is included in the computation of the outflow. Black lines correspond to when the contribution of the baryon gas  is included, while red line corresponds to assuming radiation domination. For the dashed black line we assume a PNS radius reduced by $20\%$.}
  \label{fig:YAidgas}
\end{figure}

Nucleosynthesis favors a high value of radiation entropy rather than total entropy. Quantitatively, we observe a deterioration of integrated yields in Fig.~\ref{fig:YAidgas}. Indeed, when including the baryonic gas, yields are strongly reduced in the atomic mass window $90\leq A \leq 100$, by a factor $\sim 5$ at $A=90$ and by more than one order of magnitude at $A=100$. However, the yields are influenced by multiple parameters. Here, as an explicit example we consider the impact of reducing the PNS radius when including the baryon gas component. A more compact PNS would naturally increase the entropy, counterbalancing the reduction in $S_{\rm r}$ related to the inclusion of the baryonic component. Indeed, different EoS predict different PNS radii for a given PNS mass. Here, for illustration, we reduce the PNS radius by $20\%$, which suffices to restore the radiation entropy for the $18~M_\odot$ progenitor model, as shown in the outflow depicted in dashed black in Fig.~\ref{fig:profiles-idgas}. In turn, Fig.~\ref{fig:YAidgas} demonstrates that the yields obtained with this heuristic prescription (in dashed black) are similar to the ones obtained assuming radiation domination (solid red line), with a factor of a few difference in the production of $\Mo$ and $\Ru$. We checked also that the production of $\Nb$ recovers in this case. It is worth mentioning that this example stresses that multiple inputs may affect the yields, and therefore a negative feedback coming from the change in one aspect of the model may be counterbalanced by changing another feature. A more comprehensive analysis of the impact of the baryon gas component on the yields is left for future work.

\section{\textcolor{black}{Observed Solar Isotopic Mass Fractions}}
\label{app:solar-ab}
To compute the production factor of an isotope $(A,Z)$
\begin{equation}
    f_{A,Z} = \frac{\langle X_{A,Z} \rangle}{X_{A,Z}^\odot}\,,
\end{equation}
we need to evaluate its observed mass fraction in the solar system $X_{A,Z}^\odot$. Following Ref.~\cite{Friedland:2023kqp}, we derive 
$X_{A,Z}^\odot$ from the isotopic abundances $N_{A,Z}^\odot$ tabulated in Ref.~\cite{Lodders:2003}. These abundances are obtained from meteoritic measurements of carbonaceous chondrites and reported on the cosmochemical scale in which the silicon abundance is normalized to $N_{\rm Si}^\odot\equiv10^6$. 

Because the tabulated values are number abundances, they must be converted to mass fractions. To perform this conversion, we use hydrogen as a reference element. Specifically, we use as normalization factor the ratio between the solar hydrogen mass fraction ($X_{\rm H}^\odot=0.7110$) and its measured cosmochemical-scale number abundance (${N}^\odot_{{\rm H}}=2.431\times 10^{10}$)~\cite{Lodders:2003}. This yields
\begin{equation}
    X_{A,Z}^\odot = A~N_{A,Z}^\odot \Bigg( \frac{X_{\rm H}^\odot}{N_{\rm H}^\odot} \Bigg)\,,
\end{equation}
This procedure provides a consistent set of solar-system isotopic mass fractions directly comparable to the time-averaged mass fractions obtained from the abundances $\langle Y_{A,Z} \rangle$ computed with {\tt SkyNet}.

\end{document}